\documentclass[11pt]{article}
\usepackage{amsmath, amsfonts, amssymb, amsthm, newlfont, graphicx}

\usepackage{amssymb,latexsym}
\usepackage[mathscr]{eucal}
\usepackage{setspace}
\usepackage{graphics}
\usepackage{array}

\begin{document}
\title{Integrable Generalized KdV and MKdV Equations with Spatiotemporally Varying Coefficients}
\author{Matthew Russo and S. Roy Choudhury\\  \small  Department of Mathematics, University of Central Florida, Orlando, FL  32816-1364 USA\\
Corresponding author email: choudhur@cs.ucf.edu }        
\date{\today}       
\maketitle

\centerline{ \noindent\Large\textbf{Abstract}}
A technique based on extended Lax Pairs is first considered to derive variable-coefficient generalizations of various Lax-integrable NLPDE hierarchies recently introduced in the literature. As illustrative examples, we consider generalizations of KdV equations and three variants of generalized MKdV equations. It is demonstrated that the techniques yield Lax- or S-integrable NLPDEs with both time- AND space-dependent
coefficients which are thus more general than almost all cases considered earlier via other methods such as the
Painlev\'e Test, Bell Polynomials, and various similarity methods. 

However, this technique, although operationally effective, has the significant disadvantage that, for any integrable system with spatiotemporally varying coefficients, one must 'guess' a generalization of the structure of the known Lax Pair for the corresponding system with constant coefficients. Motivated by the somewhat arbitrary nature of the above procedure, we embark in this paper on an attempt to systematize the derivation of Lax-integrable sytems with variable coefficients. An ideal approach would be a method which does not require knowledge of the Lax pair to an associated constant coefficient system, and also involves little to no guesswork. Hence we attempt to apply the Estabrook-Wahlquist (EW) prolongation technique, a relatively self-consistent procedure requiring little prior infomation. However, this immediately requires that the technique be significantly generalized or broadened in several different ways, including
solving matrix partial differential equations instead of algebraic ones as the structure of the Lax Pair is deduced systematically following the standard Lie-algebraic procedure of proceeding downwards from the coefficient of the highest derivative. The same is true while finding the explicit forms for the various 'coefficient' matrices which occur in the procedure, and which must satisfy the various constraint equations which result at various stages of the calculation. 

The new and extended EW technique whch results is illustrated by algorithmically deriving generalized Lax-integrable versions of the generalized fifth-order KdV, and MKdV equations. \\
\\

Key Words: Generalizing Lax or S-integrable equations, spatially and temporally-dependent coefficients, generalized Lax Pairs, extended Estabrook-Wahlquist method.

\section{Introduction}

     Variable Coefficient Korteweg de Vries (vcKdV) and Modified Korteweg de Vries (vcMKdV)equations have a long history dating from their derivation in various applications\cite{K1}-\cite{K10}. However, almost all studies, including those which derived exact solutions by a variety of techniques,
as well as those which considered integrable sub-cases and various integrability properties by methods such as Painlev\'e analysis, Hirota's method, and Bell Polynomials treat vcKdV equations with coefficients which are functions of the time only. For instance, for generalized variable coefficient NLS (vcNLS) equations, a particular coefficient is usually taken to be a function of $x$ \cite{K11}, as has also been sometimes done for vcMKdV equations\cite{K12}. The papers
\cite{K13}-\cite{Khawaja} are somewhat of an exception in that they treat vcNLS equations having coefficients with general $x$ and $t$ dependences. Variational principles, solutions, and other integrability properties have also been considered for some of the above variable coefficient NLPDEs in cases with time-dependent coefficients.

In applications, the coefficients of vcKdV equations may include spatial dependence, in addition to the temporal variations that have been extensively considered using a variety of techniques. Both for this reason, as well as
for their general mathematical interest, extending integrable hierarchies of nonlinear PDEs (NLPDEs) to include {\it both} spatial and temporal dependence of the coefficients is worthwhile. 

Given the above, we compare two methods for deriving the integrability conditions of both a general form of variable-coefficient MKdV (vcMKdV) equation, as well as a general, variable-coefficient KdV (vcKdV) equation. In both cases,the coefficients are allowed to vary in space AND time.

The first method employed here is based on directly establishing Lax integrability (or S-integrability to use the technical term) as detailed in the following sections. As such, it is fairly 
general, although subject to the ensuing equations being solvable. We should stress that the computer algebra involved is quite challenging, and an order of magnitude beyond that encountered for integrable, constant coefficient NLPDEs.

However, this first technique, although operationally effective, has the significant disadvantage that, for any integrable system with spatiotemporally varying coefficients, one must 'guess' a generalization of the structure of the known Lax Pair for the corresponding system with constant coefficients. This involves
replacing constants in the Lax Pair for the constant coefficient integrable system, including powers of the spectral parameter, by functions. Provided that one has guessed correctly and generalized the constant coefficient system's Lax Pair sufficiently, and this is  of course hard to be sure of 'a priori', one may then
proceed to systematically deduce the Lax Pair for the corresponding variable-coefficient integrable system.

Motivated by the somewhat arbitrary nature of the above procedure, we next attempt to systematize the derivation of Lax-integrable sytems with variable coefficients. Of the many techniques which have been employed for constant coefficient integrable systems, the Estabrook-Wahlquist (EW) prolongation technique \cite{EW1}-\cite{EW4}is among the most self-contained. The method directly proceeds to attempt construction of the Lax Pair or linear spectral problem, whose compatibility condition is the integrable system under discussion. While not at all guaranteed to work, any successful implementation of the technique means that Lax-integrability has already been verified during the procedure, and in addition the Lax Pair is algorithmically obtained. If the technique fails, that does not necessarily imply non-integrability  of the equation contained in the compatibility condition of the assumed Lax Pair. It may merely mean that some of the starting assumptions may not be appropriate or general enough.

Hence we attempt to apply the Estabrook-Wahlquist (EW) method as a second, more algorithmic, technique to generate a variety of such integrable systems with such spatiotemporally varying coefficients. However, this immediately requires that the technique be significantly generalized or broadened in several different ways which we then develop and outline, before illustrating this new and extended method with examples.

The outline of this paper is as follows. In Section 2, we briefly review the Lax Pair method and its modifications for variable-coefficient NLPDEs, and then apply it to three classes of generalized vcMKdV equations. In section 3 we consider an analogous treatment of a generalized vcKdV equation.  In section 4, we lay out the extensions required to apply the EW procedure to Lax-integrable systems
with spatiotemporally varying coefficients. Sections 5 and 6 then illustrate this new, extended EW method in detail for Lax-integrable versions of the 
MKdV and generalized Korteweg-deVries (KdV) equations respectively, each with spatiotemporally varying coefficients. We also illustrate that this generalized
EW procedure algorithmically generates the same results as those obtained in a more ad hoc manner in Sections 2 and 3.
Some solutions of the generalized vcKdV equation considered in Sections 3 and 6 are then obtained in Section 7 via
the use of truncated Painlev\'e expansions. Section 8 briefly reviews the results and conclusions, and directions for possible future work.

More involved algebraic details, which are integral to both the procedures employed here, are relegated to the appendices.

\section{Extended Lax Pair method and application to three generalized Variable Coefficient MKdV (vcMKdV) equations}

In the Lax pair method \cite{K15} - \cite{K16} for solving and determining the integrability conditions for nonlinear partial differential equations (NLPDEs) a pair of $n\times n$ matrices, $\textbf{U}$ and $\textbf{V}$ needs to be derived or constructed. The key component of this construction is that the integrable nonlinear PDE under consideration must be contained in, or result from, the compatibility of the following two linear Lax equations (the Lax Pair)
\begin{eqnarray}
\Phi_{x} &=& U\Phi \\
\Phi_{t} &=& V\Phi
\end{eqnarray}
where $\Phi$ is an eigenfunction, and $\textbf{U}$ and $\textbf{V}$ are the time-evolution and spatial-evolution (eigenvalue problem) matrices. 

From the cross-derivative condition (i.e. $\Phi_{xt} = \Phi_{tx}$) we get
\begin{equation} \label{ZCC1}
U_{t}-V_{x}+[U,V] = \dot{0}
\end{equation}
known as the zero-curvature condition where $\dot{0}$ is contingent on $v(x,t)$ being a solution to the nonlinear PDE. A Darboux transformation can then be applied to the linear system to obtain solutions from known solutions and other integrability properties of the integrable NLPDE. 

We first consider the following three variants of generalized variable-coefficient MKdV \\
(vcMKdV) equations:

\begin{equation} 
v_{t} + a_{1}v_{xxx} + a_{2}v^{2}v_{x} = 0 \label{vcMKdV1}
\end{equation}

\vspace{3mm}

\begin{equation} 
v_{t} + b_{1}v_{xxxxx} + b_{2}v^{2}v_{xxx} + b_{3}vv_{x}v_{xx} + b_{4}v_{x}^{3} + b_{5}v^{4}v_{x} = 0 \label{vcMKdV2}
\end{equation}

\vspace{3mm}

\begin{eqnarray} 
&& v_{t} + c_{1}v_{xxxxxxx} + c_{2}v^{2}v_{xxxxx} + c_{4}vv_{xx}v_{xxx} + c_{5}v_{x}^{2}v_{xxx} \nonumber \\
&& + c_{6}v_{x}v_{xx}^{2} + c_{7}v^{4}v_{xxx} + c_{8}v^{3}v_{x}v_{xx} + c_{9}v^{2}v_{x}^{3} + c_{10}v^{6}v_{x} = 0 \label{vcMKdV3}
\end{eqnarray}

These equations, which we shall always call the physical (or field) NLPDEs to distinguish them from the many other NLPDEs we encounter, will be Lax-integrable or S-integrable if we can find a Lax pair whose compatibility condition $\eqref{ZCC1}$
contains the appropriate equation ($\eqref{vcMKdV1}$, $\eqref{vcMKdV2}$ or $\eqref{vcMKdV3}$). 

One expands the Lax pair $\textbf{U}$ and $\textbf{V}$ in powers of $v$ and its derivatives with unknown functions as coefficients. This results in a VERY LARGE system of coupled NLPDEs for the variable coefficient functions in $\eqref{vcMKdV1}$-$\eqref{vcMKdV3}$. Upon solving these (and a solution is not guaranteed, and may prove to be impossible to obtain in general for some physical NLPDEs), we simultaneously obtain the Lax pair and integrability conditions on the $a_{i},b_{i}$, and $c_{i}$ for which $\eqref{vcMKdV1}$-$\eqref{vcMKdV3}$ are Lax-integrable. 

The results, for which the details are given in Appendix A, are given in the following three subsections.

\subsubsection{Conditions on the $a_{i}$}

\begin{eqnarray}
&& 6a_{1}a_{2x}^{3} - 6a_{1}a_{2}a_{2x}a_{2xx} + a_{1}a_{2}^{2}a_{2xxx} - \frac{K_{t}}{K}a_{2}^{3} + a_{2}^{2}a_{2t} - a_{2}^{3}a_{1xxx} \nonumber \\
&& + 3a_{1xx}a_{2}^{2}a_{2x} - 6a_{1x}a_{2}a_{2x}^{2} + 3a_{1x}a_{2}^{2}a_{2xx} = 0
\end{eqnarray}

where $K(t)$ is an arbitrary function of $t$.

\subsubsection{Conditions on the $b_{i}$}

\begin{eqnarray}
&& b_{3} = 2(b_{2} + b_{4}) \\
&& b_{5} = H_{1}(t)(2b_{2} - b_{4})
\end{eqnarray}
\begin{eqnarray}
&& 12b_{2x}b_{2xx}b_{4}^{2} + 12b_{2}^{2}b_{2xx}b_{4x} - 3b_{2xx}b_{4x}b_{4}^{2} + 4b_{2}^{2}b_{2xxx}b_{4} - 4b_{2}b_{2xxx}b_{4}^{2} - 24b_{2}b_{2x}^{2}b_{4x} \nonumber \\ 
&& + 24b_{2}b_{2x}b_{4x}^{2} + 12b_{2}^{2}b_{2x}b_{4xx} - 12b_{2}^{2}b_{4x}b_{4xx} + 4b_{2}^{2}b_{4}b_{4xxx} - b_{2}b_{4}^{2}b_{4xxx} - 24b_{2x}^{2}b_{4}b_{4x} \nonumber \\
&& + 6b_{2x}b_{4}b_{4x}^{2} - 3b_{2x}b_{4}^{2}b_{4xx} + b_{2xxx}b_{4}^{3} - 6b_{2}b_{4x}^{3} - 4b_{2}^{3}b_{4xxx} + 24a_{2x}^{3}b_{4} \nonumber \\
&& - 24b_{2}b_{2x}b_{2xx}b_{4} + 6b_{2}b_{4}b_{4x}b_{4xx} = 0
\end{eqnarray}
\begin{eqnarray}
&& -9600b_{1}b_{2x}^{4}b_{4x} + 9600b_{1}b_{2x}^{3}b_{4x}^{2} - 4800b_{1}b_{2x}^{2}b_{4x}^{3} + 1200b_{1}b_{2x}b_{4x}^{4} - 3840b_{1x}b_{2}b_{2x}^{4} - 240b_{1x}b_{2}b_{4x}^{4} \nonumber \\
&& + 1920b_{1x}b_{2x}^{4}b_{4} + 120b_{1x}b_{4}b_{4x}^{4} + 3840b_{1}b_{2x}^{5} - 120b_{1}b_{4x}^{5} + 7680b_{1x}b_{2}b_{2x}^{3}b_{4x} - 5760b_{1x}b_{2}b_{2x}^{2}b_{4x}^{2} \nonumber \\
&& + 1920b_{1x}b_{2}b_{2x}b_{4x}^{3} - 3840b_{1x}b_{2x}^{3}b_{4}b_{4x} + 2880b_{1x}b_{2x}^{2}b_{4}b_{4x}^{2} - 960b_{1x}b_{2x}b_{4}b_{4x}^{3} - 32b_{1xxxxx}b_{2}^{5} \nonumber \\
&& + b_{1xxxxx}b_{4}^{5} + 10b_{1xxxx}b_{2x}b_{4}^{4} - 80b_{1xxxx}b_{2}^{4}b_{4x} - 5b_{1xxxx}b_{4}^{4}b_{4x} + 80b_{1xxxxx}b_{2}^{4}b_{4} - 80b_{1xxxxx}b_{2}^{3}b_{4}^{2} \nonumber \\
&& + 40b_{1xxxxx}b_{2}^{2}b_{4}^{3} - 10b_{1xxxxx}b_{2}b_{4}^{4} + 32b_{1}b_{2}^{4}b_{2xxxxx} + 2b_{1}b_{2xxxxx}b_{4}^{4} - 16b_{1}b_{2}^{4}b_{4xxxxx} - b_{1}b_{4}^{4}b_{4xxxxx} \nonumber \\
&&  - 960b_{1x}b_{2}^{3}b_{2xx}^{2} - 240b_{1x}b_{2}^{3}b_{4xx}^{2} + 30b_{1x}b_{4}^{3}b_{4xx}^{2} + 160b_{1x}b_{2}^{4}b_{2xxxx} - 80b_{1x}b_{2}^{4}b_{4xxxx} + 10b_{1x}b_{4}^{4}b_{2xxxx} \nonumber \\
&& - 5b_{1x}b_{4}^{4}b_{4xxxx} + 1920b_{1xx}b_{2}^{2}b_{2x}^{3} - 240b_{1xx}b_{2}^{2}b_{4x}^{3} + 480b_{1xx}b_{2x}^{3}b_{4}^{2} - 60b_{1xx}b_{4}^{2}b_{4x}^{3} + 320b_{1xx}b_{2}^{4}b_{2xxx} \nonumber \\
&& - 160b_{1xx}b_{2}^{4}b_{4xxx} + 20b_{1xx}b_{2xxx}b_{4}^{4} - 10b_{1xx}b_{4}^{4}b_{4xxx} - 640b_{1xxx}b_{2}^{3}b_{2x}^{2} - 160b_{1xxx}b_{2}^{3}b_{4x}^{2} \nonumber \\
&& + 80b_{1xxx}b_{2x}^{2}b_{4}^{3} + 20b_{1xxx}b_{4}^{3}b_{4x}^{2} + 320b_{1xxx}b_{2}^{4}b_{2xx} - 160b_{1xxx}b_{2}^{4}b_{4xx} + 20b_{1xxx}b_{2xx}b_{4}^{4} \nonumber \\
&& - 10b_{1xxx}b_{4}^{4}b_{4xx} + 960b_{1}b_{2}^{2}b_{2xx}b_{2xxx}b_{4} - 480b_{1}b_{2}b_{2xx}b_{4xxx}b_{4}^{2} - 480b_{1}b_{2}^{2}b_{4xx}b_{2xxx}b_{4} \nonumber \\
&& + 240b_{1}b_{2}b_{2xx}b_{4}^{2} + 240b_{1}b_{2}^{2}b_{4}b_{4xx}b_{4xxx} - 120b_{1}b_{2}b_{4}^{2}b_{4xx}b_{4xxx} + 480b_{1}b_{2}^{2}b_{2x}b_{2xxxx}b_{4} \nonumber \\
&& - 240b_{1}b_{2}b_{2x}b_{2xxxx}b_{4}^{2} - 240b_{1}b_{2}^{2}b_{2x}b_{4}b_{4xxxx} + 120b_{1}b_{2}b_{2x}b_{4}^{2}b_{4xxxx} - 240b_{1}b_{2}^{2}b_{2xxxx}b_{4}b_{4x} \nonumber \\
&& + 120b_{1}b_{2}b_{2xxxx}b_{4}^{2}b_{4x} + 120b_{1}b_{2}^{2}b_{4}b_{4x}b_{4xxxx} - 60b_{1}b_{2}b_{4}^{2}b_{4x}b_{4xxxx} - 5760b_{1x}b_{2}^{2}b_{2x}b_{2xx}b_{4x} \nonumber \\
&& + 2880b_{1x}b_{2}^{2}b_{2x}b_{4x}b_{4xx} - 5760b_{1x}b_{2}b_{2x}^{2}b_{2xx}b_{4} + 2880b_{1x}b_{2}b_{2x}^{2}b_{4}b_{4xx} - 1440b_{1x}b_{2}b_{2xx}b_{4}b_{4x}^{2} \nonumber \\
&& + 720b_{1x}b_{2}b_{4}b_{4x}^{2}b_{4xx} - 1440b_{1x}b_{2x}b_{2xx}b_{4}^{2}b_{4x} + 720b_{1x}b_{2}b_{4}b_{4x}^{2}b_{4xx} - 1440b_{1x}b_{2x}b_{2xx}b_{4}^{2}b_{4x} \nonumber \\
&& + 720b_{1x}b_{2x}b_{4}^{2}b_{4x}b_{4xx} + 1920b_{1x}b_{2}^{2}b_{2x}b_{2xxx}b_{4} - 960b_{1x}b_{2}^{2}b_{2x}b_{4}b_{4xxx} - 960b_{1x}b_{2}^{2}b_{2xxx}b_{4}b_{4x} \nonumber \\
&& + 480b_{1x}b_{2}^{2}b_{4}b_{4x}b_{4xxx} - 960b_{1x}b_{2}b_{2x}b_{2xxx}b_{4}^{2} + 480b_{1x}b_{2}b_{2x}b_{4}^{2}b_{4xxx} + 480b_{1x}b_{2}b_{2xxx}b_{4}^{2}b_{4x} \nonumber \\
&& - 240b_{1x}b_{2}b_{4}^{2}b_{4x}b_{4xxx} + 2880b_{1xx}b_{2}^{2}b_{2x}b_{2xx}b_{4} - 1440b_{1xx}b_{2}^{2}b_{2x}b_{4}b_{4xx} - 1440b_{1xx}b_{2}^{2}b_{2xx}b_{4}b_{4x} \nonumber \\
&& + 720b_{1xx}b_{2}^{2}b_{4}b_{4x}b_{4xx} - 1440b_{1xx}b_{2}b_{2x}b_{2xx}b_{4}^{2} + 720b_{1xx}b_{2}b_{2x}b_{4}^{2}b_{4xx} + 720b_{1xx}b_{2}b_{2xx}b_{4}^{2}b_{4x} \nonumber \\
&& - 360b_{1xx}b_{2}b_{4}^{2}b_{4x}b_{4xx} + 11520b_{1}b_{2}b_{2x}^{2}b_{2xx}b_{4x} - 5760b_{1}b_{2x}^{2}b_{2xx}b_{4}b_{4x} - 5760b_{1}b_{2}b_{2x}^{2}b_{4x}b_{4xx} \nonumber \\
&& + 2880b_{1}b_{2x}^{2}b_{4}b_{4x}b_{4xx} - 5760b_{1}b_{2}b_{2x}b_{2xx}b_{4x}^{2} + 2880b_{1}b_{2x}b_{2xx}b_{4}b_{4x}^{2} + 2880b_{1}b_{2}b_{2x}b_{4x}^{2}b_{4xx} \nonumber \\
&& - 1440b_{1}b_{2x}b_{4}b_{4x}^{2}b_{4xx} - 2880b_{1}b_{2}b_{2x}b_{2xx}^{2}b_{4} - 2880b_{1}b_{2}^{2}b_{2x}b_{2xx}b_{4xx} - 720b_{1}b_{2x}b_{2xx}b_{4}^{2}b_{4xx} \nonumber \\
&& - 720b_{1}b_{2}b_{2x}b_{4}b_{4xx}^{2} + 1440b_{1}b_{2}b_{2xx}^{2}b_{4}b_{4x} + 1440b_{1}b_{2}^{2}b_{2xx}b_{4x}b_{4xx} + 360b_{1}b_{2xx}b_{4}^{2}b_{4x}b_{4xx} \nonumber \\
&& + 360b_{1}b_{2}b_{4}b_{4x}b_{4xx}^{2} - 1920b_{1}b_{2}b_{2x}^{3}b_{2xxx}b_{4} + 960b_{1}b_{2}b_{2x}^{3}b_{4}b_{4xxx} - 1920b_{1}b_{2}^{2}b_{2x}b_{2xxx}b_{4x} \nonumber \\
&& - 480b_{1}b_{2x}b_{2xxx}b_{4}^{2}b_{4x} + 960b_{1}b_{2}^{2}b_{2x}b_{4x}b_{4xxx} + 240b_{1}b_{2x}b_{4}^{2}b_{4x}b_{4xxx} - 480b_{1}b_{2}b_{2xxx}b_{4}b_{4x}^{2} \nonumber \\
&& + 240b_{1}b_{2}b_{4}b_{4x}^{2}b_{4xxx} - 1440b_{1x}b_{2}^{2}b_{2xx}b_{4}b_{4xx} + 720b_{1x}b_{2}b_{2xx}b_{4}^{2}b_{4xx} + 2880b_{1xx}b_{2}b_{2x}^{2}b_{4}b_{4x} \nonumber \\
&& - 1440b_{1xx}b_{2}b_{2x}b_{4}b_{4x}^{2} - 960b_{1xxx}b_{2}^{2}b_{2x}b_{4}b_{4x} + 480b_{1xxx}b_{2}b_{2x}b_{4}^{2}b_{4x} + 5760b_{1x}b_{2}b_{2x}b_{2xx}b_{4}b_{4x} \nonumber \\
&& - 2880b_{1x}b_{2}b_{2x}b_{4}b_{4x}b_{4xx} + 2880b_{1}b_{2}b_{2x}b_{2xx}b_{4}b_{4xx} - 1440b_{1}b_{2}b_{2xx}b_{4}b_{4x}b_{4xx} + 80b_{1}b_{2xx}b_{2xxx}b_{4}^{3} \nonumber \\
&& + 1920b_{1}b_{2}b_{2x}b_{2xxx}b_{4}b_{4x} - 960b_{1}b_{2}b_{2x}b_{4}b_{4x}b_{4xxx} - 640b_{1}b_{2}^{3}b_{2xx}b_{2xxx} + 320b_{1}b_{2}^{3}b_{2xx}b_{4xxx} \nonumber \\
&& - 40b_{1}b_{2xx}b_{4}^{3}b_{4xxx} + 320b_{1}b_{2}^{3}b_{2xxx}b_{4xx} - 40b_{1}b_{2xxx}b_{4}^{3}b_{4xx} - 160b_{1}b_{2}^{3}b_{4xx}b_{4xxx} + 20b_{1}b_{4}^{3}b_{4xx}b_{4xxx} \nonumber \\
&& - 320b_{1}b_{2}^{3}b_{2x}b_{2xxxx} + 40b_{1}b_{2x}b_{2xxxx}b_{4}^{3} + 160b_{1}b_{2}^{3}b_{2x}b_{4xxxx} - 20b_{1}b_{2x}b_{4}^{3}b_{4xxxx} + 160b_{1}b_{2}^{3}b_{2xxxx}b_{4x} \nonumber \\
&& - 20b_{1}b_{2xxxx}b_{4}^{3}b_{4x} - 80b_{1}b_{2}^{3}b_{4x}b_{4xxxx} + 10b_{1}b_{4}^{3}b_{4x}b_{4xxxx} + 5760b_{1x}b_{2}^{2}b_{2x}^{2}b_{2xx} - 2880b_{1x}b_{2}^{2}b_{2x}^{2}b_{4xx} \nonumber
\end{eqnarray}

\begin{eqnarray}
&& + 360b_{1x}b_{2xx}b_{4}^{2}b_{4x}^{2} - 320b_{1x}b_{2}^{3}b_{4x}b_{4xxx} - 1920b_{1xx}b_{2}^{3}b_{2x}b_{2xx} - 120b_{1xx}b_{2x}b_{4}^{3}b_{4xx} \nonumber \\
&& + 1440b_{1x}b_{2}^{2}b_{2xx}b_{4x}^{2} - 720b_{1x}b_{2}^{2}b_{4x}^{2}b_{4xx} + 1440b_{1x}b_{2x}^{2}b_{2xx}b_{4}^{2} - 720b_{1x}b_{2x}^{2}b_{4}^{2}b_{4xx} \nonumber \\
&& - 180b_{1x}b_{4}^{2}b_{4x}^{2}b_{4xx} - 1280b_{1x}b_{2}^{3}b_{2x}b_{2xxx} + 640b_{1x}b_{2}^{3}b_{2x}b_{4xxx} + 640b_{1x}b_{2}^{3}b_{2xxx}b_{4x} \nonumber \\
&& + 160b_{1x}b_{2x}b_{2xxx}b_{4}^{3} - 80b_{1x}b_{2x}b_{4}^{3}b_{4xxx} - 80b_{1x}b_{2xxx}b_{4}^{3}b_{4x} + 40b_{1x}b_{4}^{3}b_{4x}b_{4xxx} - 80b_{1xxxx}b_{2}b_{2x}b_{4}^{3} \nonumber \\
&& + 960b_{1xx}b_{2}^{3}b_{2x}b_{4xx} + 960b_{1xx}b_{2}^{3}b_{2xx}b_{4x} - 480b_{1xx}b_{2}^{3}b_{4x}b_{4xx} + 240b_{1xx}b_{2x}b_{2xx}b_{4}^{3} \nonumber \\
&& - 120b_{1xx}b_{2xx}b_{4}^{3}b_{4x} + 60b_{1xx}b_{4}^{3}b_{4x}b_{4xx} - 320b_{1xxxx}b_{2}^{3}b_{2x}b_{4} + 240b_{1xxxx}b_{2}^{2}b_{2x}b_{4}^{2} \nonumber \\
&& + 160b_{1xxxx}b_{2}^{3}b_{4}b_{4x} - 120b_{1xxxx}b_{2}^{2}b_{4}^{2}b_{4x} + 40b_{1xxxx}b_{2}b_{4}^{3}b_{4x} - 7680b_{1}b_{2}b_{2x}^{3}b_{2xx} + 3840b_{1}b_{2x}^{3}b_{2xx}b_{4} \nonumber \\
&& + 3840b_{1}b_{2}b_{2x}^{3}b_{4xx} - 1920b_{1}b_{2x}^{3}b_{4}b_{4xx} + 960b_{1}b_{2}b_{2xx}b_{4x}^{3} - 480b_{1}b_{2xx}b_{4}b_{4x}^{3} - 480b_{1}b_{2}b_{4x}^{3}b_{4xx} \nonumber \\
&& + 240b_{1}b_{4}b_{4x}^{3}b_{4xx} + 2880b_{1}b_{2}^{2}b_{2x}b_{2xx}^{2} + 720b_{1}b_{2x}b_{2xx}^{2}b_{4}^{2} + 720b_{1}b_{2}^{2}b_{2x}b_{4xx}^{2} + 180b_{1}b_{2x}b_{4}^{2}b_{4xx}^{2} \nonumber \\
&& - 1440b_{1}b_{2}^{2}b_{2xx}^{2}b_{4x} - 360b_{1}b_{2xx}^{2}b_{4}^{2}b_{4x} - 360b_{1}b_{2}^{2}b_{4x}b_{4xx}^{2} - 90b_{1}b_{4}^{2}b_{4x}b_{4xx}^{2} + 1920b_{1}b_{2}^{2}b_{2x}^{2}b_{2xxx} \nonumber \\
&& + 480b_{1}b_{2x}^{2}b_{2xxx}b_{4}^{2} - 960b_{1}b_{2}^{2}b_{2x}^{2}b_{4xxx} - 240b_{1}b_{2x}^{2}b_{4}^{2}b_{4xxx} + 480b_{1}b_{2}^{2}b_{2xxx}b_{4x}^{2} + 120b_{1}b_{2xxx}b_{4}^{2}b_{4x}^{2} \nonumber \\
&& - 240b_{1}b_{2}^{2}b_{4x}^{2}b_{4xxx} - 60b_{1}b_{4}^{2}b_{4x}^{2}b_{4xxx} - 64b_{1}b_{2}^{3}b_{2xxxxx}b_{4} + 48b_{1}b_{2}^{2}b_{2xxxxx}b_{4}^{2} - 16b_{1}b_{2}b_{2xxxxx}b_{4}^{3} \nonumber \\
&& + 32b_{1}b_{2}^{3}b_{4}b_{4xxxxx} - 24b_{1}b_{2}^{2}b_{4}^{2}b_{4xxxxx} + 8b_{1}b_{2}b_{4}^{3}b_{4xxxxx} + 960b_{1x}b_{2}^{3}b_{2xx}b_{4xx} + 1440b_{1x}b_{2}^{2}b_{2xx}^{2}b_{4} \nonumber \\
&& + 360b_{1x}b_{2}^{2}b_{4}b_{4xx}^{2} - 720b_{1x}b_{2}b_{2xx}^{2}b_{4}^{2} - 180b_{1x}b_{2}b_{4}^{2}b_{4xx}^{2} - 120b_{1x}b_{2xx}b_{4}^{3}b_{4xx} - 320b_{1x}b_{2}^{3}b_{2xxxx}b_{4} \nonumber \\
&& + 160b_{1x}b_{2}^{3}b_{4}b_{4xxxx} + 240b_{1x}b_{2}^{2}b_{4}^{2}b_{2xxxx} - 120b_{1x}b_{2}^{2}b_{4}^{2}b_{4xxxx} - 80b_{1x}b_{2}b_{2xxxx}b_{4}^{3} + 40b_{1x}b_{2}b_{4}^{3}b_{4xxxx} \nonumber \\
&& - 2880b_{1xx}b_{2}^{2}b_{2x}^{2}b_{4x} + 1440b_{1xx}b_{2}^{2}b_{2x}b_{4x}^{2} - 1920b_{1xx}b_{2}b_{2x}^{3}b_{4} + 240b_{1xx}b_{2}b_{4}b_{4x}^{3} - 720b_{1xx}b_{2x}^{2}b_{4}^{2}b_{4x} \nonumber \\
&& + 360b_{1xx}b_{2x}b_{4}^{2}b_{4x}^{2} - 640b_{1xx}b_{2}^{3}b_{2xxx}b_{4} + 320b_{1xx}b_{2}^{3}b_{4}b_{4xxx} + 480b_{1xx}b_{2}^{2}b_{2xxx}b_{4}^{2} - 240b_{1xx}b_{2}^{2}b_{4}^{2}b_{4xxx} \nonumber \\
&& - 160b_{1xx}b_{2}b_{2xxx}b_{4}^{3} + 80b_{1xx}b_{2}b_{4}^{3}b_{4xxx} + 640b_{1xxx}b_{2}^{3}b_{2x}b_{4x} + 960b_{1xxx}b_{2}^{2}b_{2x}^{2}b_{4} + 240b_{1xxx}b_{2}^{2}b_{4}b_{4x}^{2} \nonumber \\
&& - 480b_{1xxx}b_{2}b_{2x}^{2}b_{4}^{2} - 120b_{1xxx}b_{2}b_{4}^{2}b_{4x}^{2} - 80b_{1xxx}b_{2x}b_{4}^{3}b_{4x} - 640b_{1xxx}b_{2}^{3}b_{2xx}b_{4} + 320b_{1xxx}b_{2}^{3}b_{4}b_{4xx} \nonumber \\
&& + 480b_{1xxx}b_{2}^{2}b_{2xx}b_{4}^{2} - 240b_{1xxx}b_{2}^{2}b_{4}^{2}b_{4xx} - 160b_{1xxx}b_{2}b_{2xx}b_{4}^{3} + 80b_{1xxx}b_{2}b_{4}^{3}b_{4xx} + 160b_{1xxxx}b_{2}^{4}b_{2x} \nonumber \\
&& + 120b_{1x}b_{2xx}^{2}b_{4}^{3} = 0 
\end{eqnarray}

\vspace{3mm}
\noindent
where $H_{1}(t)$ is an arbitrary function of $t$ and $b_{2}$ or $b_{4}$ are arbitrary. The latter equation may look a bit daunting but is quite easily managed with the aid of a CAS (in this case MAPLE) once $b_{2}$ and $b_{4}$ are given.

\subsubsection{Conditions on the $c_{i}$}

\begin{eqnarray}
&& c_{4} = -10c_{2} + 5c_{3} - 4c_{5} + 2c_{6} \\
&& c_{8} = \frac{2}{3}c_{9} + 4c_{7}
\end{eqnarray}
\begin{eqnarray}
&& c_{10x}c_{6} - c_{6}c_{10x} - 2c_{5}c_{10x} + 2c_{5x}c_{10} + c_{10x}c_{3} - c_{3x}c_{10} = 0 \\
&& -12c_{10x}c_{7} + 12c_{7x}c_{10} + c_{10x}c_{9} - c_{9x}c_{10} = 0 \\
&& 5Gc_{10}^{3} - 2Hc_{10x}c_{7x}c_{10} + Hc_{10}^{2}c_{7xx} + 2Hc_{10x}^{2}c_{7} - Hc_{7}c_{10}c_{10xx} = 0
\end{eqnarray}
\begin{eqnarray}
&& 14c_{3}c_{10x}^{2} - 7c_{3}c_{10}c_{10xx} - 60c_{2}c_{10x}^{2} + 30c_{2}c_{10}c_{10xx} + 60c_{2x}c_{10}c_{10x} - 30c_{2xx}c_{10}^{2} \nonumber \\
&& + 4c_{5x}c_{10}c_{10x} - 14c_{3x}c_{10}c_{10x} + 7c_{3xx}c_{10}^{2} - 4c_{5}c_{10x}^{2} + 2c_{5}c_{10}c_{10xx} - 2c_{5xx}c_{10}^{2} = 0
\end{eqnarray}
\begin{eqnarray}
&& 18c_{3}c_{10x}^{2} - 18c_{3}c_{10}c_{10x}c_{10xx} + 3a_{3}c_{10}^{3}c_{10xxx} - 18c_{3x}c_{10}c_{10x}^{2} + 9c_{3x}c_{10}^{2}c_{10xx} - 60c_{2}c_{10x}^{3} \nonumber \\
&& + 60c_{2}c_{10}c_{10x}c_{10xx} - 10c_{2}c_{10}^{2}c_{10xxx} + 60c_{2x}c_{10}c_{10x}^{2} - 30c_{2x}c_{10}^{2}c_{10xx} - 30c_{2xx}c_{10}^{2}c_{10x} \nonumber \\
&& + 10c_{2xxx}c_{10}^{3} + 6c_{5x}c_{10}c_{10x}^{2} - 3c_{5x}c_{10}^{2}c_{10xx} - 3c_{5xx}c_{10}^{2}c_{10x} + 9c_{3xx}c_{10}^{2}c_{10x} - 3c_{3xxx}c_{10}^{3} \nonumber \\
&& - 6c_{5}c_{10x}^{3} + c_{5}c_{10}c_{10x}c_{10xx} - c_{5}c_{10}^{2}c_{10xxx} + c_{5xxx}c_{10}^{3} = 0
\end{eqnarray}

\begin{eqnarray}
&& c_{1xxxxxxx}c_{10}^{7} + 5040c_{1x}c_{10}c_{10x}^{6} - 7c_{1xxxxxx}c_{10}^{6}c_{10x} - c_{1}c_{10}^{6}c_{10xxxxxxx} - 1260c_{1xxx}c_{10}^{4}c_{10x}^{2}c_{10xx} \nonumber \\
&& + 140c_{1x}c_{10}^{5}c_{10xxx}^{2} - 7c_{1x}c_{10}^{6}c_{10xxxxxx} - 2520c_{1xx}c_{10}^{2}c_{10x}^{5} - 21c_{1xx}c_{10}^{6}c_{10xxxxx} + 840c_{1xxx}c_{10}^{3}c_{10x}^{4} \nonumber \\
&& + 210c_{1xxx}c_{10}^{5}c_{10xx}^{2} - 35c_{1xxx}c_{10}^{6}c_{10xxxx} - 210c_{1xxxx}c_{10}^{4}c_{10x}^{3} - 35c_{1xxxx}c_{10}^{6}c_{10xxx} \nonumber \\
&& - 21c_{1xxxxx}c_{10}^{6}c_{10xx} + 5040c_{1}c_{10}^{3}c_{10x}^{2}c_{10xx}c_{10xxx} - 630c_{1}c_{10}^{4}c_{10x}c_{10xx}c_{10xxxx} - 630c_{1x}c_{10}^{4}c_{10xx}^{3} \nonumber \\
&& + \frac{H_{t}}{H}c_{10}^{7} - c_{10}^{6}c_{10t} - 5040c_{1}c_{10x}^{7} + 15210c_{1}c_{10}c_{10x}^{5}c_{10xx} - 4200c_{1}c_{10}^{2}c_{10x}^{4}c_{10xxx} + 42c_{1xxxxx}c_{10}^{5}c_{10x}^{2} \nonumber \\
&& - 630c_{1}c_{10}^{4}c_{10xx}^{2}c_{10xxx} - 126c_{1}c_{10}^{4}c_{10x}^{2}c_{10xxxxx} + 70c_{1}c_{10}^{5}c_{10xxx}c_{10xxxx} + 42c_{1}c_{10}^{5}c_{10xx}c_{10xxxxx} \nonumber \\
&& + 14c_{1}c_{10}^{5}c_{10x}c_{10xxxxxx} - 12600c_{1}c_{10}^{2}c_{10x}^{3}c_{10xx}^{2} + 2520c_{1}c_{10}^{3}c_{10x}c_{10xx}^{3} - 420c_{1}c_{10}^{4}c_{10x}c_{10xxx}^{2} \nonumber \\
&& - 12600c_{1x}c_{10}^{2}c_{10x}^{4}c_{10xx} + 7560c_{1x}c_{10}^{3}c_{10x}^{2}c_{10xx}^{2} + 3360c_{1x}c_{10}^{3}c_{10x}^{3}c_{10xxx} - 630c_{1x}c_{10}^{4}c_{10x}^{2}c_{10xxxx} \nonumber \\
&& + 210c_{1x}c_{10}^{5}c_{10xx}c_{10xxxx} + 84c_{1x}c_{10}^{5}c_{10x}c_{10xxxxx} + 5040c_{1xx}c_{10}^{3}c_{10x}^{3}c_{10xx} - 1890c_{1xx}c_{10}^{4}c_{10x}c_{10xx}^{2} \nonumber \\
&& - 1260c_{1xx}c_{10}^{4}c_{10x}^{2}c_{10xxx} + 420c_{1xx}c_{10}^{5}c_{10xx}c_{10xxx} + 210c_{1xx}c_{10}^{5}c_{10x}c_{10xxxx} \nonumber \\
&& + 280c_{1xxx}c_{10}^{5}c_{10x}c_{10xxx} + 210c_{1xxxx}c_{10}^{5}c_{10x}c_{10xx} - 2520c_{1x}c_{10}^{4}c_{10x}c_{10xx}c_{10xxx} \nonumber \\
&& + 840c_{1}c_{10}^{3}c_{10x}^{3}c_{10xxxx} = 0
\end{eqnarray}

\begin{eqnarray}
&& -240c_{3}c_{10x}^{5} + 720c_{2}c_{10x}^{5} + 120c_{5}c_{10x}^{5} + 40c_{3xxx}c_{10}^{3}c_{10x}^{2} - 20c_{3xxx}c_{10}^{4}c_{10xx} + 6c_{2}c_{10}^{4}c_{10xxxxx} \nonumber \\
&& - 180c_{2x}c_{10}^{3}c_{10xx}^{2} + 30c_{2x}c_{10}^{4}c_{10xxxx} + 360c_{2xx}c_{10}^{2}c_{10x}^{3} + 60c_{2xx}c_{10}^{4}c_{10xxx} - 120c_{2xxx}c_{10}^{3}c_{10x}^{2} \nonumber \\
&& + 60c_{2xxx}c_{10}^{4}c_{10xx} - 30c_{5x}c_{10}^{3}c_{10xx}^{2} + 5c_{5x}c_{10}^{4}c_{10xxxx} + 60c_{5xx}c_{10}^{2}c_{10x}^{3} + 10c_{5xx}c_{10}^{4}c_{10xxx} \nonumber \\
&& + 10c_{5xxx}c_{10}^{4}c_{10xx} + c_{5}c_{10}^{4}c_{10xxxxx} - 120c_{3xx}c_{10}^{2}c_{10x}^{3} - 20c_{3xx}c_{10}^{4}c_{10xxx} - 2c_{3}c_{10}^{4}c_{10xxxxx} \nonumber \\
&& - 10c_{3x}c_{10}^{4}c_{10xxxx} + 30c_{2xxxx}c_{10}^{4}c_{10x} + 5c_{5xxxx}c_{10}^{4}c_{10x} - 10c_{3xxxx}c_{10}^{4}c_{10x} - 6c_{2xxxxx}c_{10}^{5} \nonumber \\
&& - c_{5xxxxx}c_{10}^{5} - 720c_{2x}c_{10}c_{10x}^{4} - 120c_{5x}c_{10}c_{10x}^{4} + 240c_{3x}c_{10}c_{10x}^{4} + 480c_{3}c_{10}c_{10x}^{3}c_{10xx} \nonumber \\
&& - 120c_{3}c_{10}^{2}c_{10x}^{2}c_{10xxx} + 40c_{3}c_{10}^{3}c_{10xx}c_{10xxx} + 20c_{3}c_{10}^{3}c_{10x}c_{10xxxx} - 360c_{3x}c_{10}^{2}c_{10x}^{2}c_{10xx} \nonumber \\
&& + 120c_{3xx}c_{10}^{3}c_{10x}c_{10xx} - 1440c_{2}c_{10}c_{10x}^{3}c_{10xx} + 540c_{2}c_{10}^{2}c_{10x}c_{10xx}^{2} + 360c_{2}c_{10}^{2}c_{10x}^{2}c_{10xxx} \nonumber \\
&& - 20c_{5xxx}c_{10}^{3}c_{10x}^{2} + 60c_{3x}c_{10}^{3}c_{10xx}^{2} + 2c_{3xxxxx}c_{10}^{5} - 180c_{3}c_{10}^{2}c_{10x}c_{10xx}^{2} - 120c_{2}c_{10}^{3}c_{10xx}c_{10xxx} \nonumber \\
&& + 80c_{3x}c_{10}^{3}c_{10x}c_{10xxx} + 180c_{5x}c_{10}^{2}c_{10x}^{2}c_{10xx} + 60c_{5}c_{10}^{2}c_{10x}^{2}c_{10xxx} - 20c_{5}c_{10}^{3}c_{10xx}c_{10xxx} \nonumber \\
&& - 60c_{2}c_{10}^{3}c_{10x}c_{10xxxx} + 1080c_{2x}c_{10}^{2}c_{10x}^{2}c_{10xx} - 240c_{2x}c_{10}^{3}c_{10x}c_{10xxx} - 360c_{2xx}c_{10}^{3}c_{10x}c_{10xx} \nonumber \\
&& - 40c_{5x}c_{10}^{3}c_{10x}c_{10xxx} - 60c_{5xx}c_{10}^{3}c_{10x}c_{10xx} - 240c_{5}c_{10}c_{10x}^{3}c_{10xx} + 90c_{5}c_{10}^{2}c_{10x}c_{10xx}^{2} \nonumber \\
&& - 10c_{5}c_{10}^{3}c_{10x}c_{10xxxx} = 0
\end{eqnarray}

\noindent
where $G(t)$ and $H(t)$ are arbitrary functions of $t$.

\section{The Generalized Variable Coefficient Fifth-order KdV (vcKdV) Equation}

Here, we will apply the technique of the last section in exactly the same fashion to generalized vcKdV equations, but will omit the details for the sake of brevity. Please note that {\it the coefficients $a_i$ in this section are totally distinct or different from those given the same symbols in the
previous section. All equations in this section are thus to be read independently of those in the previous one.}

Consider the generalized fifth-order vcKdV equation in the form 

\begin{equation} \label{genkdv}
u_{t} + a_{1}uu_{xxx} + a_{2}u_{x}u_{xx} + a_{3}u^{2}u_{x} + a_{4}uu_{x} + a_{5}u_{xxx} + a_{6}u_{xxxxx} + a_{7}u + a_{8}u_{x} = 0
\end{equation}

As before, we consider the generalized variable-coefficient KDV equation to be integrable if we can find a Lax pair which satisfies $\eqref{ZCC1}$. In the method given in (cite Khwaja) one expands the Lax pair $\textbf{U}$ and $\textbf{V}$ in powers of $u$ its derivatives with unknown function coefficients and require $\eqref{ZCC1}$ to be equivalent to the nonlinear system. This results in a system of coupled PDEs for the unknown coefficients for which upon solving we simultaneously obtain the Lax pair and integrability conditions on the $a_{i}$. The results, for which the details are given in Appendix B, are as follows

\begin{eqnarray}
a_{2-4} &=& H_{2-4}a_{1} \\
a_{7} &=& \frac{a_{1}}{H_{1}}\left(\left(\frac{H_{1}}{a_{1}}\right)_{t} + \left(\frac{H_{1}a_{5}}{a_{1}}\right)_{xxx} + \left(\frac{H_{1}a_{6}}{a_{1}}\right)_{xxxxx} + \left(\frac{H_{1}a_{8}}{a_{1}}\right)_{x}\right)
\end{eqnarray}
where $H_{1-4}(t)$ are arbitrary functions of $t$ and $a_{1},a_{5},a_{6}$ and $a_{8}$ are taken to be arbitrary functions of $x$ and $t$. This form helps to give integrability conditions to sub-equations of $\eqref{genkdv}$. For example an integrable variable-coefficient KDV equation would require that $H_{2-4}(t) = a_{6-8}(x,t) = 0$. For $a_{7}(x,t) = 0$ we would need to further require (through a little algebraic manipulation) that the following be satisfied

\begin{equation}
\left(\frac{H(t)}{a_{1}}\right)_{t} + H(t)\left(\frac{a_{5}}{a_{1}}\right)_{xxx} = 0
\end{equation}

Having considered these two examples of various vcMKdV equations, as well as the vcKdV equation, we shall now proceed to consider
whether these results may be recovered in a more algorithmic manner. As discussed in Section 1, {\it it would be
advantageous if they could be obtained without: a. requiring to know the form of the Lax Pair for the corresponding
constant-coefficient Lax-integrable equation, and b. requiring to generalize this constant-coefficient Lax Pair
by guesswork.} Towards that end, we now proceed to consider how this may be accomplished by generalizing and extending
the Estabrook-Wahlquist technique to Lax-integrable systems with variable coefficients.

\section{The Extended Estabrook-Wahlquist Technique}

In the standard Estabrook-Wahlquist method one begins with a constant coefficient NLPDE and assumes an implicit dependence on $u(x,t)$ and its partial derivatives of the spatial and time evolution matrices ($\mathbb{F},\mathbb{G}$) involved in the linear scattering problem 
\[ \psi_{x} = \mathbb{F}\psi, \ \ \ \psi_{t} = \mathbb{G}\psi \]
The evolution matrices $\mathbb{F}$ and $\mathbb{G}$ are connected via a zero-curvature condition (independence of path in spatial and time evolution) derived by mandating $\psi_{xt} = \psi_{tx}$. That is, it requires
\[ \mathbb{F}_{t} - \mathbb{G}_{x} + [\mathbb{F},\mathbb{G}] = 0 \]
provided $u(x,t)$ satisfies the NLPDE. 

Considering the forms $\mathbb{F} = \mathbb{F}(u,u_{x},u_{t},\ldots,u_{mx,nt})$ and $\mathbb{G} = \mathbb{G}(u,u_{x},u_{t},\ldots,u_{kx,jt})$ for the space and time evolution matrices where $u_{px,qt} = \frac{\partial^{p+q}u}{\partial x^{p}\partial t^{q}}$ we see that this condition is equivalent to

\[ \sum_{m,n}{\mathbb{F}_{u_{mx,nt}}u_{mx,(n+1)t}} - \sum_{j,k}{\mathbb{G}_{u_{jx,kt}}u_{(j+1)x,kt}} + [\mathbb{F},\mathbb{G}] = 0 \]

\noindent
From here there is often a systematic approach\cite{EW1}-\cite{EW4} to determining the form for $\mathbb{F}$ and $\mathbb{G}$ which is outlined in \cite{EW3} and will be utilized in the examples to follow. 

Typically a valid choice for dependence on $u(x,t)$ and its partial derivatives is to take $\mathbb{F}$ to depend on all terms in the NLPDE for which there is a partial time derivative present. Similarly we may take $\mathbb{G}$ to depend on all terms for which there is a partial space derivative present. For example, given the Camassa-Holm equation,
\[ u_{t} + 2ku_{x} - u_{xxt} + 3uu_{x} - 2u_{x}u_{xx} - uu_{xxx} = 0, \]
one would consider $\mathbb{F} = \mathbb{F}(u,u_{xx})$ and $\mathbb{G} = \mathbb{G}(u,u_{x},u_{xx})$. Imposing compatibility allows one to determine the explicit form of $\mathbb{F}$ and $\mathbb{G}$ in a very algorithmic way. Additionally the compatibility condition induces a set of constraints on the coefficient matrices in $\mathbb{F}$ and $\mathbb{G}$. These coefficient matrices subject to the constraints generate a finite dimensional matrix Lie algebra.

In the extended Estabrook-Wahlquist method we allow for $\mathbb{F}$ and $\mathbb{G}$ to be functions of $t$, $x$, $u$ and the partial derivatives of $u$. Although the details change, the general procedure will remain essentially the same. We will begin by equating the coefficient of the highest partial derivative of the unknown function(s) to zero and work our way down until we have eliminated all partial derivatives of the unknown function(s). 

{\it This typically results in a large partial differential equation (in the standard Estabrook-Wahlquist method, this is an algebraic equation) which can be solved by equating the coefficients of the different powers of the unknown function(s) to zero.} This final step induces a set of constraints on the coefficient matrices in $\mathbb{F}$ and $\mathbb{G}$. {\it Another big difference which we will see in the examples comes in the final and, arguably, the hardest step. In the standard Estabrook-Wahlquist method the final step involves finding explicit forms for the set of coefficient matrices such that they satisfy the contraints derived in the procedure. Note these constraints are nothing more than a system of algebraic matrix equations. In the extended Estabrook-Wahlquist method these constraints will be in the form of matrix partial differential equations which can be used to derive an integrability condition on the coefficients in the NLPDE.}

As we are now letting $\mathbb{F}$ and $\mathbb{G}$ have explicit dependence on $x$ and $t$ and for notational clarity, it will be more convenient to consider the following version of the zero-curvature condition

\begin{equation}\label{ZCC}
\mbox{D}_{t}\mathbb{F} - \mbox{D}_{x}\mathbb{G} + [\mathbb{F},\mathbb{G}] = 0
\end{equation}

\noindent
where $\mbox{D}_{t}$ and $\mbox{D}_{x}$ are the total derivative operators on time and space, respectively. Recall the definition of the total derivative

\[ \mbox{D}_{y}f(y,z,u_{1}(y,z),u_{2}(y,z),\ldots,u_{n}(y,z)) = \frac{\partial f}{\partial y} + \frac{\partial f}{\partial u_{1}}\frac{\partial u_{1}}{\partial y} + \frac{\partial f}{\partial u
_{2}}\frac{\partial u_{2}}{\partial y} + \cdots + \frac{\partial f}{\partial u_{n}}\frac{\partial u_{n}}{\partial y} \]

\noindent
Thus we can write the compatibility condition as

\[ \mathbb{F}_{t} + \sum_{m,n}{\mathbb{F}_{u_{mx,nt}}u_{mx,(n+1)t}} - \mathbb{G}_{x} - \sum_{j,k}{\mathbb{G}_{u_{jx,kt}}u_{(j+1)x,kt}} + [\mathbb{F},\mathbb{G}] = 0 \]

\noindent
It is important to note that the subscripted $x$ and $t$ denotes the partial derivative with respect to only the $x$ and $t$ elements, respectively. That is, although $u$ and it's derivatives depend on $x$ and $t$ this will not invoke use of the chain rule as they are treated as independent variables. This will become more clear in the examples of the next section. 

Note that compatibility of the time and space evolution matrices will yield a set of constraints which contain the constant coefficient constraints as a subset. In fact, taking the variable coefficients to be the appropriate constants will yield exactly the Estabrook-Wahlquist results for the constant coefficient version of the NLPDE. That is, the constraints given by the Estabrook-Wahlquist method for a constant coefficient NLPDE are always a proper subset of the constraints given by a variable-coefficient version of the NLPDE. This can easily be shown. Letting $\mathbb{F}$ and $\mathbb{G}$ not depend explicitly on $x$ and $t$ and taking the coefficients in the NLPDE to be constant the zero-curvature condition as it is written above becomes

\[ \sum_{m,n}{\mathbb{F}_{u_{mx,nt}}u_{mx,(n+1)t}} - \sum_{j,k}{\mathbb{G}_{u_{jx,kt}}u_{(j+1)x,kt}} + [\mathbb{F},\mathbb{G}] = 0 \]

\noindent
which is exactly the standard Estabrook-Wahlquist method. 

The conditions derived via mandating $\eqref{ZCC}$ be satisfied upon solutions of the vc-NLPDE may be used to determine conditions on the coefficient matrices and variable-coefficients (present in the NLPDE). Successful closure of these conditions is equivalent to the system being S-integrable. A major advantage to using the Estabrook-Wahlquist method that carries forward with the extension is the fact that it requires little guesswork and yields quite general results. 

In Khawaja's method\cite{Khawaja}-\cite{Lecce} an educated guess is made for the structure of the variable-coefficient pde Lax pair based on the associated constant coefficient Lax pair. That is, Khawaja considered the matrices
\[ \mathbb{F} = U = \begin{bmatrix}
f_{1} + f_{2}q & f_{3} + f_{4}q \\
f_{5} + f_{6}r & f_{7} + f_{8}r
\end{bmatrix} \]
and
\[ \mathbb{G} = V = \begin{bmatrix}
g_{1} + g_{2}q + g_{3}q_{x} + g_{4}rq & g_{5} + g_{6}q + g_{7}q_{x} + g_{8}rq \\
g_{9} + g_{10}r + g_{11}r_{x} + g_{12}rq & g_{13} + g_{14}r + g_{15}r_{x} + g_{16}rq
\end{bmatrix} \]
where $f_{i}$ and $g_{i}$ unknown functions of $x$ and $t$ which satisfy conditions derived by mandating the zero-curvature condition be satisfied on solutions of the variable-coefficient NLPDE. In fact, in a previous paper Khawaja derives the associated Lax pair via a similar means where he begins with an even weaker assumption on the structure of the Lax pair. This Lax pair is omitted from the paper as it becomes clear the zero-curvature condition mandates many of the coefficients be zero. 

{\it An ideal approach would be a method which does not require knowledge of the Lax pair to an associated constant coefficient system and involves little to no guesswork. The extended Estabrook-Wahlquist does exactly this. It will be shown that the results obtained from Khawaja's method are in fact a special case of the extended Estabrook-Wahlquist method. }

We now proceed with the variable coefficient MKdV equation as our first example of this extended Estabrook Wahlquist method.

\section{The Variable Coefficient MKdV (vcMKdV) Equation Reconsidered}

For this example we consider the mKdV equation given by

\begin{equation}
v_{t} + b_{1}v_{xxx} + b_{2}v^{2}v_{x} = 0 \label{MKDV}
\end{equation}

\noindent
where $b_{1}$ and $b_{2}$ are arbitrary functions of $x$ and $t$. Following the procedure we let 

\[ \mathbb{F} = \mathbb{F}(x,t,u) ,\ \ \ \mathbb{G} = \mathbb{G}(x,t,u,u_{x},u_{xx}) \]

Plugging this into $\eqref{ZCC}$ we obtain

\begin{equation}
\mathbb{F}_{t} + \mathbb{F}_{v}v_{t} - \mathbb{G}_{x} - \mathbb{G}_{v}v_{x} - \mathbb{G}_{v_{x}}v_{xx} - \mathbb{G}_{v_{xx}}v_{xxx} + [\mathbb{F},\mathbb{G}] = 0
\end{equation}

\noindent
Using $\eqref{MKDV}$ to substitute for $v_{t}$ we have

\begin{equation}
\mathbb{F}_{t} - \mathbb{G}_{x} - (\mathbb{G}_{v}+b_{2}\mathbb{F}_{v}v^{2})v_{x} - \mathbb{G}_{v_{x}}v_{xx} - (\mathbb{G}_{v_{xx}}+b_{1}\mathbb{F}_{v})v_{xxx} + [\mathbb{F},\mathbb{G}] = 0 \label{MKDVstep1}
\end{equation} 

\noindent
Since $\mathbb{F}$ and $\mathbb{G}$ do not depend on $v_{xxx}$ we can set the coefficient of the $v_{xxx}$ term to zero from which we have

\[ \mathbb{G}_{v_{xx}}+b_{1}\mathbb{F}_{v} = 0 \Rightarrow \mathbb{G} = -b_{1}\mathbb{F}_{v}v_{xx} + \mathbb{K}^{0}(x,t,v,v_{x}) \]

\noindent
Substituting this into $\eqref{MKDVstep1}$ we have

\begin{equation}
\mathbb{F}_{t} + (b_{1}\mathbb{F}_{v})_{x}v_{xx} - \mathbb{K}^{0}_{x} + b_{1}\mathbb{F}_{vv}v_{x}v_{xx} - \mathbb{K}^{0}_{v}v_{x} - b_{2}\mathbb{F}_{v}v^{2}v_{x} - \mathbb{K}^{0}_{v_{x}}v_{xx} - b_{1}[\mathbb{F},\mathbb{F}_{v}]v_{xx} + [\mathbb{F},\mathbb{K}^{0}] = 0 \label{MKDVstep2}
\end{equation}

Since $\mathbb{F}$ and $\mathbb{K}^{0}$ do not depend on $v_{xx}$ we can equate the coefficient of the $v_{xx}$ term to zero from which we require

\begin{equation}
(b_{1}\mathbb{F}_{v})_{x} + b_{1}\mathbb{F}_{vv}v_{x} - \mathbb{K}^{0}_{v_{x}} - b_{1}[\mathbb{F},\mathbb{F}_{v}] = 0
\end{equation}

\noindent
Solving for $\mathbb{K}^{0}$ we have

\[ \mathbb{K}^{0} = (b_{1}\mathbb{F}_{v})_{x}v_{x} + \frac{1}{2}b_{1}\mathbb{F}_{vv}v_{x}^{2} - b_{1}[\mathbb{F},\mathbb{F}_{v}]v_{x} + \mathbb{K}^{1}(x,t,v) \]

\noindent
Substituting this expression into $\eqref{MKDVstep2}$ we have

\begin{eqnarray}
&& \mathbb{F}_{t} - (b_{1}\mathbb{F}_{v})_{xx}v_{x} - \frac{1}{2}(b_{1}\mathbb{F}_{vv})_{x}v_{x}^{2} + (b_{1}[\mathbb{F},\mathbb{F}_{v}])_{x}v_{x} - \mathbb{K}^{1}_{x} - (b_{1}\mathbb{F}_{vv})_{x}v_{x}^{2} - \frac{1}{2}b_{1}\mathbb{F}_{vvv}v_{x}^{3} \nonumber \\
&& - \mathbb{K}^{1}_{v}v_{x} + b_{1}[\mathbb{F},\mathbb{F}_{vv}]v_{x}^{2} - b_{2}\mathbb{F}_{v}v^{2}v_{x} + [\mathbb{F},(b_{1}\mathbb{F}_{v})_{x}]v_{x} + \frac{1}{2}b_{1}[\mathbb{F},\mathbb{F}_{vv}]v_{x}^{2} - b_{1}[\mathbb{F},[\mathbb{F},\mathbb{F}_{v}]]v_{x} \nonumber \\
&& + [\mathbb{F},\mathbb{K}^{1}] = 0 \label{MKDVstep3}
\end{eqnarray}

Since $\mathbb{F}$ and $\mathbb{K}^{1}$ do not depend on $v_{x}$ we can equate the coefficients of the $v_{x}$, $v_{x}^{2}$ and $v_{x}^{3}$ terms to zero from which we obtain the system

\begin{eqnarray}
O(v_{x}^{3}) &:& \mathbb{F}_{vvv} = 0 \\
O(v_{x}^{2}) &:&  \frac{1}{2}(b_{1}\mathbb{F}_{vv})_{x} + (b_{1}\mathbb{F}_{vv})_{x} - b_{1}[\mathbb{F},\mathbb{F}_{vv}] - \frac{1}{2}b_{1}[\mathbb{F},\mathbb{F}_{vv}] = 0 \\
O(v_{x}) &:&  (b_{1}\mathbb{F}_{v})_{xx} - (b_{1}[\mathbb{F},\mathbb{F}_{v}])_{x} + \mathbb{K}^{1}_{v} + b_{2}\mathbb{F}_{v}v^{2} - [\mathbb{F},(b_{1}\mathbb{F}_{v})_{x}] + b_{1}[\mathbb{F},[\mathbb{F},\mathbb{F}_{v}]] = 0
\end{eqnarray}

Since the MKdV equation does not contain a $vv_{t}$ term and for ease of computation we take require $\mathbb{F}_{vv} = 0$ from which we have $\mathbb{F} = \mathbb{X}_{1}(x,t) + \mathbb{X}_{2}(x,t)v$. For the $O(v_{x})$ equation we solve for $\mathbb{K}^{1}$ and thus have

\begin{eqnarray}
\mathbb{K}^{1} &=& -(b_{1}\mathbb{X}_{2})_{xx}v + (b_{1}[\mathbb{X}_{1},\mathbb{X}_{2}])_{x}v + [\mathbb{X}_{1},(b_{1}\mathbb{X}_{2})_{x}]v + \frac{1}{2}[\mathbb{X}_{2},(b_{1}\mathbb{X}_{2})_{x}]v^{2} - b_{1}[\mathbb{X}_{1},[\mathbb{X}_{1},\mathbb{X}_{2}]]v \nonumber \\
&& - \frac{1}{3}b_{2}\mathbb{X}_{2}v^{3} - \frac{1}{2}b_{1}[\mathbb{X}_{2},[\mathbb{X}_{1},\mathbb{X}_{2}]]v^{2} + \mathbb{X}_{0}(x,t)
\end{eqnarray}

\noindent
Substituting this expression for $\mathbb{K}^{1}$ into $\eqref{MKDVstep3}$ we obtain

\begin{eqnarray}
&& \mathbb{X}_{1,t} + (b_{1}\mathbb{X}_{2})_{xxx}v - (b_{1}[\mathbb{X}_{1},\mathbb{X}_{2}])_{xx}v + \frac{1}{3}(b_{2}\mathbb{X}_{2})_{x}v^{3} - ([\mathbb{X}_{1},(b_{1}\mathbb{X}_{2})_{x}])_{x}v - \frac{1}{2}([\mathbb{X}_{2},(b_{1}\mathbb{X}_{2})_{x}])_{x}v^{2} \nonumber \\
&& + (b_{1}[\mathbb{X}_{1},[\mathbb{X}_{1},\mathbb{X}_{2}]])_{x}v + \frac{1}{2}(b_{1}[\mathbb{X}_{2},[\mathbb{X}_{1},\mathbb{X}_{2}]])_{x}v^{2} - \mathbb{X}_{0,x} - [\mathbb{X}_{1},(b_{1}\mathbb{X}_{2})_{xx}]v + [\mathbb{X}_{1},(b_{1}[\mathbb{X}_{1},\mathbb{X}_{2}])_{x}]v \nonumber \\
&& + \mathbb{X}_{2,t}v - \frac{1}{3}b_{2}[\mathbb{X}_{1},\mathbb{X}_{2}]v^{3} + [\mathbb{X}_{1},[\mathbb{X}_{1},(b_{1}\mathbb{X}_{2})_{x}]]v + \frac{1}{2}[\mathbb{X}_{1},[\mathbb{X}_{2},(b_{1}\mathbb{X}_{2})_{x}]]v^{2} - b_{1}[\mathbb{X}_{1},[\mathbb{X}_{1},[\mathbb{X}_{1},\mathbb{X}_{2}]]]v \nonumber \\
&& - \frac{1}{2}b_{1}[\mathbb{X}_{1},[\mathbb{X}_{2},[\mathbb{X}_{1},\mathbb{X}_{2}]]]v^{2} - [\mathbb{X}_{2},(b_{1}\mathbb{X}_{2})_{xx}]v^{2} + [\mathbb{X}_{2},(b_{1}[\mathbb{X}_{1},\mathbb{X}_{2}])_{x}]v^{2} + [\mathbb{X}_{2},[\mathbb{X}_{1},(b_{1}\mathbb{X}_{2})_{x}]]v^{2} \nonumber \\
&& + [\mathbb{X}_{1},\mathbb{X}_{0}] + \frac{1}{2}[\mathbb{X}_{2},[\mathbb{X}_{2},(b_{1}\mathbb{X}_{2})_{x}]]v^{3} - b_{1}[\mathbb{X}_{2},[\mathbb{X}_{1},[\mathbb{X}_{1},\mathbb{X}_{2}]]]v^{2} - \frac{1}{2}b_{1}[\mathbb{X}_{2},[\mathbb{X}_{2},[\mathbb{X}_{1},\mathbb{X}_{2}]]]v^{3} \nonumber \\
&& + [\mathbb{X}_{2},\mathbb{X}_{0}]v = 0
\end{eqnarray}

Since the $\mathbb{X}_{i}$ do not depend on $v$ we can equate the coefficients of the different powers of $v$ to zero. We thus obtain the constraints

\begin{eqnarray}
O(1) &:& \mathbb{X}_{1,t} - \mathbb{X}_{0,x} + [\mathbb{X}_{1},\mathbb{X}_{0}] \\
O(v) &:& \mathbb{X}_{2,t} - ([\mathbb{X}_{1},(b_{1}\mathbb{X}_{2})_{x}])_{x} + (b_{1}[\mathbb{X}_{1},[\mathbb{X}_{1},\mathbb{X}_{2}]])_{x} - [\mathbb{X}_{1},(b_{1}\mathbb{X}_{2})_{xx}] + [\mathbb{X}_{1},(b_{1}[\mathbb{X}_{1},\mathbb{X}_{2}])_{x}] \nonumber \\
&& - (b_{1}[\mathbb{X}_{1},\mathbb{X}_{2}])_{xx} + (b_{1}\mathbb{X}_{2})_{xxx} + [\mathbb{X}_{1},[\mathbb{X}_{1},(b_{1}\mathbb{X}_{2})_{x}]] - b_{1}[\mathbb{X}_{1},[\mathbb{X}_{1},[\mathbb{X}_{1},\mathbb{X}_{2}]]] \nonumber \\
&& + [\mathbb{X}_{2},\mathbb{X}_{0}] = 0 \\
O(v^{2}) &:&  - \frac{1}{2}([\mathbb{X}_{2},(b_{1}\mathbb{X}_{2})_{x}])_{x} + \frac{1}{2}(b_{1}[\mathbb{X}_{2},[\mathbb{X}_{1},\mathbb{X}_{2}]])_{x} + \frac{1}{2}[\mathbb{X}_{1},[\mathbb{X}_{2},(b_{1}\mathbb{X}_{2})_{x}]] - [\mathbb{X}_{2},(b_{1}\mathbb{X}_{2})_{xx}] \nonumber \\
&& - \frac{1}{2}b_{1}[\mathbb{X}_{1},[\mathbb{X}_{2},[\mathbb{X}_{1},\mathbb{X}_{2}]]] + [\mathbb{X}_{2},(b_{1}[\mathbb{X}_{1},\mathbb{X}_{2}])_{x}] - b_{1}[\mathbb{X}_{2},[\mathbb{X}_{1},[\mathbb{X}_{1},\mathbb{X}_{2}]]] \nonumber \\
&& + [\mathbb{X}_{2},[\mathbb{X}_{1},(b_{1}\mathbb{X}_{2})_{x}]] = 0 \\
O(v^{3}) &:& \frac{1}{3}(b_{2}\mathbb{X}_{2})_{x} - \frac{1}{3}b_{2}[\mathbb{X}_{1},\mathbb{X}_{2}] + \frac{1}{2}[\mathbb{X}_{2},[\mathbb{X}_{2},(b_{1}\mathbb{X}_{2})_{x}]] - \frac{1}{2}b_{1}[\mathbb{X}_{2},[\mathbb{X}_{2},[\mathbb{X}_{1},\mathbb{X}_{2}]]] = 0
\end{eqnarray}

\noindent
Note that if we decouple the $O(v^{3})$ equation into the following equations

\begin{eqnarray}
&& b_{1}[\mathbb{X}_{1},\mathbb{X}_{2}] - (b_{1}\mathbb{X}_{2})_{x} = 0 \\
&& b_{2}[\mathbb{X}_{1},\mathbb{X}_{2}] - (b_{2}\mathbb{X}_{2})_{x} = 0
\end{eqnarray}

\noindent
we find that the $O(v^{2})$ equation is immediately satisfied and the $O(v)$ equation reduces to

\begin{equation}
[\mathbb{X}_{2},\mathbb{X}_{0}] + \mathbb{X}_{2,t} = 0
\end{equation}

{\it Again we should note that had we opted instead for the forms}

\[ \mathbb{X}_{0} = \begin{bmatrix}
g_{1}(x,t) & g_{4}(x,t) \\
g_{10}(x,t) & g_{16}(x,t)
\end{bmatrix}, \ \ \mathbb{X}_{1} = \begin{bmatrix}
f_{1}(x,t) & f_{3}(x,t) \\
f_{5}(x,t) & f_{7}(x,t)
\end{bmatrix}, \ \ \mathbb{X}_{2} = \begin{bmatrix}
f_{2}(x,t) & f_{4}(x,t) \\
f_{6}(x,t) & f_{8}(x,t)
\end{bmatrix} \]

\noindent
{\it we would obtain an equivalent system to that obtained in $\cite{Lecce}$ for the mKdV. The additional unknown functions which appear in Khawaja's method ($\cite{Lecce}$) can again be introduced with the proper substitutions via their functional dependence on the twelve unknown functions given above.}

Therefore utilizing the same generators as in the generalized KdV equation we obtain the system of equations

\begin{eqnarray}
&& f_{1}f_{4} - f_{2}f_{3} = 0 \\
&& (b_{1}f_{j})_{x} = (b_{2}f_{j})_{x} = 0, \ \ \ j=3,4 \\
&& f_{3}g_{3} - f_{4}g_{2} = 0 \\
&& f_{jt} + (-1)^{j}f_{j}(g_{1} - g_{4}) = 0, \ \ \ j=3,4 \\
&& g_{jx} + (-1)^{j}(f_{1}g_{3} - f_{2}g_{2}) = 0, \ \ \ j=1,4 \\
&& f_{jt} - g_{(j+1)x} + (-1)^{j}f_{j}(g_{4} - g_{1}) = 0, \ \ \ j=2,3
\end{eqnarray}

Solving this system yields the following

\begin{eqnarray}
f_{j}(x,t) &=& \frac{F_{j}(t)}{b_{1}(x,t)}, \ \ \ j=3,4 \\
g_{3}(x,t) &=& \frac{F_{4}(t)g_{2}(x,t)}{F_{3}(t)} \\
f_{2}(x,t) &=& \frac{F_{4}(t)f_{1}(x,t)}{F_{3}(t)} \\
g_{1}(x,t) &=& G_{1}(t) \\
g_{4}(x,t) &=& G_{4}(t)
\end{eqnarray}

Subject to the constraints

\begin{eqnarray}
&& \left(\frac{F_{j}}{b_{1}}\right)_{t} + (-1)^{j}(G_{1} - G_{4}) = 0, \ \ \ j=3,4 \label{MKDVFINAL1} \\
&& \left(\frac{b_{2}F_{j}}{b_{1}}\right)_{x} = 0 \ \ \ j=3,4 \label{MKDVFINAL2} \\
&& f_{jt} - g_{(j+1)x} + (-1)^{j}f_{j}(G_{4} - G_{1}) = 0, \ \ \ j=2,3
\end{eqnarray}

Solving $\eqref{MKDVFINAL1}$ and $\eqref{MKDVFINAL2}$ for $F_{4},b_{1},b_{2}$ and $G_{4}$ we obtain

\begin{eqnarray}
F_{4}(t) &=& \frac{c_{1}F_{2}(t)^{2}}{F_{3}(t)} \\
b_{1}(x,t) &=& F_{1}(x)F_{2}(t) \\
b_{2}(x,t) &=& F_{1}(x)F_{5}(t) \\
G_{4}(t) &=& \frac{F_{3}(t)F_{2}'(t) - F_{2}(t)F_{3}'(t) + G_{1}(t)F_{2}(t)F_{3}(t)}{F_{2}(t)F_{3}(t)} \\
g_{2}(x,t) &=& \int{\frac{[f_{1}(x,t)]_{t}F_{3}(t)F_{2}(t) - f_{1}(x,t)F_{3}'(t)F_{2}(t) + f_{1}(x,t)F_{3}(t)F_{2}'(t)}{F_{3}(t)F_{2}(t)}dx} + F_{6}(t)
\end{eqnarray}

\noindent
where $F_{1}$ and $F_{2}$ are arbitrary functions in their respective variables and $c_{1}$ is an arbitrary constant. 

{\it The Lax pair for the variable-coefficient MKdV equation with the previous integrability conditions is thus given by}

\begin{eqnarray}
F &=& X_{1} + X_{2}v \\
G &=& -b_{1}X_{2}v_{xx} - \frac{1}{3}b_{2}\mathbb{X}_{2}v^{3} + \mathbb{X}_{0}
\end{eqnarray}

Next, we illustrate our generalized Estabrook-Wahlquist technique by applying it to 
the generalized  fifth-order vcKdV equation.

\section{The Generalized Fifth-Order KdV (vcKdV) Equation Reconsidered}

As a second example of the extended EW technique, let us consider the generalized KDV equation

\begin{equation} \label{KDV}
u_{t} + a_{1}uu_{xxx} + a_{2}u_{x}u_{xx} + a_{3}u^{2}u_{x} + a_{4}uu_{x} + a_{5}u_{xxx} + a_{6}u_{xxxxx} + a_{7}u + a_{8}u_{x} = 0
\end{equation}

\noindent
where $a_{1-8}$ are arbitrary functions of $x$ and $t$. As with the last example, we will go through the procedure outlined earlier in the paper and show how one can obtain the results previously obtained for the constant coefficient cases. Running through the standard procedure we let $\mathbb{F} = \mathbb{F}(x,t,u)$ and $\mathbb{G} = \mathbb{G}(x,t,u,u_{x},u_{xx},u_{xxx},u_{xxxx})$. Plugging this into $\eqref{ZCC}$ we obtain

\begin{equation}
\mathbb{F}_{t} + \mathbb{F}_{u}u_{t} - \mathbb{G}_{x} - \mathbb{G}_{u}u_{x} - \mathbb{G}_{u_{x}}u_{xx} - \mathbb{G}_{u_{xx}}u_{xxx} - \mathbb{G}_{u_{xxx}}u_{xxxx} - \mathbb{G}_{u_{xxxx}}u_{xxxxx} + [\mathbb{F},\mathbb{G}] = 0
\end{equation}

\noindent
Next, substituting $\eqref{KDV}$ into this expression in order to eliminate the $u_{t}$ yields

\begin{eqnarray} \nonumber
&& \mathbb{F}_{t} - \mathbb{F}_{u}\left(a_{1}uu_{xxx} + a_{2}u_{x}u_{xx} + a_{3}u^{2}u_{x} + a_{4}uu_{x} + a_{5}u_{xxx} + a_{6}u_{xxxxx} + a_{7}u + a_{8}u_{x}\right) - \mathbb{G}_{x} \\ \label{KDVstep1}
&& - \mathbb{G}_{u}u_{x} - \mathbb{G}_{u_{x}}u_{xx} - \mathbb{G}_{u_{xx}}u_{xxx} - \mathbb{G}_{u_{xxx}}u_{xxxx} - \mathbb{G}_{u_{xxxx}}u_{xxxxx} + [\mathbb{F},\mathbb{G}] = 0
\end{eqnarray}

Since $\mathbb{F}$ and $\mathbb{G}$ do not depend on $u_{xxxxx}$ we can equate the coefficient of the $u_{xxxxx}$ term to zero. This requires that we must have

\[ \mathbb{G}_{u_{xxxx}} + a_{6}\mathbb{F}_{u} = 0 \Rightarrow \mathbb{G} = -a_{6}\mathbb{F}_{u}u_{xxxx} + \mathbb{K}^{0}(x,t,u,u_{x},u_{xx},u_{xxx}) \]

\noindent
Now updating $\eqref{KDVstep1}$ we obtain

\begin{eqnarray}
&& \mathbb{F}_{t} - \mathbb{F}_{u}\left(a_{1}uu_{xxx} + a_{2}u_{x}u_{xx} + a_{3}u^{2}u_{x} + a_{4}uu_{x} + a_{5}u_{xxx} + a_{7}u + a_{8}u_{x}\right) + a_{6x}\mathbb{F}_{u}u_{xxxx} \nonumber \\
&&  + a_{6}\mathbb{F}_{xu}u_{xxxx} - \mathbb{K}^{0}_{x} - \mathbb{K}^{0}_{u}u_{x} - \mathbb{K}^{0}_{u_{x}}u_{xx} - \mathbb{K}^{0}_{u_{xx}}u_{xxx} - \mathbb{K}^{0}_{u_{xxx}}u_{xxxx} + a_{6}\mathbb{F}_{uu}u_{x}u_{xxxx} \nonumber \\ 
&& - [\mathbb{F},\mathbb{F}_{u}]a_{6}u_{xxxx} + [\mathbb{F},\mathbb{K}^{0}] = 0 \label{KDVstep2}
\end{eqnarray}

Since $\mathbb{F}$ and $\mathbb{K}^{0}$ do not depend on $u_{xxxx}$ we can equate the coefficient of the $u_{xxxx}$ term to zero. This requires that we have

\begin{equation}
a_{6x}\mathbb{F}_{u} + a_{6}\mathbb{F}_{xu} + a_{6}\mathbb{F}_{uu}u_{x}- \mathbb{K}^{0}_{u_{xxx}} - [\mathbb{F},\mathbb{F}_{u}]a_{6} = 0
\end{equation}

\noindent
Thus, integrating with respect to $u_{xxx}$ and solving for $\mathbb{K}^{0}$ we have

\begin{equation}
\mathbb{K}^{0} = a_{6x}\mathbb{F}_{u}u_{xxx} + a_{6}\mathbb{F}_{xu}u_{xxx} + a_{6}\mathbb{F}_{uu}u_{x}u_{xxx} - [\mathbb{F},\mathbb{F}_{u}]a_{6}u_{xxx} + \mathbb{K}^{1}(x,t,u,u_{x},u_{xx})
\end{equation}

\noindent
Now we update $\eqref{KDVstep2}$ by plugging in our expression for $\mathbb{K}^{1}$ to obtain

\begin{eqnarray} 
&& \mathbb{F}_{t} - \mathbb{F}_{u}\left(a_{1}uu_{xxx} + a_{2}u_{x}u_{xx} + a_{3}u^{2}u_{x} + a_{4}uu_{x} + a_{5}u_{xxx} + a_{7}u + a_{8}u_{x}\right) - a_{6xx}\mathbb{F}_{u}u_{xxx} \nonumber \\
&& - 2a_{6x}\mathbb{F}_{xu}u_{xxx} - a_{6}\mathbb{F}_{xxu}u_{xxx} - a_{6x}\mathbb{F}_{uu}u_{x}u_{xxx} - a_{6}\mathbb{F}_{xuu}u_{x}u_{xxx} + [\mathbb{F}_{x},\mathbb{F}_{u}]a_{6}u_{xxx} \nonumber \\
&& + [\mathbb{F},\mathbb{F}_{xu}]a_{6}u_{xxx} + [\mathbb{F},\mathbb{F}_{u}]a_{6x}u_{xxx} - \mathbb{K}^{1}_{x} - a_{6x}\mathbb{F}_{uu}u_{x}u_{xxx} - a_{6}\mathbb{F}_{xuu}u_{x}u_{xxx} \nonumber \\
&& - a_{6}\mathbb{F}_{uuu}u_{x}^{2}u_{xxx} + [\mathbb{F},\mathbb{F}_{uu}]a_{6}u_{x}u_{xxx} - \mathbb{K}^{1}_{u}u_{x} - a_{6}\mathbb{F}_{uu}u_{xx}u_{xxx} - \mathbb{K}^{1}_{u_{x}}u_{xx} - \mathbb{K}^{1}_{u_{xx}}u_{xxx} \nonumber \\ 
&&  + a_{6x}[\mathbb{F},\mathbb{F}_{u}]u_{xxx} + a_{6}[\mathbb{F},\mathbb{F}_{xu}]u_{xxx} + a_{6}[\mathbb{F},\mathbb{F}_{uu}]u_{x}u_{xxx} - [\mathbb{F},[\mathbb{F},\mathbb{F}_{u}]]a_{6}u_{xxx} + [\mathbb{F},\mathbb{K}^{1}] = 0 \label{KDV3}
\end{eqnarray}

Since $\mathbb{F}$ and $\mathbb{K}^{1}$ do not depend on $u_{xxx}$ we can equate the coefficient of the $u_{xxx}$ term to zero. This requires that we have

\begin{eqnarray} 
&& - \mathbb{F}_{u}(a_{1}u + a_{5}) - a_{6xx}\mathbb{F}_{u} - 2a_{6x}\mathbb{F}_{xu} - a_{6}\mathbb{F}_{xxu} - a_{6x}\mathbb{F}_{uu}u_{x} - a_{6}\mathbb{F}_{xuu}u_{x} \nonumber \\
&& + [\mathbb{F}_{x},\mathbb{F}_{u}]a_{6} + [\mathbb{F},\mathbb{F}_{xu}]a_{6} + [\mathbb{F},\mathbb{F}_{u}]a_{6x} - a_{6x}\mathbb{F}_{uu}u_{x} - a_{6}\mathbb{F}_{xuu}u_{x} - a_{6}\mathbb{F}_{uuu}u_{x}^{2} \nonumber \\
&& + [\mathbb{F},\mathbb{F}_{uu}]a_{6}u_{x} - a_{6}\mathbb{F}_{uu}u_{xx} - \mathbb{K}^{1}_{u_{xx}} + a_{6x}[\mathbb{F},\mathbb{F}_{u}] + a_{6}[\mathbb{F},\mathbb{F}_{xu}] + a_{6}[\mathbb{F},\mathbb{F}_{uu}]u_{x} \nonumber \\ 
&& - [\mathbb{F},[\mathbb{F},\mathbb{F}_{u}]]a_{6} = 0
\end{eqnarray}

\noindent
Integrating with respect to $u_{xx}$ and solving for $\mathbb{K}^{1}$ and collecting like terms we have

\begin{eqnarray}
\mathbb{K}^{1} &=& -\mathbb{F}_{u}(a_{1}u + a_{5})u_{xx} - (a_{6}\mathbb{F}_{u})_{xx}u_{xx} - 2(a_{6}\mathbb{F}_{uu})_{x}u_{x}u_{xx} + 2(a_{6}[\mathbb{F},\mathbb{F}_{u}])_{x}u_{xx} \nonumber \\
&& - a_{6}\mathbb{F}_{uuu}u_{x}^{2}u_{xx} + 2a_{6}[\mathbb{F},\mathbb{F}_{uu}]u_{x}u_{xx} - \frac{1}{2}a_{6}\mathbb{F}_{uu}u_{xx}^{2} - a_{6}[\mathbb{F}_{x},\mathbb{F}_{u}]u_{xx} \nonumber \\
&& - a_{6}[\mathbb{F},[\mathbb{F},\mathbb{F}_{u}]]u_{xx} + \mathbb{K}^{2}(x,t,u,u_{x}) \label{KDV4}
\end{eqnarray}

Plugging $\eqref{KDV4}$ into $\eqref{KDV3}$ and simplifying a little bit we obtain

\begin{eqnarray} 
&& \mathbb{F}_{t} - \mathbb{F}_{u}(a_{2}u_{x}u_{xx} + a_{3}u^{2}u_{x} + a_{4}uu_{x} + a_{7}u + a_{8}u_{x}) + (a_{1}\mathbb{F}_{u})_{x}uu_{xx} + (a_{5}\mathbb{F}_{u})_{x}u_{xx} \nonumber \\
&& + (a_{6}\mathbb{F}_{u})_{xxx}u_{xx} + 2(a_{6}\mathbb{F}_{uu})_{xx}u_{x}u_{xx} - (a_{6}[\mathbb{F},\mathbb{F}_{u}])_{xx}u_{xx} + (a_{6}\mathbb{F}_{uuu})_{x}u_{x}^{2}u_{xx} \nonumber \\
&& + \frac{1}{2}(a_{6}\mathbb{F}_{uu})_{x}u_{xx}^{2}  - ([\mathbb{F},(a_{6}\mathbb{F}_{u})_{x}])_{x}u_{xx} + (a_{6}[\mathbb{F},[\mathbb{F},\mathbb{F}_{u}]])_{x}u_{xx} - \mathbb{K}^{2}_{x} + \mathbb{F}_{uu}a_{1}uu_{x}u_{xx} \nonumber \\
&& + \mathbb{F}_{u}a_{1}u_{x}u_{xx} + \mathbb{F}_{uu}a_{5}u_{x}u_{xx} + (a_{6}\mathbb{F}_{uu})_{xx}u_{x}u_{xx} + 2(a_{6}\mathbb{F}_{uuu})_{x}u_{x}^{2}u_{xx} + a_{6}\mathbb{F}_{uuuu}u_{x}^{3}u_{xx} \nonumber \\
&& - a_{6}[\mathbb{F}_{u},\mathbb{F}_{uu}]u_{x}^{2}u_{xx} - a_{6}[\mathbb{F},\mathbb{F}_{uuu}]u_{x}^{2}u_{xx} + \frac{5}{2}a_{6}\mathbb{F}_{uuu}u_{xx}^{2}u_{x} - [\mathbb{F}_{u},(a_{6}\mathbb{F}_{u})_{x}]u_{x}u_{xx} \nonumber \\
&& - 2[\mathbb{F},(a_{6}\mathbb{F}_{uu})_{x}]u_{x}u_{xx} - a_{6}[\mathbb{F}_{u},\mathbb{F}_{uu}]u_{x}^{2}u_{xx} - a_{6}[\mathbb{F},\mathbb{F}_{uuu}]u_{x}^{2}u_{xx} + a_{6}[\mathbb{F}_{u},[\mathbb{F},\mathbb{F}_{u}]]u_{x}u_{xx} \nonumber \\
&& + a_{6}[\mathbb{F},[\mathbb{F},\mathbb{F}_{uu}]]u_{x}u_{xx} - \mathbb{K}^{2}_{u}u_{x} + 2(a_{6}\mathbb{F}_{uu})_{x}u_{xx}^{2} - \frac{3}{2}a_{6}[\mathbb{F},\mathbb{F}_{uu}]u_{xx}^{2} - \mathbb{K}^{2}_{u_{x}}u_{xx} - a_{1}[\mathbb{F},\mathbb{F}_{u}]uu_{xx} \nonumber \\
&& - a_{5}[\mathbb{F},\mathbb{F}_{u}]u_{xx} - [\mathbb{F},(a_{6}\mathbb{F}_{u})_{xx}]u_{xx} - [\mathbb{F},(a_{6}\mathbb{F}_{uu})_{x}]u_{x}u_{xx} + [\mathbb{F},(a_{6}[\mathbb{F},\mathbb{F}_{u}])_{x}]u_{xx} \nonumber \\
&& - a_{6}[\mathbb{F},\mathbb{F}_{uuu}]u_{x}^{2}u_{xx} + 2a_{6}[\mathbb{F},[\mathbb{F},\mathbb{F}_{uu}]]u_{x}u_{xx} + [\mathbb{F},[\mathbb{F},(a_{6}\mathbb{F}_{u})_{x}]]u_{xx} - 3(a_{6}[\mathbb{F},\mathbb{F}_{uu}])_{x}u_{x}u_{xx} \nonumber \\
&& - a_{6}[\mathbb{F},[\mathbb{F},[\mathbb{F},\mathbb{F}_{u}]]]u_{xx} + [\mathbb{F},\mathbb{K}^{2}] = 0 \label{KDV5}
\end{eqnarray}

Now, since $\mathbb{K}^{2}$ and $\mathbb{F}$ do not depend on $u_{xx}$ we can start by setting the coefficients of the $u_{xx}^{2}$ and the $u_{xx}$ terms to zero. Note the difference here that we have multiple powers of $u_{xx}$ present in the $\eqref{KDV5}$. Setting the $O(u_{xx}^{2})$ term to zero requires

\begin{equation}
\frac{3}{2}(a_{6}\mathbb{F}_{uu})_{x} + \frac{5}{2}a_{6}\mathbb{F}_{uuu}u_{x} - \frac{3}{2}a_{6}[\mathbb{F},\mathbb{F}_{uu}]= 0 \label{DetermF}
\end{equation}

\noindent
Since $\mathbb{F}$ does not depend on $u_{x}$ we must have that the coefficient of the $u_{x}$ term in this previous expression is zero. This is equivalent to

\[ \mathbb{F}_{uuu} = 0 \Rightarrow \mathbb{F} = \mathbb{X}_{1}(x,t) + \mathbb{X}_{2}(x,t)u + \frac{1}{2}\mathbb{X}_{3}(x,t)u^{2} \]

\noindent
Plugging this into $\eqref{DetermF}$ we obtain

\begin{equation}
3(a_{6}\mathbb{X}_{3})_{x} - 3a_{6}([\mathbb{X}_{1},\mathbb{X}_{3}] + [\mathbb{X}_{2},\mathbb{X}_{3}]u) = 0 \label{CondX3}
\end{equation}

\noindent
Now since the $\mathbb{X}_{i}$ do not depend on $u$ we can set the coefficient of the $u$ to zero. That is, we require that $\mathbb{X}_{2}$ and $\mathbb{X}_{3}$ commute. We find now that $\eqref{CondX3}$ reduces to the condition

\begin{equation}
(a_{6}\mathbb{X}_{3})_{x} - a_{6}[\mathbb{X}_{1},\mathbb{X}_{3}] = 0 \label{CondX3final}
\end{equation}

\noindent
For ease of computation and in order to immediately satisfy $\eqref{CondX3final}$ we set $\mathbb{X}_{3} = 0$. Plugging into $\eqref{KDV5}$ our expression for $\mathbb{F}$ we obtain

\begin{eqnarray} 
&& \mathbb{X}_{1,t} + \mathbb{X}_{2,t}u - \mathbb{X}_{2}(a_{2}u_{x}u_{xx} + a_{3}u^{2}u_{x} + a_{4}uu_{x} + a_{7}u + a_{8}u_{x}) + (a_{1}\mathbb{X}_{2})_{x}uu_{xx} + (a_{5}\mathbb{X}_{2})_{x}u_{xx} \nonumber \\
&& + (a_{6}\mathbb{X}_{2})_{xxx}u_{xx} - (a_{6}[\mathbb{X}_{1},\mathbb{X}_{2}])_{xx}u_{xx} - \mathbb{K}^{2}_{x} + \mathbb{X}_{2}a_{1}u_{x}u_{xx} - [\mathbb{X}_{2},(a_{6}\mathbb{X}_{2})_{x}]u_{x}u_{xx} \nonumber \\
&& - ([\mathbb{X}_{1},(a_{6}\mathbb{X}_{2})_{x}])_{x}u_{xx} - ([\mathbb{X}_{2},(a_{6}\mathbb{X}_{2})_{x}])_{x}uu_{xx} + (a_{6}[\mathbb{X}_{1},[\mathbb{X}_{1},\mathbb{X}_{2}]])_{x}u_{xx} + (a_{6}[\mathbb{X}_{2},[\mathbb{X}_{1},\mathbb{X}_{2}]])_{x}uu_{xx} \nonumber \\
&& - \mathbb{K}^{2}_{u}u_{x} - \mathbb{K}^{2}_{u_{x}}u_{xx} - a_{1}[\mathbb{X}_{1},\mathbb{X}_{2}]uu_{xx} + a_{6}[\mathbb{X}_{2},[\mathbb{X}_{1},\mathbb{X}_{2}]]u_{x}u_{xx} + [\mathbb{X}_{2},(a_{6}[\mathbb{X}_{1},\mathbb{X}_{2}])_{x}]uu_{xx} \nonumber \\
&& - a_{5}[\mathbb{X}_{1},\mathbb{X}_{2}]u_{xx} - [\mathbb{X}_{1},(a_{6}\mathbb{X}_{2})_{xx}]u_{xx} - [\mathbb{X}_{2},(a_{6}\mathbb{X}_{2})_{xx}]uu_{xx} + [\mathbb{X}_{1},(a_{6}[\mathbb{X}_{1},\mathbb{X}_{2}])_{x}]u_{xx} \nonumber \\
&& + [\mathbb{X}_{1},[\mathbb{X}_{1},(a_{6}\mathbb{X}_{2})_{x}]]u_{xx} + [\mathbb{X}_{1},[\mathbb{X}_{2},(a_{6}\mathbb{X}_{2})_{x}]]uu_{xx} + [\mathbb{X}_{2},[\mathbb{X}_{1},(a_{6}\mathbb{X}_{2})_{x}]]uu_{xx} + [\mathbb{X}_{1},\mathbb{K}^{2}] \nonumber \\
&& - a_{6}[\mathbb{X}_{1},[\mathbb{X}_{1},[\mathbb{X}_{1},\mathbb{X}_{2}]]]u_{xx} - a_{6}[\mathbb{X}_{1},[\mathbb{X}_{2},[\mathbb{X}_{1},\mathbb{X}_{2}]]]uu_{xx} - a_{6}[\mathbb{X}_{2},[\mathbb{X}_{1},[\mathbb{X}_{1},\mathbb{X}_{2}]]]uu_{xx} \nonumber \\
&& + [\mathbb{X}_{2},[\mathbb{X}_{2},(a_{6}\mathbb{X}_{2})_{x}]]u^{2}u_{xx} - a_{6}[\mathbb{X}_{2},[\mathbb{X}_{2},[\mathbb{X}_{1},\mathbb{X}_{2}]]]u^{2}u_{xx} + [\mathbb{X}_{2},\mathbb{K}^{2}]u = 0 \label{KDV6}
\end{eqnarray}

Now again using the fact that the $\mathbb{X}_{i}$ and $\mathbb{K}^{2}$ do not depend on $u_{xx}$ we can set the coefficient of the $u_{xx}$ term in $\eqref{KDV6}$ to zero. This requires

\begin{eqnarray}
&& (a_{6}\mathbb{X}_{2})_{xxx} - (a_{6}[\mathbb{X}_{1},\mathbb{X}_{2}])_{xx} + a_{1}\mathbb{X}_{2}u_{x} - [\mathbb{X}_{2},(a_{6}\mathbb{X}_{2})_{x}]u_{x} - a_{2}\mathbb{X}_{2}u_{x} \nonumber \\
&& - ([\mathbb{X}_{1},(a_{6}\mathbb{X}_{2})_{x}])_{x} - ([\mathbb{X}_{2},(a_{6}\mathbb{X}_{2})_{x}])_{x}u + (a_{6}[\mathbb{X}_{1},[\mathbb{X}_{1},\mathbb{X}_{2}]])_{x} + (a_{6}[\mathbb{X}_{2},[\mathbb{X}_{1},\mathbb{X}_{2}]])_{x}u \nonumber \\
&& - \mathbb{K}^{2}_{u_{x}} - a_{1}[\mathbb{X}_{1},\mathbb{X}_{2}]u + a_{6}[\mathbb{X}_{2},[\mathbb{X}_{1},\mathbb{X}_{2}]]u_{x} + [\mathbb{X}_{2},(a_{6}[\mathbb{X}_{1},\mathbb{X}_{2}])_{x}]u + (a_{5}\mathbb{X}_{2})_{x} \nonumber \\
&& - a_{5}[\mathbb{X}_{1},\mathbb{X}_{2}] - [\mathbb{X}_{1},(a_{6}\mathbb{X}_{2})_{xx}] - [\mathbb{X}_{2},(a_{6}\mathbb{X}_{2})_{xx}]u + [\mathbb{X}_{1},(a_{6}[\mathbb{X}_{1},\mathbb{X}_{2}])_{x}] \nonumber \\
&& + [\mathbb{X}_{1},[\mathbb{X}_{1},(a_{6}\mathbb{X}_{2})_{x}]] + [\mathbb{X}_{1},[\mathbb{X}_{2},(a_{6}\mathbb{X}_{2})_{x}]]u + [\mathbb{X}_{2},[\mathbb{X}_{1},(a_{6}\mathbb{X}_{2})_{x}]]u + (a_{1}\mathbb{X}_{2})_{x}u \nonumber \\
&& - a_{6}[\mathbb{X}_{1},[\mathbb{X}_{1},[\mathbb{X}_{1},\mathbb{X}_{2}]]] - a_{6}[\mathbb{X}_{1},[\mathbb{X}_{2},[\mathbb{X}_{1},\mathbb{X}_{2}]]]u - a_{6}[\mathbb{X}_{2},[\mathbb{X}_{1},[\mathbb{X}_{1},\mathbb{X}_{2}]]]u \nonumber \\
&& + [\mathbb{X}_{2},[\mathbb{X}_{2},(a_{6}\mathbb{X}_{2})_{x}]]u^{2} - a_{6}[\mathbb{X}_{2},[\mathbb{X}_{2},[\mathbb{X}_{1},\mathbb{X}_{2}]]]u^{2} = 0 \label{KDV7}
\end{eqnarray}

\noindent
Thus integrating with respect to $u_{x}$ and solving for $\mathbb{K}^{2}$ we have

\begin{eqnarray} 
\mathbb{K}^{2} &=& (a_{6}\mathbb{X}_{2})_{xxx}u_{x} + \frac{1}{2}a_{1}\mathbb{X}_{2}u_{x}^{2} - \frac{1}{2}[\mathbb{X}_{2},(a_{6}\mathbb{X}_{2})_{x}]u_{x}^{2} - \frac{1}{2}a_{2}\mathbb{X}_{2}u_{x}^{2} + (a_{6}[\mathbb{X}_{2},[\mathbb{X}_{1},\mathbb{X}_{2}]])_{x}uu_{x} \nonumber \\
&& - (a_{6}[\mathbb{X}_{1},\mathbb{X}_{2}])_{xx}u_{x} - ([\mathbb{X}_{1},(a_{6}\mathbb{X}_{2})_{x}])_{x}u_{x} - ([\mathbb{X}_{2},(a_{6}\mathbb{X}_{2})_{x}])_{x}uu_{x} + (a_{6}[\mathbb{X}_{1},[\mathbb{X}_{1},\mathbb{X}_{2}]])_{x}u_{x} \nonumber \\
&& - a_{1}[\mathbb{X}_{1},\mathbb{X}_{2}]uu_{x} + \frac{1}{2}a_{6}[\mathbb{X}_{2},[\mathbb{X}_{1},\mathbb{X}_{2}]]u_{x}^{2} + [\mathbb{X}_{2},(a_{6}[\mathbb{X}_{1},\mathbb{X}_{2}])_{x}]uu_{x} + (a_{5}\mathbb{X}_{2})_{x}u_{x} \nonumber \\
&& - a_{5}[\mathbb{X}_{1},\mathbb{X}_{2}]u_{x} - [\mathbb{X}_{1},(a_{6}\mathbb{X}_{2})_{xx}]u_{x} - [\mathbb{X}_{2},(a_{6}\mathbb{X}_{2})_{xx}]uu_{x} + [\mathbb{X}_{1},(a_{6}[\mathbb{X}_{1},\mathbb{X}_{2}])_{x}]u_{x} \nonumber \\
&& + [\mathbb{X}_{1},[\mathbb{X}_{1},(a_{6}\mathbb{X}_{2})_{x}]]u_{x} + [\mathbb{X}_{1},[\mathbb{X}_{2},(a_{6}\mathbb{X}_{2})_{x}]]uu_{x} + [\mathbb{X}_{2},[\mathbb{X}_{1},(a_{6}\mathbb{X}_{2})_{x}]]uu_{x} + (a_{1}\mathbb{X}_{2})_{x}uu_{x} \nonumber \\
&& - a_{6}[\mathbb{X}_{1},[\mathbb{X}_{1},[\mathbb{X}_{1},\mathbb{X}_{2}]]]u_{x} - a_{6}[\mathbb{X}_{1},[\mathbb{X}_{2},[\mathbb{X}_{1},\mathbb{X}_{2}]]]uu_{x} - a_{6}[\mathbb{X}_{2},[\mathbb{X}_{1},[\mathbb{X}_{1},\mathbb{X}_{2}]]]uu_{x} \nonumber \\
&& + [\mathbb{X}_{2},[\mathbb{X}_{2},(a_{6}\mathbb{X}_{2})_{x}]]u^{2}u_{x} - a_{6}[\mathbb{X}_{2},[\mathbb{X}_{2},[\mathbb{X}_{1},\mathbb{X}_{2}]]]u^{2}u_{x} + \mathbb{K}^{3}(x,t,u) \label{KDV8}
\end{eqnarray}

It is helpful at this stage to define the following new matrices

\begin{eqnarray}
&& \mathbb{X}_{4} = [\mathbb{X}_{1},\mathbb{X}_{2}], \ \ \mathbb{X}_{5} = [\mathbb{X}_{1},\mathbb{X}_{4}], \ \ \mathbb{X}_{6} = [\mathbb{X}_{2},\mathbb{X}_{4}] \\
&& \mathbb{X}_{7} = [\mathbb{X}_{1},\mathbb{X}_{5}], \ \ \mathbb{X}_{8} = [\mathbb{X}_{2},\mathbb{X}_{5}], \ \ \mathbb{X}_{9} = [\mathbb{X}_{1},\mathbb{X}_{6}], \ \ \mathbb{X}_{10} = [\mathbb{X}_{2},\mathbb{X}_{6}]
\end{eqnarray}

Now we update $\eqref{KDV6}$ by plugging in $\eqref{KDV8}$. This yields a long expression which is \eqref{KDV9} in Appendix C.

Now since $\mathbb{K}^{3}$ and the $\mathbb{X}_{i}$ do not depend on $u_{x}$ we can set the coefficient of the $u_{x}^{2}$ term to zero \eqref{KDV9}. Therefore we require

\begin{eqnarray}
&& - \frac{1}{2}(a_{1}\mathbb{X}_{2})_{x} + \frac{1}{2}([\mathbb{X}_{2},(a_{6}\mathbb{X}_{2})_{x}])_{x} + \frac{1}{2}(a_{2}\mathbb{X}_{2})_{x} - (a_{6}\mathbb{X}_{6})_{x} + a_{1}\mathbb{X}_{4} \nonumber \\
&& - \frac{1}{2}(a_{6}\mathbb{X}_{6})_{x} + ([\mathbb{X}_{2},(a_{6}\mathbb{X}_{2})_{x}])_{x} + a_{6}\mathbb{X}_{9} + [\mathbb{X}_{2},(a_{6}\mathbb{X}_{2})_{xx}] + \frac{1}{2}a_{6}\mathbb{X}_{10}u \nonumber \\
&& - [\mathbb{X}_{1},[\mathbb{X}_{2},(a_{6}\mathbb{X}_{2})_{x}]] - [\mathbb{X}_{2},[\mathbb{X}_{1},(a_{6}\mathbb{X}_{2})_{x}]] - (a_{1}\mathbb{X}_{2})_{x} - [\mathbb{X}_{2},(a_{6}\mathbb{X}_{4})_{x}] \nonumber \\
&& - 2[\mathbb{X}_{2},[\mathbb{X}_{2},(a_{6}\mathbb{X}_{2})_{x}]]u + 2a_{6}\mathbb{X}_{10}u - \frac{1}{2}a_{2}\mathbb{X}_{4} + a_{6}\mathbb{X}_{8} + \frac{1}{2}a_{1}\mathbb{X}_{4} \nonumber \\
&& - \frac{1}{2}[\mathbb{X}_{1},[\mathbb{X}_{2},(a_{6}\mathbb{X}_{2})_{x}]] + \frac{1}{2}a_{6}\mathbb{X}_{9} - \frac{1}{2}[\mathbb{X}_{2},[\mathbb{X}_{2},(a_{6}\mathbb{X}_{2})_{x}]]u = 0 \label{KDV10}
\end{eqnarray}

Further since we know that the $\mathbb{X}_{i}$ do not depend on $u$ we can decouple this condition as follows.

\begin{eqnarray}
&& \frac{3}{2}([\mathbb{X}_{2},(a_{6}\mathbb{X}_{2})_{x}])_{x} - \frac{3}{2}(a_{1}\mathbb{X}_{2})_{x} + \frac{1}{2}(a_{2}\mathbb{X}_{2})_{x} - \frac{3}{2}(a_{6}\mathbb{X}_{6})_{x} + \frac{3}{2}a_{1}\mathbb{X}_{4} + \frac{3}{2}a_{6}\mathbb{X}_{9} \nonumber \\
&& - \frac{3}{2}[\mathbb{X}_{1},[\mathbb{X}_{2},(a_{6}\mathbb{X}_{2})_{x}]] - [\mathbb{X}_{2},[\mathbb{X}_{1},(a_{6}\mathbb{X}_{2})_{x}]] - [\mathbb{X}_{2},(a_{6}\mathbb{X}_{4})_{x}] + [\mathbb{X}_{2},(a_{6}\mathbb{X}_{2})_{xx}] \nonumber \\
&& - \frac{1}{2}a_{2}\mathbb{X}_{4} + a_{6}\mathbb{X}_{8} = 0  \label{KDV11} \\
&& a_{6}\mathbb{X}_{10} - [\mathbb{X}_{2},[\mathbb{X}_{2},(a_{6}\mathbb{X}_{2})_{x}]] = 0 \label{KDV12}
\end{eqnarray}

Taking these conditions into account and once again noting the fact that $\mathbb{K}^{3}$ and the $\mathbb{X}_{i}$ are not independent of $u_{x}$ we can simplify and equate the coefficient of the $u_{x}$ in $\eqref{KDV9}$ to zero. Thus we now obtain the condition \eqref{KDV13} in Appendix C.

Now we update $\eqref{KDV9}$ by plugging in $\eqref{KDV14}$. Upon doing this we will have a rather large expression in which is no more than a algebraic equation in $u$. We will find our remaining constraints by equating the coefficients of the different powers of $u$ in this expression to zero. This updated version of $\eqref{KDV9}$ is very lengthy, and omitted here.

Now, in the final step, as the $\mathbb{X}_{i}$ do not depend on $u$ we can set the coefficients of the different powers of $u$ in this last, lengthy expression to zero. Thus we have

\begin{eqnarray}
O(1) &:& \mathbb{X}_{1,t} - \mathbb{X}_{0,x} + [\mathbb{X}_{1},\mathbb{X}_{0}] = 0 \\
O(u) &:& [\mathbb{X}_{2},\mathbb{X}_{0}] - a_{6}[\mathbb{X}_{1},[\mathbb{X}_{1},\mathbb{X}_{7}]] + [\mathbb{X}_{1},[\mathbb{X}_{1},[\mathbb{X}_{1},(a_{6}\mathbb{X}_{4})_{x}]]] + [\mathbb{X}_{1},[\mathbb{X}_{1},[\mathbb{X}_{1},[\mathbb{X}_{1},(a_{6}\mathbb{X}_{2})_{x}]]]] \nonumber \\
&& + [\mathbb{X}_{1},[\mathbb{X}_{1},(a_{5}\mathbb{X}_{2})_{x}]] - [\mathbb{X}_{1},[\mathbb{X}_{1},[\mathbb{X}_{1},(a_{6}\mathbb{X}_{2})_{xx}]]] + [\mathbb{X}_{1},(a_{6}\mathbb{X}_{7})_{x}] - [\mathbb{X}_{1},[\mathbb{X}_{1},(a_{6}\mathbb{X}_{4})_{xx}]] \nonumber \\
&& - a_{5}\mathbb{X}_{7} - [\mathbb{X}_{1},[\mathbb{X}_{1},([\mathbb{X}_{1},(a_{6}\mathbb{X}_{2})_{x}])_{x}]] + [\mathbb{X}_{1},[\mathbb{X}_{1},(a_{6}\mathbb{X}_{2})_{xxx}]] - [\mathbb{X}_{1},([\mathbb{X}_{1},(a_{6}\mathbb{X}_{4})_{x}])_{x}] \nonumber \\
&& - [\mathbb{X}_{1},([\mathbb{X}_{1},[\mathbb{X}_{1},(a_{6}\mathbb{X}_{2})_{x}]])_{x}] + [\mathbb{X}_{1},(a_{5}\mathbb{X}_{4})_{x}] + [\mathbb{X}_{1},([\mathbb{X}_{1},(a_{6}\mathbb{X}_{2})_{xx}])_{x}] - [\mathbb{X}_{1},(a_{5}\mathbb{X}_{2})_{xx}] \nonumber \\
&& - [\mathbb{X}_{1},(a_{6}\mathbb{X}_{5})_{xx}] + [\mathbb{X}_{1},(a_{6}\mathbb{X}_{4})_{xxx}] + [\mathbb{X}_{1},([\mathbb{X}_{1},(a_{6}\mathbb{X}_{2})_{x}])_{xx}] - a_{8}\mathbb{X}_{4} - [\mathbb{X}_{1},(a_{6}\mathbb{X}_{2})_{xxxx}] \nonumber \\
&& + [\mathbb{X}_{1},[\mathbb{X}_{1},(a_{6}\mathbb{X}_{5})_{x}]] + (a_{6}[\mathbb{X}_{1},\mathbb{X}_{7}])_{x} - ([\mathbb{X}_{1},[\mathbb{X}_{1},[\mathbb{X}_{1},(a_{6}\mathbb{X}_{2})_{x}]]])_{x} - ([\mathbb{X}_{1},[\mathbb{X}_{1},(a_{6}\mathbb{X}_{4})_{x}]])_{x} \nonumber \\
&& + (a_{8}\mathbb{X}_{2})_{x} + ([\mathbb{X}_{1},[\mathbb{X}_{1},(a_{6}\mathbb{X}_{2})_{xx}]])_{x} - ([\mathbb{X}_{1},(a_{5}\mathbb{X}_{2})_{x}])_{x} - (a_{6}\mathbb{X}_{7})_{xx} + ([\mathbb{X}_{1},(a_{6}\mathbb{X}_{4})_{xx}])_{x} \nonumber \\
&& + ([\mathbb{X}_{1},([\mathbb{X}_{1},(a_{6}\mathbb{X}_{2})_{x}])_{x}])_{x} - ([\mathbb{X}_{1},(a_{6}\mathbb{X}_{2})_{xxx}])_{x} + ([\mathbb{X}_{1},[\mathbb{X}_{1},(a_{6}\mathbb{X}_{2})_{x}]])_{xx} + (a_{5}\mathbb{X}_{5})_{x} \nonumber \\
&& + \mathbb{X}_{2,t} - ([\mathbb{X}_{1},(a_{6}\mathbb{X}_{2})_{xx}])_{xx} - (a_{5}\mathbb{X}_{4})_{xx} + (a_{5}\mathbb{X}_{2})_{xxx} + (a_{6}\mathbb{X}_{5})_{xxx} - ([\mathbb{X}_{1},(a_{6}\mathbb{X}_{2})_{x}])_{xxx} \nonumber \\
&& - ([\mathbb{X}_{1},(a_{6}\mathbb{X}_{5})_{x}])_{x} + (a_{6}\mathbb{X}_{2})_{xxxxx} - (a_{6}\mathbb{X}_{4})_{xxxx} - a_{7}\mathbb{X}_{2} + ([\mathbb{X}_{1},(a_{6}\mathbb{X}_{4})_{x}])_{xx} = 0
\end{eqnarray}

\begin{eqnarray}
O(u^{2}) &:& - a_{5}\mathbb{X}_{8} - a_{6}[\mathbb{X}_{2},[\mathbb{X}_{1},\mathbb{X}_{7}]] + [\mathbb{X}_{2},[\mathbb{X}_{1},[\mathbb{X}_{1},[\mathbb{X}_{1},(a_{6}\mathbb{X}_{2})_{x}]]]] + [\mathbb{X}_{2},[\mathbb{X}_{1},[\mathbb{X}_{1},(a_{6}\mathbb{X}_{4})_{x}]]] \nonumber \\
&& - [\mathbb{X}_{2},[\mathbb{X}_{1},[\mathbb{X}_{1},(a_{6}\mathbb{X}_{2})_{xx}]]] + [\mathbb{X}_{2},[\mathbb{X}_{1},(a_{5}\mathbb{X}_{2})_{x}]] - [\mathbb{X}_{2},[\mathbb{X}_{1},([\mathbb{X}_{1},(a_{6}\mathbb{X}_{2})_{x}])_{x}]] - \frac{1}{2}a_{4}\mathbb{X}_{4} \nonumber \\
&& + [\mathbb{X}_{2},(a_{6}\mathbb{X}_{7})_{x}] - [\mathbb{X}_{2},[\mathbb{X}_{1},(a_{6}\mathbb{X}_{4})_{xx}]] + [\mathbb{X}_{2},[\mathbb{X}_{1},(a_{6}\mathbb{X}_{2})_{xxx}]] - [\mathbb{X}_{2},([\mathbb{X}_{1},(a_{6}\mathbb{X}_{4})_{x}])_{x}] \nonumber \\
&& - [\mathbb{X}_{2},([\mathbb{X}_{1},[\mathbb{X}_{1},(a_{6}\mathbb{X}_{2})_{x}]])_{x}] + [\mathbb{X}_{2},(a_{5}\mathbb{X}_{4})_{x}] + [\mathbb{X}_{2},([\mathbb{X}_{1},(a_{6}\mathbb{X}_{2})_{xx}])_{x}] - [\mathbb{X}_{2},(a_{5}\mathbb{X}_{2})_{xx}] \nonumber \\
&& + [\mathbb{X}_{2},(a_{6}\mathbb{X}_{4})_{xxx}] + [\mathbb{X}_{2},([\mathbb{X}_{1},(a_{6}\mathbb{X}_{2})_{x}])_{xx}] - [\mathbb{X}_{2},(a_{6}\mathbb{X}_{2})_{xxxx}] + [\mathbb{X}_{2},[\mathbb{X}_{1},(a_{6}\mathbb{X}_{5})_{x}]] \nonumber \\
&& - [\mathbb{X}_{2},(a_{6}\mathbb{X}_{5})_{xx}] - \frac{1}{2}a_{5}\mathbb{X}_{9} + \frac{1}{2}[\mathbb{X}_{1},[\mathbb{X}_{2},(a_{5}\mathbb{X}_{2})_{x}]] + \frac{1}{2}[\mathbb{X}_{1},[\mathbb{X}_{2},[\mathbb{X}_{1},[\mathbb{X}_{1},(a_{6}\mathbb{X}_{2})_{x}]]]] \nonumber \\
&& - \frac{1}{2}[\mathbb{X}_{1},[\mathbb{X}_{2},[\mathbb{X}_{1},(a_{6}\mathbb{X}_{2})_{xx}]]] + \frac{1}{2}[\mathbb{X}_{1},[\mathbb{X}_{1},(a_{1}\mathbb{X}_{2})_{x}]] - \frac{1}{2}[\mathbb{X}_{1},[\mathbb{X}_{2},([\mathbb{X}_{1},(a_{6}\mathbb{X}_{2})_{x}])_{x}]] \nonumber \\
&& + \frac{1}{2}[\mathbb{X}_{1},[\mathbb{X}_{1},[\mathbb{X}_{1},[\mathbb{X}_{2},(a_{6}\mathbb{X}_{2})_{x}]]]] - \frac{1}{2}[\mathbb{X}_{1},[\mathbb{X}_{2},(a_{6}\mathbb{X}_{4})_{xx}]] + \frac{1}{2}[\mathbb{X}_{1},[\mathbb{X}_{2},[\mathbb{X}_{1},(a_{6}\mathbb{X}_{4})_{x}]]] \nonumber \\
&& - \frac{1}{2}a_{6}[\mathbb{X}_{1},[\mathbb{X}_{1},\mathbb{X}_{9}]] + \frac{1}{2}[\mathbb{X}_{1},[\mathbb{X}_{2},(a_{6}\mathbb{X}_{2})_{xxx}]] + \frac{1}{2}[\mathbb{X}_{1},[\mathbb{X}_{1},[\mathbb{X}_{2},[\mathbb{X}_{1},(a_{6}\mathbb{X}_{2})_{x}]]]] \nonumber \\
&& + \frac{1}{2}[\mathbb{X}_{1},[\mathbb{X}_{2},(a_{6}\mathbb{X}_{5})_{x}]] - \frac{1}{2}[\mathbb{X}_{1},[\mathbb{X}_{1},[\mathbb{X}_{2},(a_{6}\mathbb{X}_{2})_{xx}]]] - \frac{1}{2}[\mathbb{X}_{1},[\mathbb{X}_{1},([\mathbb{X}_{2},(a_{6}\mathbb{X}_{2})_{x}])_{x}]] \nonumber \\
&& - \frac{1}{2}a_{6}[\mathbb{X}_{1},[\mathbb{X}_{1},\mathbb{X}_{8}]] + \frac{1}{2}[\mathbb{X}_{1},[\mathbb{X}_{1},[\mathbb{X}_{2},(a_{6}\mathbb{X}_{4})_{x}]]] - \frac{1}{2}a_{1}\mathbb{X}_{7} + \frac{1}{2}[\mathbb{X}_{1},[\mathbb{X}_{1},(a_{6}\mathbb{X}_{6})_{x}]] \nonumber \\
&& - \frac{1}{2}[\mathbb{X}_{1},([\mathbb{X}_{2},[\mathbb{X}_{1},(a_{6}\mathbb{X}_{2})_{x}]])_{x}] + \frac{1}{2}[\mathbb{X}_{1},(a_{6}\mathbb{X}_{9})_{x}] + \frac{1}{2}[\mathbb{X}_{1},(a_{6}\mathbb{X}_{8})_{x}] - \frac{1}{2}[\mathbb{X}_{1},(a_{6}\mathbb{X}_{6})_{xx}] \nonumber \\
&& - \frac{1}{2}[\mathbb{X}_{1},([\mathbb{X}_{1},[\mathbb{X}_{2},(a_{6}\mathbb{X}_{2})_{x}]])_{x}] + \frac{1}{2}[\mathbb{X}_{1},([\mathbb{X}_{2},(a_{6}\mathbb{X}_{2})_{xx}])_{x}] - \frac{1}{2}[\mathbb{X}_{1},([\mathbb{X}_{2},(a_{6}\mathbb{X}_{4})_{x}])_{x}] \nonumber \\
&& - \frac{1}{2}[\mathbb{X}_{1},(a_{1}\mathbb{X}_{2})_{xx}] + \frac{1}{2}[\mathbb{X}_{1},(a_{1}\mathbb{X}_{4})_{x}] + \frac{1}{2}[\mathbb{X}_{1},([\mathbb{X}_{2},(a_{6}\mathbb{X}_{2})_{x}])_{xx}] - \frac{1}{2}a_{6}[\mathbb{X}_{1},[\mathbb{X}_{2},\mathbb{X}_{7}]] \nonumber \\
&& - \frac{1}{2}([\mathbb{X}_{2},(a_{5}\mathbb{X}_{2})_{x}])_{x} - \frac{1}{2}([\mathbb{X}_{2},[\mathbb{X}_{1},[\mathbb{X}_{1},(a_{6}\mathbb{X}_{2})_{x}]]])_{x} + \frac{1}{2}([\mathbb{X}_{2},[\mathbb{X}_{1},(a_{6}\mathbb{X}_{2})_{xx}]])_{x} \nonumber \\
&& + \frac{1}{2}(a_{5}\mathbb{X}_{6})_{x} + \frac{1}{2}([\mathbb{X}_{2},([\mathbb{X}_{1},(a_{6}\mathbb{X}_{2})_{x}])_{x}])_{x} - \frac{1}{2}([\mathbb{X}_{2},(a_{6}\mathbb{X}_{5})_{x}])_{x} - \frac{1}{2}([\mathbb{X}_{1},(a_{1}\mathbb{X}_{2})_{x}])_{x} \nonumber \\
&& + \frac{1}{2}([\mathbb{X}_{2},(a_{6}\mathbb{X}_{4})_{xx}])_{x} + \frac{1}{2}(a_{6}[\mathbb{X}_{1},\mathbb{X}_{9}])_{x} + \frac{1}{2}(a_{6}[\mathbb{X}_{1},\mathbb{X}_{8}])_{x} - \frac{1}{2}([\mathbb{X}_{2},[\mathbb{X}_{1},(a_{6}\mathbb{X}_{4})_{x}]])_{x} \nonumber \\
&& - \frac{1}{2}([\mathbb{X}_{2},(a_{6}\mathbb{X}_{2})_{xxx}])_{x} - \frac{1}{2}([\mathbb{X}_{1},[\mathbb{X}_{1},[\mathbb{X}_{2},(a_{6}\mathbb{X}_{2})_{x}]]])_{x} - \frac{1}{2}([\mathbb{X}_{1},[\mathbb{X}_{2},[\mathbb{X}_{1},(a_{6}\mathbb{X}_{2})_{x}]]])_{x} \nonumber \\
&& + \frac{1}{2}([\mathbb{X}_{1},[\mathbb{X}_{2},(a_{6}\mathbb{X}_{2})_{xx}]])_{x} - \frac{1}{2}([\mathbb{X}_{1},[\mathbb{X}_{2},(a_{6}\mathbb{X}_{4})_{x}]])_{x} + \frac{1}{2}([\mathbb{X}_{1},([\mathbb{X}_{2},(a_{6}\mathbb{X}_{2})_{x}])_{x}])_{x} \nonumber \\
&& + \frac{1}{2}(a_{1}\mathbb{X}_{5})_{x} - \frac{1}{2}([\mathbb{X}_{1},(a_{6}\mathbb{X}_{6})_{x}])_{x} - \frac{1}{2}(a_{6}\mathbb{X}_{9})_{xx} - \frac{1}{2}(a_{6}\mathbb{X}_{8})_{xx} + \frac{1}{2}([\mathbb{X}_{1},[\mathbb{X}_{2},(a_{6}\mathbb{X}_{2})_{x}]])_{xx} \nonumber \\
&& + \frac{1}{2}(a_{6}\mathbb{X}_{6})_{xxx} + \frac{1}{2}([\mathbb{X}_{2},[\mathbb{X}_{1},(a_{6}\mathbb{X}_{2})_{x}]])_{xx} - \frac{1}{2}([\mathbb{X}_{2},(a_{6}\mathbb{X}_{2})_{xx}])_{xx} + \frac{1}{2}([\mathbb{X}_{2},(a_{6}\mathbb{X}_{4})_{x}])_{xx} \nonumber \\
&& + \frac{1}{2}(a_{1}\mathbb{X}_{2})_{xxx} - \frac{1}{2}(a_{1}\mathbb{X}_{4})_{xx} - \frac{1}{2}([\mathbb{X}_{2},(a_{6}\mathbb{X}_{2})_{x}])_{xxx} + \frac{1}{2}(a_{6}[\mathbb{X}_{2},\mathbb{X}_{7}])_{x} + \frac{1}{2}(a_{4}\mathbb{X}_{2})_{x} = 0
\end{eqnarray}

\begin{eqnarray}
O(u^{3}) &:& - \frac{1}{3}([\mathbb{X}_{2},(a_{6}\mathbb{X}_{6})_{x}])_{x} + \frac{1}{2}[\mathbb{X}_{2},[\mathbb{X}_{2},[\mathbb{X}_{1},[\mathbb{X}_{1},(a_{6}\mathbb{X}_{2})_{x}]]]] - \frac{1}{2}[\mathbb{X}_{2},[\mathbb{X}_{2},[\mathbb{X}_{1},(a_{6}\mathbb{X}_{2})_{xx}]]] \nonumber \\
&& + \frac{1}{3}(a_{3}\mathbb{X}_{2})_{x} - \frac{1}{2}[\mathbb{X}_{2},[\mathbb{X}_{2},(a_{6}\mathbb{X}_{4})_{xx}]] + \frac{1}{2}[\mathbb{X}_{2},[\mathbb{X}_{2},(a_{6}\mathbb{X}_{5})_{x}]] + \frac{1}{2}[\mathbb{X}_{2},[\mathbb{X}_{1},(a_{1}\mathbb{X}_{2})_{x}]] \nonumber \\
&& - \frac{1}{2}[\mathbb{X}_{2},[\mathbb{X}_{2},([\mathbb{X}_{1},(a_{6}\mathbb{X}_{2})_{x}])_{x}]] - \frac{1}{2}a_{6}[\mathbb{X}_{2},[\mathbb{X}_{1},\mathbb{X}_{9}]] + \frac{1}{2}[\mathbb{X}_{2},[\mathbb{X}_{1},[\mathbb{X}_{2},[\mathbb{X}_{1},(a_{6}\mathbb{X}_{2})_{x}]]]] \nonumber \\
&& - \frac{1}{2}a_{6}[\mathbb{X}_{2},[\mathbb{X}_{1},\mathbb{X}_{8}]] + \frac{1}{2}[\mathbb{X}_{2},[\mathbb{X}_{2},[\mathbb{X}_{1},(a_{6}\mathbb{X}_{4})_{x}]]] + \frac{1}{2}[\mathbb{X}_{2},[\mathbb{X}_{1},[\mathbb{X}_{1},[\mathbb{X}_{2},(a_{6}\mathbb{X}_{2})_{x}]]]] \nonumber \\
&& + \frac{1}{2}[\mathbb{X}_{2},[\mathbb{X}_{2},(a_{6}\mathbb{X}_{2})_{xxx}]] - \frac{1}{2}[\mathbb{X}_{2},[\mathbb{X}_{1},[\mathbb{X}_{2},(a_{6}\mathbb{X}_{2})_{xx}]]] - \frac{1}{2}[\mathbb{X}_{2},[\mathbb{X}_{1},([\mathbb{X}_{2},(a_{6}\mathbb{X}_{2})_{x}])_{x}]] \nonumber \\
&& - \frac{1}{2}a_{1}\mathbb{X}_{8} + \frac{1}{2}[\mathbb{X}_{2},[\mathbb{X}_{1},[\mathbb{X}_{2},(a_{6}\mathbb{X}_{4})_{x}]]] + \frac{1}{2}[\mathbb{X}_{2},[\mathbb{X}_{1},(a_{6}\mathbb{X}_{6})_{x}]] - \frac{1}{2}[\mathbb{X}_{2},(a_{6}\mathbb{X}_{6})_{xx}] \nonumber \\
&& - \frac{1}{2}[\mathbb{X}_{2},([\mathbb{X}_{2},[\mathbb{X}_{1},(a_{6}\mathbb{X}_{2})_{x}]])_{x}] + \frac{1}{2}[\mathbb{X}_{2},(a_{6}\mathbb{X}_{9})_{x}] - \frac{1}{2}[\mathbb{X}_{2},([\mathbb{X}_{1},[\mathbb{X}_{2},(a_{6}\mathbb{X}_{2})_{x}]])_{x}] \nonumber \\
&& + \frac{1}{2}[\mathbb{X}_{2},([\mathbb{X}_{2},(a_{6}\mathbb{X}_{2})_{xx}])_{x}] - \frac{1}{2}[\mathbb{X}_{2},(a_{1}\mathbb{X}_{2})_{xx}] + \frac{1}{2}[\mathbb{X}_{2},([\mathbb{X}_{2},(a_{6}\mathbb{X}_{2})_{x}])_{xx}] + \frac{1}{2}[\mathbb{X}_{2},(a_{1}\mathbb{X}_{4})_{x}] \nonumber \\
&& + \frac{1}{2}[\mathbb{X}_{2},(a_{6}\mathbb{X}_{8})_{x}] - \frac{1}{2}[\mathbb{X}_{2},([\mathbb{X}_{2},(a_{6}\mathbb{X}_{4})_{x}])_{x}] - \frac{1}{2}a_{6}[\mathbb{X}_{2},[\mathbb{X}_{2},\mathbb{X}_{7}]] - \frac{1}{3}a_{6}[\mathbb{X}_{1},[\mathbb{X}_{2},\mathbb{X}_{9}]] \nonumber \\
&& + \frac{1}{3}[\mathbb{X}_{1},[\mathbb{X}_{2},(a_{1}\mathbb{X}_{2})_{x}]] + \frac{1}{3}[\mathbb{X}_{1},[\mathbb{X}_{2},[\mathbb{X}_{2},[\mathbb{X}_{1},(a_{6}\mathbb{X}_{2})_{x}]]]] - \frac{1}{3}[\mathbb{X}_{1},[\mathbb{X}_{2},([\mathbb{X}_{2},(a_{6}\mathbb{X}_{2})_{x}])_{x}]] \nonumber \\
&& + \frac{1}{3}[\mathbb{X}_{1},[\mathbb{X}_{2},[\mathbb{X}_{1},[\mathbb{X}_{2},(a_{6}\mathbb{X}_{2})_{x}]]]] - \frac{1}{3}[\mathbb{X}_{1},[\mathbb{X}_{2},[\mathbb{X}_{2},(a_{6}\mathbb{X}_{2})_{xx}]]] - \frac{1}{3}a_{6}[\mathbb{X}_{1},[\mathbb{X}_{2},\mathbb{X}_{8}]] \nonumber \\
&& + \frac{1}{3}[\mathbb{X}_{1},[\mathbb{X}_{2},(a_{6}\mathbb{X}_{6})_{x}]] + \frac{1}{3}[\mathbb{X}_{1},[\mathbb{X}_{2},[\mathbb{X}_{2},(a_{6}\mathbb{X}_{4})_{x}]]] - \frac{1}{3}a_{1}\mathbb{X}_{9} - \frac{1}{3}([\mathbb{X}_{2},(a_{1}\mathbb{X}_{2})_{x}])_{x} \nonumber \\
&& - \frac{1}{3}a_{3}\mathbb{X}_{4} + \frac{1}{3}(a_{6}[\mathbb{X}_{2},\mathbb{X}_{9}])_{x} + \frac{1}{3}(a_{6}[\mathbb{X}_{2},\mathbb{X}_{8}])_{x} - \frac{1}{3}([\mathbb{X}_{2},[\mathbb{X}_{1},[\mathbb{X}_{2},(a_{6}\mathbb{X}_{2})_{x}]]])_{x} \nonumber \\
&& - \frac{1}{3}([\mathbb{X}_{2},[\mathbb{X}_{2},[\mathbb{X}_{1},(a_{6}\mathbb{X}_{2})_{x}]]])_{x} + \frac{1}{3}([\mathbb{X}_{2},[\mathbb{X}_{2},(a_{6}\mathbb{X}_{2})_{xx}]])_{x} + \frac{1}{3}([\mathbb{X}_{2},([\mathbb{X}_{2},(a_{6}\mathbb{X}_{2})_{x}])_{x}])_{x} \nonumber \\
&& - \frac{1}{3}([\mathbb{X}_{2},[\mathbb{X}_{2},(a_{6}\mathbb{X}_{4})_{x}]])_{x} + \frac{1}{3}(a_{1}\mathbb{X}_{6})_{x} = 0
\end{eqnarray}

\begin{eqnarray}
O(u^{4}) &:& [\mathbb{X}_{2},[\mathbb{X}_{2},[\mathbb{X}_{2},[\mathbb{X}_{1},(a_{6}\mathbb{X}_{2})_{x}]]]] - a_{6}[\mathbb{X}_{2},[\mathbb{X}_{2},\mathbb{X}_{8}]] + [\mathbb{X}_{2},[\mathbb{X}_{2},[\mathbb{X}_{1},[\mathbb{X}_{2},(a_{6}\mathbb{X}_{2})_{x}]]]] \nonumber \\
&& - a_{6}[\mathbb{X}_{2},[\mathbb{X}_{2},\mathbb{X}_{9}]] - [\mathbb{X}_{2},[\mathbb{X}_{2},[\mathbb{X}_{2},(a_{6}\mathbb{X}_{2})_{xx}]]] - [\mathbb{X}_{2},[\mathbb{X}_{2},([\mathbb{X}_{2},(a_{6}\mathbb{X}_{2})_{x}])_{x}]] \nonumber \\
&& + [\mathbb{X}_{2},[\mathbb{X}_{2},(a_{6}\mathbb{X}_{6})_{x}]] + [\mathbb{X}_{2},[\mathbb{X}_{2},[\mathbb{X}_{2},(a_{6}\mathbb{X}_{4})_{x}]]] = 0
\end{eqnarray}

   Note that if we decouple $\eqref{KDV11}$ into the following conditions

\begin{eqnarray}
&& ([\mathbb{X}_{2},(a_{6}\mathbb{X}_{2})_{x}])_{x} - (a_{6}\mathbb{X}_{6})_{x} + a_{6}\mathbb{X}_{9} - [\mathbb{X}_{1},[\mathbb{X}_{2},(a_{6}\mathbb{X}_{2})_{x}]] = 0 \label{NEWKDVCON1} \\
&& [\mathbb{X}_{2},[\mathbb{X}_{1},(a_{6}\mathbb{X}_{2})_{x}]] + [\mathbb{X}_{2},(a_{6}\mathbb{X}_{4})_{x}] - [\mathbb{X}_{2},(a_{6}\mathbb{X}_{2})_{xx}] - a_{6}\mathbb{X}_{8} = 0 \label{NEWKDVCON2} \\
&& ((a_{2} - 3a_{1})\mathbb{X}_{2})_{x} - (a_{2} - 3a_{1})\mathbb{X}_{4} = 0 \label{NEWKDVCON3}
\end{eqnarray}

\noindent
then the $O(u^{4})$ equation is identically satisfied. To reduce the complexity of the $O(u^{3})$ equation we can decouple it into the following equations

\begin{eqnarray}
&& [\mathbb{X}_{2},[\mathbb{X}_{1},[\mathbb{X}_{1},(a_{6}\mathbb{X}_{2})_{x}]]] - [\mathbb{X}_{2},[\mathbb{X}_{1},(a_{6}\mathbb{X}_{2})_{xx}]] - [\mathbb{X}_{2},(a_{6}\mathbb{X}_{4})_{xx}] + [\mathbb{X}_{2},(a_{6}\mathbb{X}_{5})_{x}] \nonumber \\
&& + [\mathbb{X}_{1},(a_{1}\mathbb{X}_{2})_{x}] - [\mathbb{X}_{2},([\mathbb{X}_{1},(a_{6}\mathbb{X}_{2})_{x}])_{x}] + [\mathbb{X}_{2},[\mathbb{X}_{1},(a_{6}\mathbb{X}_{4})_{x}]] + [\mathbb{X}_{2},(a_{6}\mathbb{X}_{2})_{xxx}] \nonumber \\
&& - a_{1}\mathbb{X}_{5} - (a_{1}\mathbb{X}_{2})_{xx} + (a_{1}\mathbb{X}_{4})_{x} - a_{6}[\mathbb{X}_{2},\mathbb{X}_{7}] = 0 \label{NEWKDVCON4} \\
&& (a_{3}\mathbb{X}_{2})_{x} + [\mathbb{X}_{1},[\mathbb{X}_{2},(a_{1}\mathbb{X}_{2})_{x}]] - a_{1}\mathbb{X}_{9} - ([\mathbb{X}_{2},(a_{1}\mathbb{X}_{2})_{x}])_{x} - a_{3}\mathbb{X}_{4} + (a_{1}\mathbb{X}_{6})_{x} = 0 \label{NEWKDVCON5}
\end{eqnarray}

\noindent
From this last condition, we can use $\eqref{NEWKDVCON1}-\eqref{NEWKDVCON5}$ to reduce the $O(u^{2})$ condition to the following

\begin{eqnarray}
&& - a_{5}\mathbb{X}_{8} - a_{6}[\mathbb{X}_{2},[\mathbb{X}_{1},\mathbb{X}_{7}]] + [\mathbb{X}_{2},[\mathbb{X}_{1},[\mathbb{X}_{1},[\mathbb{X}_{1},(a_{6}\mathbb{X}_{2})_{x}]]]] + [\mathbb{X}_{2},[\mathbb{X}_{1},[\mathbb{X}_{1},(a_{6}\mathbb{X}_{4})_{x}]]] \nonumber \\
&& - [\mathbb{X}_{2},[\mathbb{X}_{1},[\mathbb{X}_{1},(a_{6}\mathbb{X}_{2})_{xx}]]] + [\mathbb{X}_{2},[\mathbb{X}_{1},(a_{5}\mathbb{X}_{2})_{x}]] - [\mathbb{X}_{2},[\mathbb{X}_{1},([\mathbb{X}_{1},(a_{6}\mathbb{X}_{2})_{x}])_{x}]] \nonumber \\
&& + [\mathbb{X}_{2},(a_{6}\mathbb{X}_{7})_{x}] - [\mathbb{X}_{2},[\mathbb{X}_{1},(a_{6}\mathbb{X}_{4})_{xx}]] + [\mathbb{X}_{2},[\mathbb{X}_{1},(a_{6}\mathbb{X}_{2})_{xxx}]] - [\mathbb{X}_{2},([\mathbb{X}_{1},(a_{6}\mathbb{X}_{4})_{x}])_{x}] \nonumber \\
&& - [\mathbb{X}_{2},([\mathbb{X}_{1},[\mathbb{X}_{1},(a_{6}\mathbb{X}_{2})_{x}]])_{x}] + [\mathbb{X}_{2},(a_{5}\mathbb{X}_{4})_{x}] + [\mathbb{X}_{2},([\mathbb{X}_{1},(a_{6}\mathbb{X}_{2})_{xx}])_{x}] - [\mathbb{X}_{2},(a_{5}\mathbb{X}_{2})_{xx}] \nonumber \\
&& + [\mathbb{X}_{2},(a_{6}\mathbb{X}_{4})_{xxx}] + [\mathbb{X}_{2},([\mathbb{X}_{1},(a_{6}\mathbb{X}_{2})_{x}])_{xx}] - [\mathbb{X}_{2},(a_{6}\mathbb{X}_{2})_{xxxx}] + [\mathbb{X}_{2},[\mathbb{X}_{1},(a_{6}\mathbb{X}_{5})_{x}]] \nonumber \\
&& - [\mathbb{X}_{2},(a_{6}\mathbb{X}_{5})_{xx}] - \frac{1}{2}a_{5}\mathbb{X}_{9} + \frac{1}{2}[\mathbb{X}_{1},[\mathbb{X}_{2},(a_{5}\mathbb{X}_{2})_{x}]] - \frac{1}{2}([\mathbb{X}_{2},(a_{5}\mathbb{X}_{2})_{x}])_{x} + \frac{1}{2}(a_{5}\mathbb{X}_{6})_{x} \nonumber \\
&& - \frac{1}{2}a_{4}\mathbb{X}_{4} + \frac{1}{2}(a_{4}\mathbb{X}_{2})_{x} = 0
\end{eqnarray}

Decoupling this equation allows for the simplification of the $O(u)$ equation. Thus we write the previous condition as the following system of equations

\begin{eqnarray}
&& - [\mathbb{X}_{1},(a_{6}\mathbb{X}_{4})_{x}] - [\mathbb{X}_{1},[\mathbb{X}_{1},(a_{6}\mathbb{X}_{2})_{x}]] + a_{5}\mathbb{X}_{4} + [\mathbb{X}_{1},(a_{6}\mathbb{X}_{2})_{xx}] - (a_{5}\mathbb{X}_{2})_{x} \nonumber \\
&& + (a_{6}\mathbb{X}_{4})_{xx} + ([\mathbb{X}_{1},(a_{6}\mathbb{X}_{2})_{x}])_{x} - (a_{6}\mathbb{X}_{2})_{xxx} - (a_{6}\mathbb{X}_{5})_{x} = 0 \label{NEWKDVCON6} \\
&& - a_{5}\mathbb{X}_{9} + [\mathbb{X}_{1},[\mathbb{X}_{2},(a_{5}\mathbb{X}_{2})_{x}]] - ([\mathbb{X}_{2},(a_{5}\mathbb{X}_{2})_{x}])_{x} + (a_{5}\mathbb{X}_{6})_{x} - \frac{1}{2}a_{4}\mathbb{X}_{4} + \frac{1}{2}(a_{4}\mathbb{X}_{2})_{x} = 0 \label{NEWKDVCON7}
\end{eqnarray}

Using this the $O(u)$ equation is reduced to

\begin{equation}
\mathbb{X}_{2,t} + [\mathbb{X}_{2},\mathbb{X}_{0}] - a_{8}\mathbb{X}_{4} + (a_{8}\mathbb{X}_{2})_{x} - a_{7}\mathbb{X}_{2} = 0 \label{NEWKDVCON8}
\end{equation}

We therefore find that the final, reduced constraints are given by $\eqref{KDV12}, \eqref{NEWKDVCON1}-\eqref{NEWKDVCON5}$ and $\eqref{NEWKDVCON6}-\eqref{NEWKDVCON8}$. In order to satisfy these constraints we begin with the following rather simple forms for our generators,

\[ \mathbb{X}_{0} = \begin{bmatrix}
g_{1}(x,t) & g_{2}(x,t) \\
g_{3}(x,t) & g_{4}(x,t)
\end{bmatrix}, \ \ \mathbb{X}_{1} = \begin{bmatrix}
0 & f_{1}(x,t) \\
f_{2}(x,t) & 0
\end{bmatrix}, \ \ \mathbb{X}_{2} = \begin{bmatrix}
0 & f_{3}(x,t) \\
f_{4}(x,t) & 0
\end{bmatrix} \]

{\it To get more general results we will assume $a_{2} \neq 3a_{1}$. Note that had we instead opted for the forms}

\[ \mathbb{X}_{0} = \begin{bmatrix}
g_{1}(x,t) & g_{12}(x,t) \\
g_{23}(x,t) & g_{34}(x,t)
\end{bmatrix}, \ \ \mathbb{X}_{1} = \begin{bmatrix}
f_{1}(x,t) & f_{3}(x,t) \\
f_{5}(x,t) & f_{7}(x,t)
\end{bmatrix}, \ \ \mathbb{X}_{2} = \begin{bmatrix}
f_{2}(x,t) & f_{4}(x,t) \\
f_{6}(x,t) & f_{8}(x,t)
\end{bmatrix} \]

\noindent
{\it we would obtain an equivalent system to that obtained in $\cite{Lecce}$. The additional unknown functions which appear in Khawaja's method $\cite{Lecce}$ can be introduced with the proper substitutions via their functional dependence on the twelve unknown functions given above.}

Taking the naive approach of beginning with the smaller conditions first we begin with $\eqref{NEWKDVCON3}$ which, utilizing the given forms for $\mathbb{X}_{0}, \mathbb{X}_{1}$, and $\mathbb{X}_{2}$, becomes

\begin{eqnarray}
&& (a_{2}-3a_{1})(f_{1}f_{4}-f_{2}f_{3}) = 0 \\
&& ((a_{2}-3a_{1})f_{j})_{x} = 0 \ \ \ j=3,4
\end{eqnarray}

\noindent
Solving this system for $f_{2},f_{3}$ and $f_{4}$ yields

\begin{eqnarray}
f_{3}(x,t) &=& \frac{F_{3}(t)}{a_{2}(x,t)-3a_{1}(x,t)} \\
f_{4}(x,t) &=& \frac{F_{4}(t)}{a_{2}(x,t)-3a_{1}(x,t)} \\
f_{2}(x,t) &=& \frac{f_{1}(x,t)F_{4}(t)}{F_{3}(t)} \\
\end{eqnarray}

\noindent
where $F_{3,4}(t)$ are arbitrary functions of $t$. With these choices we've elected to satisfy $\mathbb{X}_{4} = 0$ rather than $a_{2} = 3a_{1}$. Looking next at $\eqref{NEWKDVCON8}$ we obtain the system 

\begin{eqnarray}
&& \left(\frac{F_{j}}{a_{2}-3a_{1}}\right)_{t} - \frac{F_{j}a_{7}}{a_{2}-3a_{1}} + \left(\frac{F_{j}a_{8}}{a_{2}-3a_{1}}\right)_{x} + \frac{1}{2}\left(\frac{F_{j}a_{4}}{a_{2}-3a_{1}}\right)_{x} \nonumber \\
&& + (-1)^{j}\frac{F_{j}(g_{4}-g_{1})}{a_{2}-3a_{1}} = 0 \ \ \ j=3,4 \label{KDVFINAL1}\\
&& \frac{F_{3}g_{3}}{a_{2}-3a_{1}} - \frac{F_{4}g_{2}}{a_{2}-3a_{1}} = 0
\end{eqnarray}

\noindent
Solving the second equation for $g_{3}$ yields

\[ g_{3} = \frac{F_{4}(t)g_{2}(x,t)}{F_{3}(t)} \]

Considering the $O(1)$ equation next, we have the following system of equations

\begin{eqnarray}
&& g_{1x} = g_{4x} = 0 \\
&& f_{1t} - g_{2x} + f_{1}(g_{4} - g_{1}) = 0 \label{KDVFINAL2} \\
&& F_{3}(F_{4}f_{1})_{t} - f_{1}F_{4}F_{3t} - g_{2x}F_{4}F_{3} + F_{3}F_{4}f_{1}(g_{1} - g_{4}) = 0 \label{KDVFINAL3}
\end{eqnarray}

It follows that we must have $g_{1}(x,t) = G_{1}(t)$ and $g_{4}(x,t) = G_{4}(t)$ where $G_{1}$ and $G_{4}$ are arbitrary functions of $t$. Since $\eqref{KDVFINAL2}$ and $\eqref{KDVFINAL3}$ do not depend on the $a_{i}$ we will postpone solving them until the end. At this point the remaining conditions have been reduced to conditions involving soley the $a_{i}$ and the previously introduced arbitrary functions of $t$. The remaining conditions are given by

\begin{eqnarray}
&& \left(\frac{a_{5}}{a_{2}-3a_{1}}\right)_{x} + \left(\frac{a_{6}}{a_{2}-3a_{1}}\right)_{xxx} = 0 \label{KDVFINAL4} \\
&& \left(\frac{a_{3}}{a_{2}-3a_{1}}\right)_{x} = 0 \label{KDVFINAL5} \\
&& \left(\frac{a_{1}}{a_{2}-3a_{1}}\right)_{xx} = 0 \label{KDVFINAL6} \\
&& \left(\frac{a_{4}}{a_{2}-3a_{1}}\right)_{x} = 0 \label{KDVFINAL7}
\end{eqnarray}

One can easily solve the system of equations given by $\eqref{KDVFINAL1}$, $\eqref{KDVFINAL4}-\eqref{KDVFINAL7}$ yielding

\begin{eqnarray}
F_{4} &=& c_{1}F_{3}e^{2\int{(G_{4} - G_{1})dt}} \\
g_{2} &=& \int{(f_{1t}+f_{1}(G_{4}-G_{1}))dx} + F_{10} \\
a_{2} &=& -\frac{(3F_{1} - 1 - 3F_{2}x)a_{1}}{F_{2}x - F_{1}} \\
a_{3} &=& \frac{F_{5}a_{1}}{F_{2}x - F_{1}} \\
a_{4} &=& \frac{F_{6}a_{1}}{F_{2}x - F_{1}} \\
a_{6} &=& \frac{(F_{7} + F_{8}x + F_{9}x^{2})a_{1}}{F_{2}x - F_{1}} - \int^{x}{\int^{y}{\frac{a_{5}(z,t) \ dz\ dy}{a_{2}(z,t)-3a_{1}(z,t)}}} \\
a_{7} &=& \frac{a_{2}-3a_{1}}{F_{3}}\left(\frac{F_{3}}{a_{2}-3a_{1}}\right)_{t} + (a_{2}-3a_{1})\left(\frac{a_{8}}{a_{2}-3a_{1}}\right)_{x} + G_{4}-G_{1}
\end{eqnarray}

\noindent
where $F_{5-10}$ are arbitrary functions of $t$. Note that $a_{1},a_{5}$ and $a_{8}$ have no restrictions beyond the appropriate differentiability and integrability conditions. 

{\it The Lax pair for the generalized variable-coefficient KdV equation with the previous integrability conditions is therefore given by}

\begin{eqnarray}
F &=& X_{1} + X_{2}u \\
G &=& -a_{6}\mathbb{X}_{2}u_{xxxx} + (a_{6}\mathbb{X}_{2})_{x}u_{xxx} - \mathbb{X}_{2}(a_{1}u + a_{5})u_{xx} - (a_{6}\mathbb{X}_{2})_{xx}u_{xx} - a_{8}\mathbb{X}_{2}u \nonumber \\
&& + \frac{1}{2}a_{1}\mathbb{X}_{2}u_{x}^{2} - \frac{1}{2}a_{2}\mathbb{X}_{2}u_{x}^{2} + (a_{1}\mathbb{X}_{2})_{x}uu_{x} - \frac{1}{3}a_{3}\mathbb{X}_{2}u^{3} - \frac{1}{2}a_{4}\mathbb{X}_{2}u^{2} + \mathbb{X}_{0}
\end{eqnarray}

     This completes the extended EW analysis of the generalized fifth-order vcKdV equation. Next, we consider some solutions of these new
integrable equations derived by two different methods in the preceding sections.

\section{Painleve Analysis Method}

Next we shall consider methods to derive some solutions of the generalized integrable hierarchies of NLPDEs
derived in the previous sections.

Given a nonlinear partial differential equation in $(n+1)$-dimensions, without specifying initial or boundary conditions, we may find a solution about a movable singular manifold
\[ \phi - \phi_{0} = 0 \]
as an infinite series given by
\begin{equation} 
u(x_{1},\ldots,x_{n},t) = \phi^{-\alpha}\sum_{m=0}^{\infty}{u_{m}\phi^{m}} \label{series}
\end{equation}

Note that when $m\in (\mathbb{Q}-\mathbb{Z})$ $\eqref{series}$ is more commonly known as a Puiseux series. One can avoid dealing with Puiseux series if proper substitutions are made, as we will see a little later on. Plugging this infinite series into the NLPDE yields a recurrence relation for the $u_{m}$'s. As with most series-type solution methods for NLPDEs we will seek a solution to our NLPDE as $\eqref{series}$ truncated at the constant term. Plugging this truncated series into our original NLPDE and collecting terms in decreasing order of $\phi$ will give us a set of determining equations for our unknown coefficients $u_{0},\ldots,u_{\alpha}$ known as the Painleve-Backlund equations. We now define new functions
\begin{eqnarray}
C_{0}(x_{0},\ldots,x_{n},t) &=& \frac{\phi_{t}}{\phi_{x_{0}}} \\
C_{1}(x_{0},\ldots,x_{n},t) &=& \frac{\phi_{x_{1}}}{\phi_{x_{0}}} \\
&\vdots& \\
C_{n}(x_{0},\ldots,x_{n},t) &=& \frac{\phi_{x_{n}}}{\phi_{x_{0}}} \\
V(x_{0},\ldots,x_{n},t) &=& \frac{\phi_{x_{0}x_{0}}}{\phi_{x_{0}}}
\end{eqnarray}

\noindent
which will allow us to eliminate all derivatives of $\phi$ other than $\phi_{x_{0}}$. For simplicity it is common to allow $C_{i}(x_{0},\ldots,x_{n},t)$ and $V(x_{0},\ldots,x_{n},t)$ to be constants, thereby reducing a system of PDEs (more than likely nonlinear) in $\left\{C_{i}(x,t),V(x,t)\right\}$ to an algebraic system in $\left\{C_{i},V\right\}$ for $(i=0,\ldots,n)$.

\subsection{Analytic solutions for generalized inhomogeneous KdV equation of form $\eqref{KDV}$}
Consider the following example
\begin{eqnarray}
&& u_{t} + \frac{H_{1}}{F(\frac{x}{C}+t)}uu_{xxx} + \left(\frac{c_{1}H_{3}-2H_{1}+1}{F(\frac{x}{C}+t)}\right)u_{x}u_{xx} + \frac{H_{3}}{F(\frac{x}{C}+t)}u^{2}u_{x}+\frac{H_{4}(t)}{F(\frac{x}{C}+t)}uu_{x} \nonumber \\
&& + \left(\frac{c_{1}(c_{1}H_{3}(t)+2H_{4}-2H_{1})}{8F(\frac{x}{C}+t)}\right)u_{xxx} + \left(\frac{c_{1}(c_{1}H_{3}-2)}{40F(\frac{x}{C}+t)}\right)u_{xxxxx} \nonumber \\
&& + \left(\frac{(5c_{1}(H_{1}-H_{4})-c_{1}-3c_{1}^{2}H_{3}-20F(\frac{x}{C}+t)C}{20F(\frac{x}{C}+t)}\right)u_{x}
\end{eqnarray}

Note that in this example we have $a_{7} = 0$. The leading order analysis yields $\alpha = 2$. Therefore we seek a solution of the form
\begin{equation}
u(x,t) = \frac{u_{0}}{\phi(x,t)^{2}} + \frac{u_{1}}{\phi(x,t)} + u_{2}(x,t)
\end{equation}

As this forms an auto-Backlund transformation for our For simplicity we will allow our initial solution $u_{2}(x,t)$ to be $0$. Plugging this in to our pde yields the following determining equations for $\phi(x,t),u_{0}(x,t),u_{1}(x,t),V$ and $C_{1}$ :

\begin{equation}
O(\phi^{-7}) : 2u_{0}\phi_{x}[(9c_{1}^{2}H_{3} + 18c_{1})\phi_{x}^{4} + (6H_{3}c_{1} + 6)u_{0}\phi_{x}^{2} + H_{3}u_{0}^{2}] = 0
\end{equation}
\begin{eqnarray}
O(\phi^{-6}) &:& -10H_{1}u_{0}u_{1}\phi_{x}^{3} - 5H_{3}u_{0}^{2}u_{1}\phi_{x} + 60c_{1}u_{0}\phi_{x}^{3}\phi_{xx} + 15c_{1}^{2}H_{3}u_{0x}\phi_{x}^{4} - 3c_{1}^{2}H_{3}u_{1}\phi_{x}^{5} \nonumber \\
&& + 10H_{1}u_{0}^{2}\phi_{x}\phi_{xx} - 10H_{1}u_{0}u_{0x}\phi_{x}^{2} + 30c_{1}^{2}H_{3}u_{0}\phi_{x}^{3}\phi_{xx} + 14c_{1}H_{3}u_{0}u_{0x}\phi_{x}^{2} \nonumber \\
&& - 10c_{1}H_{3}u_{0}u_{1}\phi_{x}^{3} + 30c_{1}u_{0x}\phi_{x}^{4} - 6c_{1}u_{1}\phi_{x}^{4} + 14u_{0}u_{0x}\phi_{x}^{2} + 4u_{0}^{2}\phi_{x}\phi_{xx} - 10u_{0}u_{1}\phi_{x}^{3} \nonumber \\
&& + 4c_{1}H_{3}u_{0}^{2}\phi_{x}\phi_{xx} + H_{3}u_{0}^{2}u_{0x} = 0
\end{eqnarray}

\begin{eqnarray}
O(\phi^{-5}) &:& H_{3}^{2}u_{0}^{2}u_{1x} - 6H_{4}u_{0}\phi_{x}^{3} - 6u_{0}\phi_{x}^{3}(c_{1}H_{3} - 2H_{1} + 1) + 3u_{0}\phi_{x}^{3} + 2H_{3}u_{1}^{2}\phi_{x}^{3} \nonumber \\
&& - 4c_{1}H_{3}^{2}u_{0}u_{1}\phi_{x}\phi_{xx} - 16H_{1}H_{3}u_{0}u_{1}\phi_{x}\phi_{xx} + 9c_{1}^{2}H_{3}^{2}u_{0}\phi_{x}\phi_{xx}^{2} + 4H_{3}u_{0x}^{2}\phi_{x} \nonumber \\
&&- 6H_{1}H_{3}u_{0x}u_{1}\phi_{x}^{2} + 14H_{1}H_{3}u_{0}u_{1x}\phi_{x}^{2} - 10c_{1}H_{3}^{2}u_{0}u_{1x}\phi_{x}^{2} - 6c_{1}H_{3}^{2}u_{0x}u_{1}\phi_{x}^{2} \nonumber \\
&& + 12c_{1}H_{3}u_{0}\phi_{x}^{2}\phi_{xxx} + 2H_{1}H_{3}u_{0}u_{0x}\phi_{xx} - 4H_{3}u_{0}u_{1}\phi_{x}\phi_{xx} - 6c_{1}^{2}H_{3}^{2}u_{1}\phi_{x}^{3}\phi_{xx} \nonumber \\
&& + 18c_{1}^{2}H_{3}^{2}u_{0x}\phi_{x}^{2}\phi_{xx} + 2c_{1}H_{3}^{2}u_{0}u_{0x}\phi_{xx} + 2H_{1}H_{3}u_{0}u_{0xx}\phi_{x} - 12c_{1}H_{3}u_{1}\phi_{x}^{3}\phi_{xx} \nonumber \\
&& + 18c_{1}H_{3}u_{0}\phi_{x}\phi_{xx}^{2} + 2H_{1}H_{3}u_{1}^{2}\phi_{x}^{3} + 4H_{3}^{2}u_{0}u_{1}^{2}\phi_{x} + 2H_{3}H_{4}u_{0}^{2}\phi_{x} - 6c_{1}H_{3}u_{1x}\phi_{x}^{4} \nonumber \\
&& + 36c_{1}H_{3}u_{0x}\phi_{x}^{2}\phi_{xx} + 6c_{1}^{2}H_{3}^{2}u_{0}\phi_{x}^{2}\phi_{xxx} + 2c_{1}H_{3}^{2}u_{0}u_{0xx}\phi_{x} - 3c_{1}^{2}H_{3}^{2}u_{1x}\phi_{x}^{4} \nonumber \\
&& - 8H_{1}H_{3}u_{0x}^{2}\phi_{x} - 10H_{3}u_{0}u_{1x}\phi_{x}^{2} + 2c_{1}H_{3}^{2}u_{1}^{2}\phi_{x}^{3} - 6H_{3}u_{0x}u_{1}\phi_{x}^{2} + 2H_{3}u_{0xx}\phi_{x} \nonumber \\
&& + 6H_{4}u_{0}\phi_{x}^{3}(c_{1}H_{3} - 2H_{1} + 1) + 2H_{1}H_{3}u_{0}^{2}\phi_{xxx} + 12c_{1}H_{3}u_{0xx}\phi_{x}^{3} + 6c_{1}^{2}H_{3}^{2}u_{0xx}\phi_{x}^{3} \nonumber \\
&& + 12H_{1}H_{4}u_{0}\phi_{x}^{3} + 2H_{3}u_{0}u_{0x}\phi_{xx} + 3u_{0}\phi_{x}^{3}(c_{1}H_{3} - 2H_{1} + 1)^{2} - 6H_{1}u_{0}\phi_{x}^{3}) \nonumber \\
&& + 6H_{1}u_{0}\phi_{x}^{3}(c_{1}H_{3} - 2H_{1} + 1) - 2H_{3}^{2}u_{0}u_{0x}u_{1} + 4c_{1}H_{3}^{2}u_{0x}^{2}\phi_{x}
\end{eqnarray}

Upon solving the $O(\phi^{-7})$ and $O(\phi^{-6})$ equations for $u_{0}$ and $u_{1}$ respectively we find that
\begin{eqnarray}
u_{0}(x,t) &=& -3c_{1}\phi_{x}^{2} \\
u_{1}(x,t) &=& 3c_{1}\phi_{xx}
\end{eqnarray}

\noindent
which lends itself nicely to a representation of the solution as $u(x,t) = 3c_{1}\log[\phi(x,t)]_{xx}$. Further with the choice $V = 1$ the choices for coefficients the remaining orders of $\phi$ are identically satisfied. Now solving the system for $\phi$ given in the previous section we find that $\phi(x,t) = c_{2} + c_{3}e^{x+Ct}$. Therefore we have the solution
\begin{equation}
u(x,t) = \frac{3c_{1}c_{2}c_{3}e^{x+Ct}}{(c_{2} + c_{3}e^{x+Ct})^{2}}
\end{equation}

\noindent
which for the selection $c_{2} = c_{3}$ reduces to the solution $u(x,t) = \frac{3c_{1}}{4}\mbox{sech}^{2}\left(\frac{x}{2}+\frac{Ct}{2}\right)$

The next example is similar to the first however in this case we don't have $a_{7} = 0$ and we will not force the $u_{2}$ term to be the trivial solution. We thus consider the following example
\begin{eqnarray}
&& u_{t} + \frac{10H_{1}\xi(t)}{F(\int{\eta(t)dt}+x)}uu_{xxx} + \frac{2(3+2H_{1})\xi(t)}{F(\int{\eta(t)dt}+x)}u_{x}u_{xx} + \frac{6H_{1}-1}{F(\int{\eta(t)dt}+x)}u^{2}u_{x}\nonumber \\
&& + \frac{10H_{5}\xi(t)}{F(\int{\eta(t)dt}+x)}u_{xxx} + \frac{4(3H_{1}+2)\xi(t)^{2}}{5F(\int{\eta(t)dt}+x)}u_{xxxxx} - \left(\frac{1}{\xi(t)}\right)'F\left(\int{\eta(t)dt}+x\right)u \nonumber \\
&& + \left(H_{6}(t) + \frac{\xi(t)^{2}(c_{2}^{4}(8H_{1} - 3) - 2500c_{1}(H_{4} + 30c_{1}H_{1} - 5c_{1}))}{5F(\int{\eta(t)dt}+x)}\right)u_{x} \nonumber \\
&& + \frac{10H_{4}\xi(t)}{F(\int{\eta(t)dt}+x)}uu_{x}  = 0
\end{eqnarray}

\noindent
where $\xi(t) = \frac{H_{5}}{10c_{1}H_{1} - 10c_{1} + H_{4}}$ and $H_{1}(t),H_{4}(t),H_{5}(t)$ and $\eta(t)$ are arbitrary functions of $t$ and $c_{1},c_{2}$ are arbitrary constants. As with our last example the leading order analysis yields $\alpha = 2$. Unlike our last example we will not force the $u_{2}$ term to be $0$ initially. The first orders of $\phi$ which determine the $u_{i}$ are as follows :

\begin{equation}
O(\phi^{-7}) : -576(3H_{1}+2)\xi(t)^{2}u_{0}\phi_{x}^{5} - 2(6H_{1}-1)u_{0}^{3}\phi_{x} - 24H_{1}u_{0}^{2}\phi_{x}^{3} - 24(3+2H_{1})\xi(t)u_{0}^{2}\phi_{x}^{3} = 0
\end{equation}

\begin{eqnarray}
O(\phi^{-6}) : && 1440H_{1}H_{5}^{2}\phi_{x}^{4}u_{0x} - 288H_{1}H_{5}^{2}\phi_{x}^{5}u_{1} + 600c_{1}^{2}H_{1}^{3}u_{0}^{2}u_{0x} - 1300c_{1}^{2}H_{1}^{2}u_{0}^{2}u_{0x} \nonumber \\
&& + 6H_{1}H_{4}^{2}u_{0}^{2}u_{0x} + 500c_{1}^{2}\phi_{x}u_{0}^{2}u_{1} + 20c_{1}H_{4}u_{0}^{2}u_{0x} + 5H_{4}^{2}\phi_{x}u_{0}^{2}u_{1} - 840c_{1}H_{5}\phi_{x}^{2}u_{0}u_{0x} \nonumber \\
&& + 84H_{4}H_{5}\phi_{x}^{2}u_{0}u_{0x} + 600c_{1}H_{5}\phi_{x}^{3}u_{0}u_{1} - 60H_{4}H_{5}\phi_{x}^{3}u_{0}u_{1} - 3000c_{1}^{2}H_{1}^{3}\phi_{x}u_{0}^{2}u_{1} \nonumber \\
&& + 6500c_{1}^{2}H_{1}^{2}\phi_{x}u_{0}^{2}u_{1} + 120c_{1}H_{1}^{2}H_{4}u_{0}^{2}u_{0x} - 4000c_{1}^{2}H_{1}\phi_{x}u_{0}^{2}u_{1} - 140c_{1}H_{1}H_{4}u_{0}^{2}u_{0x} \nonumber \\
&& - 30H_{1}H_{4}^{2}\phi_{x}u_{0}^{2}u_{1} - 100c_{1}H_{4}\phi_{x}u_{0}^{2}u_{1} - 192H_{5}^{2}\phi_{x}^{5}u_{1} - 100c_{1}^{2}u_{0}^{2}u_{0x} - H_{4}^{2}u_{0}^{2}u_{0x} \nonumber \\
&& + 960H_{5}^{2}\phi_{x}^{4}u_{0x} + 1920H_{5}^{2}\phi_{x}^{3}\phi_{xx}u_{0} + 2880H_{1}H_{5}^{2}\phi_{x}^{3}\phi_{xx}u_{0} - 240c_{1}H_{5}\phi_{x}\phi_{xx}u_{0}^{2} \nonumber \\
&& + 24H_{4}H_{5}\phi_{x}\phi_{xx}u_{0}^{2} + 1960c_{1}H_{1}^{2}H_{5}\phi_{x}\phi_{xx}u_{0}^{2} - 1720c_{1}H_{1}H_{5}\phi_{x}\phi_{xx}u_{0}^{2} \nonumber \\
&& + 2360c_{1}H_{1}^{2}H_{5}\phi_{x}^{2}u_{0}u_{0x} - 1520c_{1}H_{1}H_{5}\phi_{x}^{2}u_{0}u_{0x} + 236H_{1}H_{4}H_{5}\phi_{x}^{2}u_{0}u_{0x} \nonumber \\
&& + 2800c_{1}H_{1}H_{5}\phi_{x}^{3}u_{0}u_{1} - 340H_{1}H_{4}H_{5}\phi_{x}^{3}u_{0}u_{1} - 600c_{1}H_{1}^{2}H_{4}\phi_{x}u_{0}^{2}u_{1} \nonumber \\
&& + 700c_{1}H_{1}H_{4}\phi_{x}u_{0}^{2}u_{1} + 800c_{1}^{2}H_{1}u_{0}^{2}u_{0x} + 196H_{1}H_{4}H_{5}\phi_{x}\phi_{xx}u_{0}^{2} \nonumber \\
&& - 3400c_{1}H_{1}^{2}H_{5}\phi_{x}^{3}u_{0}u_{1} = 0
\end{eqnarray}

\begin{eqnarray}
O(\phi^{-5}) : && -600c_{1}H_{5}\phi_{x}^{2}u_{0}u_{1x} + 60H_{4}H_{5}\phi_{x}^{2}u_{0}u_{1x} + 200c_{1}H_{4}H_{5}\phi_{x}u_{0}^{2} - 100c_{1}^{2}u_{0}^{2}u_{1x} \nonumber \\
&& - H_{4}^{2}u_{0}^{2}u_{1x} - 68H_{1}H_{4}H_{5}\phi_{x}^{3}u_{1}^{2} - 80c_{1}H_{1}H_{5}\phi_{x}u_{0x}^{2} + 204H_{1}H_{4}H_{5}\phi_{x}^{2}u_{0x}u_{1} \nonumber \\
&& + 192H_{5}^{2}\phi_{x}^{4}u_{1x} - 384H_{5}^{2}\phi_{x}^{3}u_{0xx} - 680c_{1}H_{1}^{2}H_{5}\phi_{x}^{3}u_{1}^{2} + 560c_{1}H_{1}H_{5}\phi_{x}^{3}u_{1}^{2} \nonumber \\
&& - 2400c_{1}^{2}H_{1}^{3}\phi_{x}u_{0}^{2}u_{2} + 5200c_{1}^{2}H_{1}^{2}\phi_{x}u_{0}^{2}u_{2} - 3200c_{1}^{2}H_{1}\phi_{x}u_{0}^{2}u_{2} - 2400c_{1}^{2}H_{1}^{3}\phi_{x}u_{0}u_{1}^{2} \nonumber \\
&& - 2600c_{1}^{2}H_{1}^{2}u_{0}u_{0x}u_{1} + 5200c_{1}^{2}H_{1}^{2}\phi_{x}u_{0}u_{1}^{2} - 24H_{1}H_{4}^{2}\phi_{x}u_{0}^{2}u_{2} - 80c_{1}H_{4}\phi_{x}u_{0}^{2}u_{2} \nonumber \\
&& - 2400c_{1}H_{1}H_{5}^{2}\phi_{x}^{3}u_{0} - 2320c_{1}H_{1}H_{5}\phi_{x}\phi_{xx}u_{0}u_{1} + 256H_{1}H_{4}H_{5}\phi_{x}\phi_{xx}u_{0}u_{1} \nonumber \\
&& + 2560c_{1}H_{1}^{2}H_{5}\phi_{x}\phi_{xx}u_{0}u_{1} - 400c_{1}H_{1}H_{5}\phi_{x}^{2}u_{0}u_{1x} + 100H_{1}H_{4}H_{5}\phi_{x}^{2}u_{0}u_{1x} \nonumber \\
&& - 16H_{1}H_{4}H_{5}\phi_{x}u_{0x}^{2} - 360c_{1}H_{5}\phi_{x}^{2}u_{0x}u_{1} + 120c_{1}H_{1}^{2}H_{4}u_{0}^{2}u_{1x} - 140c_{1}H_{1}H_{4}u_{0}^{2}u_{1x} \nonumber \\
&& + 1200c_{1}^{2}H_{1}^{3}u_{0}u_{0x}u_{1} + 1600c_{1}^{2}H_{1}u_{0}u_{0x}u_{1} - 3200c_{1}^{2}H_{1}\phi_{x}u_{0}u_{1}^{2} + 12H_{1}H_{4}^{2}u_{0}u_{0x}u_{1} \nonumber \\
&& - 24H_{1}H_{4}^{2}\phi_{x}u_{0}u_{1}^{2} + 40c_{1}H_{4}u_{0}u_{0x}u_{1} - 80c_{1}H_{4}\phi_{x}u_{0}u_{1}^{2} + 36H_{4}H_{5}\phi_{x}^{2}u_{0x}u_{1} \nonumber \\
&& - 160c_{1}H_{1}^{2}H_{5}\phi_{x}u_{0x}^{2} + 200c_{1}H_{1}H_{5}\phi_{xxx}u_{0}^{2} - 20H_{1}H_{4}H_{5}\phi_{xxx}u_{0}^{2} \nonumber \\
&& - 200c_{1}H_{1}^{2}H_{5}\phi_{xxx}u_{0}^{2} + 560c_{1}H_{1}H_{4}\phi_{x}u_{0}u_{1}^{2} - 576H_{1}H_{5}^{2}\phi_{x}^{2}\phi_{xxx}u_{0} \nonumber \\
&& + 120c_{1}H_{5}\phi_{x}u_{0}u_{0xx} - 1728H_{1}H_{5}^{2}\phi_{x}^{2}\phi_{xx}u_{0x} + 576H_{1}H_{5}^{2}\phi_{x}^{3}\phi_{xx}u_{1} \nonumber \\
&& - 864H_{1}H_{5}^{2}\phi_{x}\phi_{xx}^{2}u_{0} + 120c_{1}H_{5}\phi_{xx}u_{0}u_{0x} - 12H_{4}H_{5}\phi_{xx}u_{0}u_{0x} \nonumber \\
&& + 240c_{1}H_{1}^{2}H_{4}u_{0}u_{0x}u_{1} - 480c_{1}H_{1}^{2}H_{4}\phi_{x}u_{0}u_{1}^{2} - 280c_{1}H_{1}H_{4}u_{0}u_{0x}u_{1} \nonumber \\
&& - 480c_{1}H_{1}^{2}H_{4}\phi_{x}u_{0}^{2}u_{2} + 1000c_{1}H_{1}^{2}H_{5}\phi_{x}^{2}u_{0}u_{1x} - 1680c_{1}H_{1}H_{5}\phi_{x}^{2}u_{0x}u_{1} \nonumber \\
&& - 12H_{4}H_{5}\phi_{x}u_{0}u_{0xx} - 200c_{1}^{2}u_{0}u_{0x}u_{1} + 240c_{1}H_{5}\phi_{x}u_{0x}^{2} + 2400c_{1}H_{1}H_{5}\phi_{x}^{3}u_{0}u_{2} \nonumber \\
&& - 2400c_{1}H_{1}^{2}H_{5}\phi_{x}^{3}u_{0}u_{2} - 200c_{1}H_{1}H_{4}H_{5}\phi_{x}u_{0}^{2} + 2040c_{1}H_{1}^{2}H_{5}\phi_{x}^{2}u_{0x}u_{1} \nonumber \\
&& + 400c_{1}^{2}\phi_{x}u_{0}u_{1}^{2} - 2H_{4}^{2}u_{0}u_{0x}u_{1} + 4H_{4}^{2}\phi_{x}u_{0}u_{1}^{2} + 400c_{1}^{2}\phi_{x}u_{0}^{2}u_{2} + 4H_{4}^{2}\phi_{x}u_{0}^{2}u_{2} \nonumber \\
&& + 2400c_{1}H_{5}^{2}\phi_{x}^{3}u_{0} - 240H_{4}H_{5}^{2}\phi_{x}^{3}u_{0} + 120c_{1}H_{5}\phi_{x}^{3}u_{1}^{2} - 12H_{4}H_{5}\phi_{x}^{3}u_{1}^{2} \nonumber \\
&& - 24H_{4}H_{5}\phi_{x}u_{0x}^{2} + 600c_{1}^{2}H_{1}^{3}u_{0}^{2}u_{1x} - 1300c_{1}^{2}H_{1}^{2}u_{0}^{2}u_{1x} + 800c_{1}^{2}H_{1}u_{0}^{2}u_{1x} \nonumber \\
&& + 6H_{1}H_{4}^{2}u_{0}^{2}u_{1x} - 384H_{5}^{2}\phi_{x}^{2}\phi_{xxx}u_{0} + 24H_{4}H_{5}\phi_{x}\phi_{xx}u_{0}u_{1} \nonumber \\
&& + 20c_{1}H_{4}u_{0}^{2}u_{1x} - 20H_{4}^{2}H_{5}\phi_{x}u_{0}^{2} + 288H_{1}H_{5}^{2}\phi_{x}^{4}u_{1x} + 384H_{5}^{2}\phi_{x}^{3}\phi_{xx}u_{1} \nonumber \\
&& - 576H_{5}^{2}\phi_{x}\phi_{xx}^{2}u_{0} - 576H_{1}H_{5}^{2}\phi_{x}^{3}u_{0xx} - 1152H_{5}^{2}\phi_{x}^{2}\phi_{xx}u_{0x} - 68H_{1}H_{4}H_{5}\phi_{xx}u_{0}u_{0x} \nonumber \\
&& - 680c_{1}H_{1}^{2}H_{5}\phi_{x}u_{0}u_{0xx} - 680c_{1}H_{1}^{2}H_{5}\phi_{xx}u_{0}u_{0x} - 240c_{1}H_{5}\phi_{x}\phi_{xx}u_{0}u_{1} \nonumber \\
&& + 560c_{1}H_{1}H_{5}\phi_{x}u_{0}u_{0xx} - 68H_{1}H_{4}H_{5}\phi_{x}u_{0}u_{0xx} + 560c_{1}H_{1}H_{5}\phi_{xx}u_{0}u_{0x} \nonumber \\
&& - 240H_{1}H_{4}H_{5}\phi_{x}^{3}u_{0}u_{2} + 560c_{1}H_{1}H_{4}\phi_{x}u_{0}^{2}u_{2} = 0
\end{eqnarray}

Upon solving the $O(\phi^{-7}),O(\phi^{-6})$ and $O(\phi^{-5})$ equations for $u_{0},u_{1}$ and $u_{2}$ respectively we find that
\begin{eqnarray}
u_{0}(x,t) &=& -\frac{12H_{5}\phi_{x}^{2}}{10c_{1}H_{1} - 10c_{1} + H_{4}} = -12\xi(t)\phi_{x}^{2}\\
u_{1}(x,t) &=& \frac{12H_{5}\phi_{xx}}{10c_{1}H_{1} - 10c_{1} + H_{4}} = 12\xi(t)\phi_{xx} \\
u_{2}(x,t) &=& -\frac{(4\phi_{x}\phi_{xxx} - 50c_{1}\phi_{x}^{2} - 3\phi_{xx}^{2})H_{5}}{(10c_{1}H_{1} - 10c_{1} + H_{4})\phi_{x}^{2}} = \frac{(4\phi_{x}\phi_{xxx} - 50c_{1}\phi_{x}^{2} - 3\phi_{xx}^{2})\xi(t)}{\phi_{x}^{2}}
\end{eqnarray}

which similarly lends itself nicely to a representation of the solution as
\[ u(x,t) = 12\xi(t)[\log(\phi(x,t))]_{xx} + u_{2}(x,t) \]

Further, if we let $C(x,t) = B(t)$ and once again $V(x,t) = 1$ the choices for coefficients reduce the remaining orders of $\phi$ to an identically satisfied system. Solving the determining equations for $\phi(x,t)$ we have that $\phi(x,t) = c_{2} + c_{3}e^{\int{B(t)dt} + x}$. Therefore we have the solution
\begin{equation}
u(x,t) = -\frac{\xi(t)\left(c_{2}^{2}(1 - 50c_{1}) - 10c_{2}c_{3}(1 + 10c_{1})e^{\int{B(t)dt} + x} + c_{3}^{2}(1 - 50c_{1})e^{2\int{B(t)dt} + 2x}\right)}{\left(c_{2} + c_{3}e^{\int{B(t)dt} + x}\right)^{2}}
\end{equation}

\section{Analytic solutions for generalized inhomogeneous MKdV Equations of form $\eqref{vcMKdV1}$}

The $a_{i}$, ($i = 1,2$), and all other quantities in this section refer to equation $\eqref{vcMKdV1}$.

Allowing $a_{2}(x,t) = F(x)G_{0}(t)$ (separable) and using the results of Khawaja's method we find that $a_{1}$ takes the form
\begin{eqnarray}
a_{1}(x,t) &=& \left(F(x)\int{\frac{1}{2}x^{2}M(x,t)dx}\right) - \left(xF(x)\int{xM(x,t)dx}\right) + \left(x^{2}F(x)\int{\frac{1}{2}M(x,t)dx}\right) \nonumber \\
&+& F(x)(G_{1}(t) + G_{2}(t)x + G_{3}(t)x^{2})
\end{eqnarray}
where $M(x,t) = \frac{H(t)G_{0}'(t)-H'(t)G_{0}(t)}{H(t)G_{0}(t)F(x)}$ and $C,F$ and $G_{0-3}$ are aribtrary functions in their respective variables. Letting $F(x) = e^{-x}$, $G_{1-3}(t)=0$ and keeping all other functions arbitrary we have the following vcMKDV
\begin{equation} 
u_{t} + \left(\frac{H(t)G_{0}'(t)-H'(t)G_{0}(t)}{H(t)G_{0}(t)}\right)u_{xxx} + e^{-x}G_{0}(t)u^{2}u_{x} = 0 \label{MKDVEX}
\end{equation}

Leading order analysis yields $\alpha = 1$. Therefore we seek a solution of the form $u(x,t) = \frac{u_{0}(x,t)}{\phi(x,t)} + u_{1}(x,t)$. Plugging this into $\eqref{MKDVEX}$ and collecting orders of $\phi$ we have
\begin{eqnarray}
O(\phi^{-4}) &:& -u_{0}\phi_{x}\left(e^{-x}G_{0}u_{0}^{2} + 6\left(\frac{H(t)G_{0}'(t)-H'(t)G_{0}(t)}{H(t)G_{0}(t)}\right)\phi_{x}^{2}\right) \\
O(\phi^{-3}) &:& e^{-x}G_{0}u_{0}^{2}u_{0x} + 6\left(\frac{H(t)G_{0}'(t)-H'(t)G_{0}(t)}{H(t)G_{0}(t)}\right)(u_{0}\phi_{x}\phi_{xx} + u_{0x}\phi_{x}^{2}) \nonumber \\
&&  - 2e^{-x}G_{0}u_{0}^{2}u_{1}\phi_{x} \\
O(\phi^{-2}) &:& e^{-x}G_{0}u_{0}^{2}u_{1x} + 2e^{-x}G_{0}u_{0}u_{0x}u_{1} - e^{-x}G_{0}u_{0}u_{1}^{2}\phi_{x} - u_{0}\phi_{t} \\
&& - \left(\frac{H(t)G_{0}'(t)-H'(t)G_{0}(t)}{H(t)G_{0}(t)}\right)(3u_{0x}\phi_{xx} + 3u_{0xx}\phi_{x} + u_{0}\phi_{xxx}) \\
O(\phi^{-1}) &:& \left(\frac{H(t)G_{0}'(t)-H'(t)G_{0}(t)}{H(t)G_{0}(t)}\right)u_{0xxx} + 2e^{-x}G_{0}u_{0}u_{1}u_{1x} + e^{-x}G_{0}u_{0x}u_{1}^{2} + u_{0t} \\
O(\phi^{0}) &:& u_{1t} + \left(\frac{H(t)G_{0}'(t)-H'(t)G_{0}(t)}{H(t)G_{0}(t)}\right)u_{1xxx} + e^{-x}G_{0}(t)u_{1}^{2}u_{1x}
\end{eqnarray}
Solving the $O(\phi^{-4})$ and $O(\phi^{-3})$ equations for $u_{0}$ and $u_{1}$ respectively we obtain the following results
\begin{eqnarray}
u_{0}(x,t) &=& \left(-6e^{x}\frac{H(t)G_{0}'(t)-H'(t)G_{0}(t)}{H(t)G_{0}(t)^{2}}\right)^{1/2}\phi_{x} \\
u_{1}(x,t) &=& -\left(-6e^{x}\frac{H(t)G_{0}'(t)-H'(t)G_{0}(t)}{H(t)G_{0}(t)^{2}}\right)^{1/2}\phi_{x}\left(\frac{\phi_{xx}}{\phi_{x}}\right)
\end{eqnarray}
Substituting the equations for $C(x,t)$ and $V(x,t)$ into the remaining orders of $\phi$, solving the new system for $C(x,t)$ and $V(x,t)$, and mandating that $H(t)$ satisfy
\[ \sqrt{-24G_{0}'(t)G_{0}(t)M(t) + 6G_{0}(t)^{2}M(t)^{2} + 24G_{0}^{2}M'(t)} = M(t)G_{0}(t) \]
where $M(t) = \frac{H(t)G_{0}'(t)-H'(t)G_{0}(t)}{H(t)}$ we therefore have
\begin{eqnarray}
H(t) &=& \frac{c_{2}G_{0}(t)}{\left(\frac{5}{24}\int{G_{0}(t)dt}+c_{1}\right)^{24/5}} \\
C(x,t) &=& -\frac{53G_{0}(t)}{15\int{G_{0}(t)dt} + 72c_{1}} \\
V(x,t) &=& -\frac{1}{6}\tanh\left(\frac{1}{12}x - \frac{53}{180}\ln\left(\frac{5}{24}\int{G_{0}(t)dt} + c_{1}\right)\right)
\end{eqnarray}

Solving the coupled pde system for $\phi(x,t)$ we get

\begin{equation}
\phi(x,t) = \tanh\left(\frac{1}{12}x - \frac{53}{180}\ln\left(\frac{5}{24}\int{G_{0}(t)dt} + c_{1}\right)\right)
\end{equation}

Therefore, after a bit of simplification, we have the solution

\begin{equation}
u(x,t) = \frac{ie^{x/2}}{\sqrt{5\int{G_{0}(t)dt}+24c_{1}}}\coth\left(\frac{1}{12}x - \frac{53}{180}\ln\left(\frac{5}{24}\int{G_{0}(t)dt} + c_{1}\right)\right)
\end{equation}

\section{Conclusions and Future Work}

We have used two direct methods to obtain very significantly extended Lax- or S-integrable families
of generalized KdV and MKdV equations with coefficients which may in general vary in both space and time.
Of these, the second technique which was developed here is a new, significantly extended version of the
well-known Estabrook-Wahlquist technique for Lax-integrable systems with constant coefficients.
Some solutions for the generalized inhomogeneous KdV equations and one family of Lax-integrable
generalized MKdV equations have also been presented here.

Future work will address the derivation of additional solutions by various methods,
as well as detailed investigations of other integrability properties of these
novel integrable inhomogeneous NLPDEs such as Backlund
Transformations and conservation laws.

%\newpage

%\begin{appendices}
%\chapter*{APPENDIX: Lax integrability conditions for generalized MKdV and KdV Equations}

\appendix

\newpage
\setcounter{equation}{0}
\renewcommand{\theequation}{A.\arabic{equation}}

\section{Appendix: Lax integrability for generalized MKdV equations}

The Lax pair is expanded in powers of $u$ and its derivatives as follows:

\begin{eqnarray}
U &=& \begin{bmatrix}
	f_{1}+f_{2}v & f_{3}+f_{4}v  \\
	 f_{5}+f_{6}v & f_{7}+f_{8}v
\end{bmatrix} \\
V^{1} &=& \begin{bmatrix}
	V^{1}_{1} & V^{1}_{2}  \\
	V^{1}_{3} & V^{1}_{4}
\end{bmatrix} \\
V^{2} &=& \begin{bmatrix}
	V^{2}_{1} & V^{2}_{2}  \\
	V^{2}_{3} & V^{2}_{4}
\end{bmatrix} \\
V^{3} &=& \begin{bmatrix}
	V^{3}_{1} & V^{3}_{2}  \\
	V^{3}_{3} & V^{3}_{4}
\end{bmatrix}
\end{eqnarray}
where
\begin{eqnarray*}
V^{1}_{1} &=& g_{1} + g_{2}v + g_{3}v^{2} \\
V^{1}_{2} &=& g_{4} + g_{5}v + g_{6}v^{2} + g_{7}v^{3} + g_{8}v_{x} + g_{9}v_{xx} \\
V^{1}_{3} &=& g_{10} + g_{11}v + g_{12}v^{2} + g_{13}v^{3} + g_{14}v_{x} + g_{15}v_{xx} \\
V^{1}_{4} &=& g_{16} + g_{17}v + g_{18}v^{2}
\end{eqnarray*}
\begin{eqnarray*}
V^{2}_{1} &=& g_{1}v_{x}^{2} + g_{2}v^{4} + g_{3}vv_{xx} + g_{4}v^{2} + g_{5} + g_{6}v_{xxxx} + g_{7}vv_{x}^{2} + g_{8}v^{2}v_{xx} + g_{9}v^{5} + g_{10}v_{xxx} \\
&& + g_{11}v^{2}v_{x} + g_{12}v_{xx} + g_{13}v^{3} + g_{14}v_{x} + g_{15}v \\
V^{2}_{2} &=& g_{16}v_{x}^{2} + g_{17}v^{4} + g_{18}vv_{xx} + g_{19}v^{2} + g_{20} + g_{21}v_{xxxx} + g_{22}vv_{x}^{2} + g_{23}v^{2}v_{xx} + g_{24}v^{5} \\
&& + g_{25}v_{xxx} + g_{26}v^{2}v_{x} + g_{27}v_{xx} + g_{28}v^{3} + g_{29}v_{x} + g_{30}v \\
V^{2}_{3} &=& g_{31}v_{x}^{2} + g_{32}v^{4} + g_{33}vv_{xx} + g_{34}v^{2} + g_{35} + g_{36}v_{xxxx} + g_{37}vv_{x}^{2} + g_{38}v^{2}v_{xx} + g_{39}v^{5} \\
&& + g_{40}v_{xxx} + g_{41}v^{2}v_{x} + g_{42}v_{xx} + g_{43}v^{3} + g_{44}v_{x} + g_{45}v \\
V^{2}_{4} &=& g_{46}v_{x}^{2} + g_{47}v^{4} + g_{48}vv_{xx} + g_{49}v^{2} + g_{50} + g_{51}v_{xxxx} + g_{52}vv_{x}^{2} + g_{53}v^{2}v_{xx} + g_{54}v^{5} \\
&& + g_{55}v_{xxx} + g_{56}v^{2}v_{x} + g_{57}v_{xx} + g_{58}v^{3} + g_{59}v_{x} + g_{60}v \\
\end{eqnarray*}
\begin{eqnarray*}
V^{3}_{1} &=& g_{1} + g_{2}v^{2} + g_{3}vv_{xx} + g_{4}v_{x}v_{xxx} + g_{5}v^{2}v_{x}^{2} + g_{6}v^{3}v_{xx} + g_{7}v_{xx}^{2} + g_{8}vv_{xxxx} + g_{9}v_{x}^{2} + g_{10}v^{4} \\
&& + g_{11}v^{6} \\
V^{3}_{2} &=& g_{12}v^{2}v_{xxx} + g_{13}vv_{x}^{2} + g_{14}v^{2}v_{xx} + g_{15}v^{2}v_{x} + g_{16}v^{4}v_{x} + g_{17}v_{x}^{2}v_{xx} + g_{18}vv_{xx}^{2} \\
&& + g_{19}v^{2}v_{xxxx} + g_{20}v^{3}v_{x}^{2} + g_{21}v^{4}v_{xx} + g_{22}vv_{x}v_{xx} + g_{23}vv_{x}v_{xxx} + g_{24}v + g_{25}v_{x} + g_{26}v_{xx} \\
&& + g_{27}v_{xxx} + g_{28}v_{xxxx} + g_{29}v_{xxxxx} + g_{30}v_{xxxxxx} + g_{31}v_{x}^{3} + g_{32}v^{3} + g_{33}v^{5} + g_{34}v^{7} + g_{35} \\
V^{3}_{3} &=& g_{36}v^{2}v_{xxx} + g_{37}vv_{x}^{2} + g_{38}v^{2}v_{xx} + g_{39}v^{2}v_{x} + g_{40}v^{4}v_{x} + g_{41}v_{x}^{2}v_{xx} + g_{42}vv_{xx}^{2}  \\
&&+ g_{43}v^{2}v_{xxxx} + g_{44}v^{3}v_{x}^{2} + g_{45}v^{4}v_{xx} + g_{46}vv_{x}v_{xx} + g_{47}vv_{x}v_{xxx} + g_{48}v + g_{49}v_{x} \\
&& + g_{50}v_{xx} + g_{51}v_{xxx} + g_{52}v_{xxxx} + g_{53}v_{xxxxx} + g_{54}v_{xxxxxx} + g_{55}v_{x}^{3} + g_{56}v^{3} + g_{57}v^{5} \\
&& + g_{58}v^{7} + g_{59} \\
V^{3}_{4} &=& g_{60} + g_{61}v^{2} + g_{62}vv_{xx} + g_{63}v_{x}v_{xxx} + g_{64}v^{2}v_{x}^{2} + g_{65}v^{3}v_{xx} + g_{66}v_{xx}^{2} + g_{67}vv_{xxxx} + g_{68}v_{x}^{2} \\
&& + g_{70}v^{6} + g_{69}v^{4} \\
\end{eqnarray*}

The compatibility condition gives
\begin{equation}
U_{t} - V_{x} + [U,V] = \dot{0} = \begin{bmatrix}
	0 & p_{1}(x,t)F_{i}[v]  \\
	p_{2}(x,t)F_{i}[v] & 0
\end{bmatrix}
\end{equation}
where $F_{i}[v]$ represents the $i^{th}$ equation in the MKDV hierarchy ($i=1-3$).

\subsection{Determining Equations for the First Equation}

Requiring the compatability condition yield the $F_{1}[v]$ gives
$f_{4} = p_{1}, f_{6} = p_{2}, g_{7} = -\frac{1}{3}p_{1}a_{2}, g_{9} = -p_{1}a_{1}, g_{13} = -\frac{1}{3}p_{2}a_{2}, g_{15} = -p_{2}a_{1}, f_{2} = f_{8} = g_{12} = g_{6} = 0$

\begin{eqnarray}
&& f_{3}p_{2} - f_{5}p_{1} = 0 \\
&& g_{18x} + p_{1}g_{11} - p_{2}g_{5} = 0 \\
&& g_{3x} + p_{2}g_{5} - p_{1}g_{11} = 0 \\
&& g_{17} + f_{3}g_{14} - f_{5}g_{8} = 0 \\
&& 2g_{18} + p_{1}g_{14} - p_{2}g_{8} = 0 \\
&& g_{2} + f_{5}g_{8} - f_{3}g_{14} = 0
\end{eqnarray}
\begin{eqnarray}
&& 2g_{3} + p_{2}g_{8} - p_{1}g_{14} = 0 \\
&& f_{5}(g_{18} - g_{3}) + p_{2}(g_{17} - g_{2}) = 0 \\
&& f_{3}(g_{18} - g_{3}) + p_{1}(g_{17} - g_{2}) = 0 \\
&& f_{1t} - g_{1x} + f_{3}g_{10} - f_{5}g_{4} = 0 \\
&& f_{7t} - g_{16x} - f_{3}g_{10} + f_{5}g_{4} = 0 \\
&& g_{11} + g_{14x} - g_{14}(f_{7} - f_{1}) = 0
\end{eqnarray}
\begin{eqnarray}
&& g_{5} + g_{8x} + g_{8}(f_{7} - f_{1}) = 0 \\
&& g_{8} - (p_{1}a_{1})_{x} - p_{1}a_{1}(f_{7} - f_{1}) = 0 \\
&& g_{14} - (p_{2}a_{1})_{x} + p_{2}a_{1}(f_{7} - f_{1}) = 0 \\
&& g_{17x} + f_{3}g_{11} - f_{5}g_{5} + p_{1}g_{10} - p_{2}g_{4} = 0 \\
&& g_{2x} - f_{3}g_{11} + f_{5}g_{5} - p_{1}g_{10} + p_{2}g_{4} = 0 \\
&& \frac{1}{3}(p_{1}a_{2})_{x} + \frac{1}{3}p_{1}a_{2}(f_{7} - f_{1}) + p_{1}(g_{18} - g_{3}) = 0
\end{eqnarray}
\begin{eqnarray}
&& \frac{1}{3}(p_{2}a_{2})_{x} - \frac{1}{3}p_{2}a_{2}(f_{7} - f_{1}) - p_{2}(g_{18} - g_{3}) = 0 \\
&& f_{3t} - g_{4x} - g_{4}(f_{7} - f_{1}) - f_{3}(g_{1} - g_{16}) = 0 \\
&& f_{5t} - g_{10x} + g_{10}(f_{7} - f_{1}) + f_{5}(g_{1} - g_{16}) = 0 \\
&& p_{1t} - g_{5x} - g_{5}(f_{7} - f_{1}) - p_{1}(g_{1} - g_{16}) - f_{3}(g_{2} - g_{17}) = 0 \\
&& p_{2t} - g_{11x} + g_{11}(f_{7} - f_{1}) + p_{2}(g_{1} - g_{16}) + f_{5}(g_{2} - g_{17}) = 0
\end{eqnarray}

\subsection{Deriving a relation between the $a_{i}$}
In this section we reduce the previous system down to equations which depend solely on the $a_{i}$'s. We find that
\newline
$g_{16} = g_{1} = f_{7} = f_{1} = g_{10} = g_{4} = g_{17} = g_{2} = f_{5} = f_{3} = 0, p_{2} = -p_{1} = -\frac{C(t)}{a_{2}}$

\begin{eqnarray*}
&& g_{18} = -g_{3} = -\frac{C(t)g_{8}}{a_{2}} \\
&& g_{11} = -g_{5} = g_{8x} \\
&& g_{14} = -g_{8} = -C(t)(\frac{a_{1}}{a_{2}})_{x}
\end{eqnarray*}

which leads to

\begin{eqnarray}
&& 6a_{1}a_{2x}^{3} - 6a_{1}a_{2}a_{2x}a_{2xx} + a_{1}a_{2}^{2}a_{2xxx} - \frac{K_{t}}{K}a_{2}^{3} + a_{2}^{2}a_{2t} - a_{2}^{3}a_{1xxx} \nonumber \\
&& + 3a_{1xx}a_{2}^{2}a_{2x} - 6a_{1x}a_{2}a_{2x}^{2} + 3a_{1x}a_{2}^{2}a_{2xx} = 0
\end{eqnarray}

where $K(t)$ and $C(t)$ are arbitrary functions of $t$.

\subsection{Determining Equations for the Second Equation}

Requiring the compatability condition yield the $F_{2}[v]$ gives
$p_{2} = p_{1}, g_{51} = g_{6} = g_{52} = g_{7} = g_{53} = g_{8} = g_{54} = g_{9} = 0, g_{18} = -2g_{16}, g_{33} = -2g_{31}, g_{48} = -2g_{46}, g_{36} = g_{21} = -p_{1}b_{1}, g_{37} = g_{22} = -p_{1}b_{4}, g_{38} = g_{23} = -p_{1}b_{2}, g_{39} = g_{24} = -\frac{1}{5}p_{1}b_{5}$

\begin{eqnarray}
&& b_{3} = 2b_{2} + 2b_{4} \\
&& g_{1x} + f_{3}g_{16} - f_{2}g_{31} = 0 \\
&& g_{55} + p_{1}b_{1}(f_{3} - f_{2}) = 0 \\
&& g_{46x} + f_{2}g_{31} - f_{3}g_{16} = 0 \\
&& 2g_{1} - g_{25} + g_{40} = 0 \\
&& g_{10} + p_{1}b_{1}(f_{2} - f_{3}) = 0
\end{eqnarray}
\begin{eqnarray}
&& g_{11} + 4g_{17} - g_{56} = 0 \\
&& g_{11} - 4g_{32} - g_{56} = 0 \\
&& g_{14} - 2g_{34} - g_{59} = 0 \\
&& g_{14} + 2g_{19} - g_{59} = 0 \\
&& g_{26} - 4g_{47} - g_{41} = 0 \\
&& g_{44} - 2g_{4} - g_{29} = 0
\end{eqnarray}
\begin{eqnarray}
&& g_{40} - 2g_{46} - g_{25} = 0 \\
&& g_{41} - 4g_{2} - g_{26} = 0 \\
&& g_{44} + 2g_{49} - g_{29} = 0 \\
&& g_{55} - 2g_{31} - g_{10} = 0 \\
&& g_{55} + 2g_{16} - g_{10} = 0 \\
&& f_{4t} - g_{50x} + f_{3}g_{20} - f_{2}g_{35} = 0
\end{eqnarray}
\begin{eqnarray}
&& f_{1t} - g_{5x} + f_{2}g_{35} - f_{3}g_{20} = 0 \\
&& g_{59} + g_{57x} + f_{2}g_{42} - f_{3}g_{27} = 0 \\
&& g_{60} + g_{59x} + f_{2}g_{44} - f_{3}g_{29} = 0 \\
&& g_{14} + g_{12x} + f_{3}g_{27} - f_{2}g_{42} = 0 \\
&& g_{15} + g_{14x} + f_{3}g_{29} - f_{2}g_{44} = 0 \\
&& 3g_{58} + g_{56x} + f_{2}g_{41} - f_{3}g_{26} = 0
\end{eqnarray}
\begin{eqnarray}
&& g_{12} + g_{10x} + f_{3}g_{25} - f_{2}g_{40} = 0 \\
&& 3g_{13} + g_{11x} + f_{3}g_{26} - f_{2}g_{41} = 0 \\
&& g_{32} - g_{17} + \frac{1}{5}p_{1}b_{5}(f_{3} - f_{2}) = 0 \\
&& g_{57} + g_{55x} + f_{2}g_{40} - f_{3}g_{25} = 0 \\
&& g_{27} - g_{42} + 2g_{46x} + 2f_{2}g_{31} - 2f_{3}g_{16} = 0 \\
&& g_{43} - g_{28} - g_{2x} + f_{2}g_{32} - f_{3}g_{17} = 0
\end{eqnarray}
\begin{eqnarray}
&& g_{19} - g_{34} - g_{58x} + f_{3}g_{28} - f_{2}g_{43} = 0 \\
&& g_{16x} + g_{16}(f_{4} - f_{1}) + f_{2}(g_{1} - g_{46}) = 0 \\
&& g_{42} + 2g_{1x} - g_{27} + 2f_{3}g_{16} - 2f_{2}g_{31} = 0 \\
&& g_{45} - g_{4x} - g_{30} + f_{2}g_{34} - f_{3}g_{19} = 0 \\
&& g_{45} + g_{49x} - g_{30} + f_{2}g_{34} - f_{3}g_{19} = 0 \\
&& g_{34} - g_{13x} - g_{19} + f_{2}g_{43} - f_{3}g_{28} = 0
\end{eqnarray}
\begin{eqnarray}
&& g_{31x} + g_{31}(f_{4} - f_{1}) + f_{3}(g_{46} - g_{1}) = 0 \\
&& g_{25} - (p_{1}b_{1})_{x} - p_{1}b_{1}(f_{4} - f_{1}) = 0 \\
&& g_{40} - (p_{1}b_{1})_{x} + p_{1}b_{1}(f_{4} - f_{1}) = 0 \\
&& g_{11} - 2g_{16} + 2g_{31} + p_{1}b_{2}(f_{2} - f_{3}) = 0 \\
&& 2g_{11} + g_{16} - g_{31} + p_{1}b_{4}(f_{2} - f_{3}) = 0 \\
&& g_{20} - g_{60x} - g_{35} + f_{3}g_{30} - f_{2}g_{45} = 0
\end{eqnarray}
\begin{eqnarray}
&& g_{28} - g_{47x} - g_{43} + f_{3}g_{27} - f_{2}g_{32} = 0 \\
&& g_{31} + 2g_{56} - g_{16} - p_{1}b_{4}(f_{2} - f_{3}) = 0 \\
&& 2g_{31} - g_{56} - 2g_{16} + p_{1}b_{2}(f_{2} - f_{3}) = 0 \\
&& g_{35} - g_{15x} - g_{20} + f_{2}g_{45} - f_{3}g_{30} = 0 \\
&& g_{2} - g_{47} + \frac{1}{5}(p_{1}b_{5})_{x} - \frac{1}{5}p_{1}b_{5}(f_{4} - f_{1}) = 0 \\
&& g_{44} + g_{42x} - g_{42}(f_{4} - f_{1}) + f_{3}(g_{57} - g_{12}) = 0
\end{eqnarray}
\begin{eqnarray}
&& g_{42} + g_{40x} - g_{40}(f_{4} - f_{1}) - f_{3}(g_{10} - g_{55}) = 0 \\
&& g_{27} + g_{25x} + g_{25}(f_{4} - f_{1}) + f_{2}(g_{10} - g_{55}) = 0 \\
&& g_{30} + g_{29x} + g_{29}(f_{4} - f_{1}) + f_{2}(g_{14} - g_{59}) = 0 \\
&& f_{2t} - g_{20x} - g_{20}(f_{4} - f_{1}) - f_{2}(g_{5} - g_{50}) = 0 \\
&& f_{3t} - g_{35x} + g_{35}(f_{4} - f_{1}) + f_{3}(g_{5} - g_{50}) = 0 \\
&& 3g_{43} + g_{41x} - g_{41}(f_{4} - f_{1}) - f_{3}(g_{11} - g_{56}) = 0
\end{eqnarray}
\begin{eqnarray}
&& g_{45} + g_{44x} - g_{44}(f_{4} - f_{1}) - f_{3}(g_{14} - g_{59}) = 0 \\
&& 3g_{28} + g_{26x} + g_{26}(f_{4} - f_{1}) + f_{2}(g_{11} - g_{56}) = 0 \\
&& g_{29} + g_{27x} + g_{27}(f_{4} - f_{1}) + f_{2}(g_{12} - g_{57}) = 0 \\
&& g_{47} - g_{2} + \frac{1}{5}(p_{1}b_{5})_{x} + \frac{1}{5}p_{1}b_{5}(f_{4} - f_{1}) = 0 \\
&& g_{26} + 2g_{46} - 2g_{1} - (p_{1}b_{2})_{x} - p_{1}b_{2}(f_{4} - f_{1}) = 0 \\
&& g_{12} + 2g_{31x} - g_{57} - 2g_{31}(f_{4} - f_{1}) - 2f_{3}(g_{1} - g_{46}) = 0
\end{eqnarray}
\begin{eqnarray}
&& g_{49} - g_{28x} - g_{4} - g_{28}(f_{4} - f_{1}) - f_{2}(g_{13} - g_{58}) = 0 \\
&& g_{50} + g_{45x} - g_{5} - g_{45}(f_{4} - f_{1}) - f_{3}(g_{15} - g_{60}) = 0 \\
&& g_{50} - g_{30x} - g_{5} - g_{30}(f_{4} - f_{1}) - f_{2}(g_{15} - g_{60}) = 0 \\
&& g_{58} - g_{17x} - g_{13} - g_{17}(f_{4} - f_{1}) - f_{2}(g_{2} - g_{47}) = 0 \\
&& g_{60} + g_{34x} - g_{15} - g_{34}(f_{4} - f_{1}) - f_{3}(g_{4} - g_{49}) = 0 \\
&& g_{46} - 2g_{26} - g_{1} + (p_{1}b_{4})_{x} + p_{1}b_{4}(f_{4} - f_{1}) = 0 \\
\end{eqnarray}
\begin{eqnarray}
&& g_{46} + 2g_{41} - g_{1} - (p_{1}b_{4})_{x} + p_{1}b_{4}(f_{4} - f_{1}) = 0 \\
&& 2g_{46} - g_{41} - 2g_{1} + (p_{1}b_{2})_{x} - p_{1}b_{2}(f_{4} - f_{1}) = 0 \\
&& g_{57} + 2g_{16x} - g_{12} + 2g_{16}(f_{4} - f_{1}) - 2f_{2}(g_{46} - g_{1}) = 0 \\
&& g_{60} - g_{19x} - g_{15} - g_{19}(f_{4} - f_{1}) - f_{2}(g_{49} - g_{4}) = 0 \\
&& g_{49} + g_{43x} - g_{4} - g_{43}(f_{4} - f_{1}) - f_{3}(g_{13} - g_{58}) = 0 \\
&& g_{58} + g_{32x} - g_{13} - g_{32}(f_{4} - f_{1}) - f_{3}(g_{47} - g_{2}) = 0
\end{eqnarray}

\subsection{Deriving a relation between the $b_{i}$}
In this section we reduce the previous system down to equations which depend solely on the $b_{i}$'s. We find that
\newline
$g_{34} = -g_{19}, g_{42} = g_{27}, g_{56} = g_{55} = g_{31} = g_{16} = g_{10} = g_{32} = g_{17} = g_{58} = g_{57} = g_{13} = g_{47} = g_{2} = g_{46} = g_{44} = g_{49} = g_{50} = g_{1} = g_{4} = g_{5} = g_{11} = g_{12} = g_{14} = g_{15} = g_{29} = f_{4} = f_{1} = 0, f_{3} = f_{2}, p_{1} = \frac{H(t)}{b_{5}}$

\begin{eqnarray*}
&& g_{59} = 2g_{19} \\
&& g_{40} = g_{25} \\
&& g_{41} = g_{26} \\
&& g_{45} = g_{30} + 2g_{19}f_{2} \\
&& g_{19} = \frac{1}{2}f_{2}(g_{43} - g_{28}) \\
&& g_{35} = g_{20} - 2f_{2}^{2}g_{19}
\end{eqnarray*}
\begin{eqnarray*}
&& g_{60} = g_{19x} \\
&& g_{25} = H(t)\left(\frac{b_{1}}{b_{5}}\right)_{x} \\
&& g_{26} = H(t)\left(\frac{b_{2}}{b_{5}}\right)_{x} \\
&& g_{27} = -H(t)\left(\frac{b_{2}}{b_{5}}\right)_{xx} \\
&& g_{28} = \frac{4}{3}H(t)\left(\frac{b_{2}}{b_{5}}\right)_{x} - \frac{1}{3}H(t)\left(\frac{b_{2}}{b_{5}}\right)_{xx} \\
&& g_{30} = -\left(\frac{b_{1}}{b_{5}}\right)_{xxxx} \\
&& g_{43} =  -\frac{1}{3}H(t)\left(\frac{b_{2}}{b_{5}}\right)_{xx}
\end{eqnarray*}

which leads to

\begin{eqnarray*}
&& b_{3} = 2b_{2} + 2b_{4} \\
&& b_{5} = C(t)(2b_{2} - b_{4})
\end{eqnarray*}

and the long equation previously presented. Here $H(t)$ and $C(t)$ are arbitrary functions of $t$.

\subsection{Determining Equations for the Third Equation}

Requiring the compatability condition yield the $F_{3}[v]$ gives

\begin{eqnarray*}
&& f_{2} = f_{8} = 0 , g_{34} = -\frac{1}{7}p_{1}a_{10}, g_{20} = -\frac{1}{3}p_{1}a_{9}, f_{4} = p_{1}, g_{30} = -p_{1}a_{1}, g_{21} = -p_{1}a_{7}, g_{19} = -p_{1}a_{2}, \\
&& g_{54} = p_{2}a_{1}, g_{58} = \frac{1}{7}p_{2}a_{10}, g_{44} = \frac{1}{3}p_{2}a_{9}, g_{45} = p_{2}a_{7}, g_{43} = p_{2}a_{2}, f_{6} = -p_{2}, g_{63} = -g_{67} = -g_{8} \\
&& = g_{4} = G_{4}(t), g_{66} = g_{7} = G_{7}(t), g_{68} = -\frac{1}{2}g_{3} = g_{9} = G_{9}(t), f_{5} = f_{3}, f_{7} = f_{1}, p_{2} = -p_{1}
\end{eqnarray*}

\begin{eqnarray}
&& f_{3}(g_{i+24} - g_{i}) = 0 \ \ (i=12-18,22,23,25-29,31-33) \\
&& f_{3}(g_{i+59} - g_{i}) = 0 \ \ (i=2,3,5,6,8,10,11) \\
&& \frac{2}{3}p_{1}c_{9} + 4p_{1}a_{7} = p_{1}c_{8} \\
&& g_{47} - p_{1}(2c_{2} - c_{3}) = 0 \\
&& g_{23} - p_{1}(2c_{2} - c_{3}) = 0 \\
&& G_{7} = -\frac{1}{2}G_{4}
\end{eqnarray}
\begin{eqnarray}
&& G_{66} = -\frac{1}{2}G_{63} \\
&& G_{68} = -\frac{1}{2}G_{62} \\
&& g_{23} + g_{17} = -p_{1}c_{5} \\
&& g_{23} + 2g_{18} = -p_{1}c_{4} \\
&& g_{47} + g_{41} = -p_{1}c_{5} \\
&& g_{47} + 2g_{42} = -p_{1}c_{4}
\end{eqnarray}
\begin{eqnarray}
&& g_{18} + 2g_{17} = -p_{1}c_{6} \\
&& g_{42} + 2g_{41} = -p_{1}c_{6} \\
&& g_{2} - \frac{1}{2}p_{1}(g_{49} - g_{25}) = 0 \\
&& g_{3} - p_{1}(g_{51} - g_{27}) = 0 \\
&& g_{5} - \frac{1}{2}p_{1}(g_{55} - g_{31}) = 0 \\
&& g_{6} - p_{1}(g_{36} - g_{16}) = 0
\end{eqnarray}
\begin{eqnarray}
&& g_{8} - p_{1}(g_{53} - g_{29}) = 0 \\
&& g_{10} - \frac{1}{4}p_{1}(g_{39} - g_{15}) = 0 \\
&& g_{11} - \frac{1}{6}p_{1}(g_{40} - g_{16}) = 0 \\
&& g_{61} + \frac{1}{2}p_{1}(g_{49} - g_{25}) = 0 \\
&& g_{62} + p_{1}(g_{51} - g_{27}) = 0 \\
&& g_{64} + \frac{1}{2}p_{1}(g_{55} - g_{31}) = 0
\end{eqnarray}
\begin{eqnarray}
&& g_{65} + p_{1}(g_{36} - g_{12}) = 0 \\
&& g_{67} + p_{1}(g_{53} - g_{29}) = 0 \\
&& g_{69} + \frac{1}{4}p_{1}(g_{39} - g_{15}) = 0 \\
&& g_{70} + \frac{1}{6}p_{1}(g_{40} - g_{16}) = 0 \\
&& g_{2x} - p_{1}(g_{48} - g_{24}) = 0 \\
&& g_{3x} - p_{1}(g_{50} - g_{26}) = 0
\end{eqnarray}
\begin{eqnarray}
&& g_{5x} - p_{1}(g_{37} - g_{13}) = 0 \\
&& g_{6x} - p_{1}(g_{38} - g_{14}) = 0 \\
&& g_{8x} - p_{1}(g_{52} - g_{28}) = 0 \\
&& g_{10x} - p_{1}(g_{56} - g_{32}) = 0 \\
&& g_{11x} - p_{1}(g_{57} - g_{33}) = 0 \\
&& g_{61x} + p_{1}(g_{48} - g_{24}) = 0
\end{eqnarray}
\begin{eqnarray}
&& g_{62x} + p_{1}(g_{50} - g_{26}) = 0 \\
&& g_{64x} + p_{1}(g_{37} - g_{13}) = 0 \\
&& g_{65x} + p_{1}(g_{38} - g_{14}) = 0 \\
&& g_{67x} + p_{1}(g_{52} - g_{28}) = 0 \\
&& g_{69x} + p_{1}(g_{56} - g_{32}) = 0 \\
&& g_{70x} + p_{1}(g_{57} - g_{33}) = 0
\end{eqnarray}
\begin{eqnarray}
&& g_{12} - (p_{1}c_{2})_{x} - p_{1}(g_{67} - g_{8}) = 0 \\
&& 2g_{12} + g_{23x} + g_{22} = 0 \\
&& g_{13} + g_{31x} = 0 \\
&& g_{14} + g_{12x} = 0 \\
&& g_{15} + g_{14x} + p_{1}(g_{3} - g_{62}) = 0 \\
&& 2g_{15} + g_{13x} = 0
\end{eqnarray}
\begin{eqnarray}
&& g_{16} - (p_{1}c_{7})_{x} - p_{1}(g_{65} - g_{6}) = 0 \\
&& 4g_{16} - \frac{1}{3}(p_{1}c_{9})_{x} - p_{1}(g_{64} - g_{5}) = 0 \\
&& 4g_{40} - \frac{1}{3}(p_{1}c_{9})_{x} + p_{1}(g_{64} - g_{5}) = 0 \\
&& g_{22} + g_{18x} = 0 \\
&& g_{24} + g_{25x} = 0 \ \ (i=24-29)  \\
&& 3g_{32} + g_{15x} = 0
\end{eqnarray}
\begin{eqnarray}
&& g_{37} + g_{55x} = 0 \\
&& g_{38} + g_{36x} = 0 \\
&& g_{39} + g_{38x} - p_{1}(g_{3} - g_{62}) = 0 \\
&& g_{40} - (p_{1}c_{7})_{x} + p_{1}(g_{65} - g_{6}) = 0 \\
&& g_{46} + g_{42x} = 0 \\
&& g_{46} + g_{47x} + 2g_{36} = 0
\end{eqnarray}
\begin{eqnarray}
&& g_{i} + g_{i+1,x} = 0 \ \ (i=48-52) \\
&& 3g_{56} + g_{39x} = 0 \\
&& 5g_{57} + g_{40x} = 0 \\
&& p_{1}(g_{59} - g_{35}) + f_{3}(g_{48} - g_{24}) = 0 \\
&& 2g_{13} + g_{22x} + 2g_{14} = 0 \\
&& g_{16} + g_{16x} + 5g_{33} = 0
\end{eqnarray}
\begin{eqnarray}
&& g_{22} + g_{17x} + 3g_{31} = 0 \\
&& 2g_{37} + g_{46x} + 2g_{38} = 0 \\
&& g_{46} + g_{41x} + 3g_{55} = 0 \\
&& 2g_{5} + 3g_{6} - p_{1}(g_{46} - g_{22}) = 0 \\
&& 2g_{64} + 3g_{65} + p_{1}(g_{46} - g_{22}) = 0 \\
&& f_{1t} - g_{1x} - f_{3}(g_{35} + g_{59}) = 0
\end{eqnarray}
\begin{eqnarray}
&& f_{1t} - g_{60x} + f_{3}(g_{35} - g_{59}) = 0 \\
&& f_{3t} - g_{35x} + f_{3}(g_{60} - g_{1}) = 0 \\
&& f_{3t} - g_{59x} - f_{3}(g_{60} - g_{1}) = 0 \\
&& g_{32x} + p_{1}(g_{2} - g_{61}) = 0 \\
&& g_{33x} + p_{1}(g_{10} - g_{69}) = 0 \\
&& g_{56x} - p_{1}(g_{2} - g_{61}) = 0
\end{eqnarray}
\begin{eqnarray}
&& g_{57x} - p_{1}(g_{10} - g_{69}) = 0 \\
&& g_{29} - (p_{1}c_{1})_{x} = 0 \\
&& g_{36} - (p_{1}c_{2})_{x} + p_{1}(g_{67} - g_{8}) = 0 \\
&& g_{53} - (p_{1}c_{1})_{x} = 0 \\
&& 2g_{39} + g_{37x} = 0 \\
&& \pm \frac{1}{7}(p_{1}c_{10})_{x} + p_{1}(g_{70} - g_{11}) = 0
\end{eqnarray}
\begin{eqnarray}
&& p_{1t} - g_{48x} + p_{1}(g_{1} - g_{60}) = 0 \\
&& p_{1t} - g_{24x} - p_{1}(g_{1} - g_{60}) = 0
\end{eqnarray}

\subsection{Deriving a relation between the $c_{i}$}
In this section we reduce the previous system down to equations which depend solely on the $c_{i}$'s. We find that
\newline
$g_{64} = g_{5} = g_{65} = g_{6} = g_{61} = g_{2} = g_{10} = g_{69} = g_{70} = g_{11} = G_{4}(t) = G_{9}(t) = 0, g_{59} = g_{35}, g_{60} = g_{1}, p_{1} = \frac{H(t)}{c_{10}}$

\begin{eqnarray*}
&& g_{23} = -p_{1}c_{5} - g_{17} \\
&& g_{47} = -p_{1}c_{5} - g_{17} \\
&& g_{42} = g_{18} = -p_{1}c_{6} - 2g_{17} \\
&& g_{40} = g_{16} = (p_{1}c_{7})_{x} \\
&& g_{22} = g_{17x} - 2g_{12} + (p_{1}c_{5})_{x} \\
&& g_{46} = g_{17x} - 2g_{12} + (p_{1}c_{5})_{x} \\
&& g_{14} = g_{12x} - \frac{1}{2}g_{17xx} - g_{13} - \frac{1}{2}(p_{1}c_{5})_{xx} \\
&& g_{52} = g_{28} =  -(p_{1}c_{1})_{xx} \\
&& g_{53} = g_{29} = (p_{1}c_{1})_{x} \\
&& g_{36} = g_{12} = (p_{1}c_{2})_{x} \\
&& g_{55} = g_{31} = \frac{1}{3}(2g_{12} - 2g_{17x} - (p_{1}c_{5})_{x} \\
&& g_{38} = -g_{13} + g_{12x} -\frac{1}{2}g_{17xx} - \frac{1}{2}(p_{1}c_{5})_{xx} \\
&& g_{41} = g_{17} = p_{1}(c_{3} - 2c_{2} - c_{5}) \\
&& g_{i} = -g_{ix} \ \ (i=24-27,48-51) \\
&& g_{39} = g_{15} = -\frac{1}{2}g_{13x} \\
&& g_{56} = g_{32} = -\frac{1}{3}g_{15x} \\
&& g_{57} = g_{33} = G_{33}(t) \\
&& g_{37} = g_{13} = \frac{1}{3}(p_{1}(2c_{3} - c_{5} - 6c_{2}))_{xx}
\end{eqnarray*}

which leads to

\begin{eqnarray*}
&& c_{4} = -10c_{2} + 5c_{3} + 2c_{6} - 4c_{5} \\
&& c_{8} = 4c_{7} + \frac{2}{3}c_{9}
\end{eqnarray*}

\noindent
and the set of equations presented earlier. Here $H(t)$ is an arbitrary function of $t$.

\newpage
\setcounter{equation}{0}
\renewcommand{\theequation}{B.\arabic{equation}}

\section{Appendix: Lax integrability conditions for Generalized KdV Equation}

{\it As mentioned in the text, the notation and calculations here refer to the treatment of
the generalized vcKdV equation of Section 3 ONLY.}

The Lax pair for the generalized vcKdV equation is expanded in powers of $u$ and its derivatives as follows:

\begin{eqnarray}
\textbf{U} &=& \begin{bmatrix}
	f_{1}+f_{2}u & f_{3}+f_{4}u  \\
	 f_{5}+f_{6}u & f_{7}+f_{8}u
\end{bmatrix} \\
\textbf{V} &=& \begin{bmatrix}
	V_{1} & V_{2}  \\
	V_{3} & V_{4}
\end{bmatrix}
\end{eqnarray}
where $V_{i} = g_{k} + g_{k+1}u + g_{k+2}u^{2} + g_{k+3}u^{3} + g_{k+4}u_{x} + g_{k+5}u_{x}^{2} + g_{k+6}u_{xx} + g_{k+7}uu_{xx} + g_{k+8}u_{xxx} + g_{k+9}uu_{xxx} + g_{k+10}u_{xxxx}$, $k = 11(i-1) + 1$ and $f_{1-8}(x,t)$ and $g_{1-44}(x,t)$ are unknown functions. The compatibility condition
\begin{equation}
U_{t} - V_{x} + [U,V] = \dot{0} = \begin{bmatrix}
	0 & p_{1}(x,t)F[u]  \\
	p_{2}(x,t)F[u] & 0
\end{bmatrix}
\end{equation}
where $F[u]$ represents the $\eqref{KDV}$ and $p_{1-2}(x,t)$ are unknown functions, require
\begin{eqnarray*}
&& g_{21}=g_{32}=f_{2}=g_{4}=g_{10}=g_{11}=f_{8}=g_{37}=g_{43}=g_{44}=0,f_{4}=p_{1},g_{15}=-\frac{1}{3}p_{1}a_{3}, \\
&& g_{21}=-p_{1}a_{6},f_{6}=p_{2},g_{26}=-\frac{1}{3}p_{2}a_{6},g_{33}=-p_{2}a_{6},
\end{eqnarray*}
\begin{eqnarray}
&& g_{41} - g_{8} = 0 \\
&& g_{39} - g_{6} = 0 \\
&& p_{2}g_{17} - p_{1}g_{28} = 0 \\
&& p_{2}g_{19} - p_{1}g_{30} = 0 \\
&& g_{19} + 2g_{17} = -p_{1}a_{2} \\
&& g_{30} + 2g_{28} = -p_{2}a_{2} \\
&& g_{41} + 2g_{39} = 0 \\
&& g_{8} + 2g_{6} = 0 \\
&& g_{39x} + f_{3}g_{28} - f_{5}g_{17} = 0 \\
&& g_{6x} + f_{5}g_{17} - f_{3}g_{28} = 0 \\
&& 2g_{25} + p_{2}(g_{38} - g_{5}) = -p_{2}a_{4} \\
\end{eqnarray}
\begin{eqnarray}
&& g_{19} + p_{1}(g_{9} - g_{42}) = -p_{1}a_{1} \\
&& g_{8} + p_{2}g_{20} - p_{1}g_{31} = 0 \\
&& g_{41} + p_{1}g_{31} - p_{2}g_{20} = 0 \\
&& g_{30} + p_{2}(g_{42} - g_{9}) = -p_{2}a_{1} \\
&& g_{9} + a_{6}(p_{2}f_{3} - p_{1}f_{5}) = 0 \\
&& g_{42} + a_{6}(p_{1}f_{5} - p_{2}f_{3}) = 0 \\
&& 2g_{14} + p_{1}(g_{5} - g_{38}) = -p_{1}a_{4} \\
&& 2g_{3} + p_{2}g_{16} - p_{1}g_{27} = 0 \\
&& 2g_{36} + p_{1}g_{27} - p_{2}g_{16} = 0 \\
&& g_{2} + g_{5x} + f_{5}g_{16} - f_{3}g_{27} = 0 \\
&& g_{35} + g_{38x} + f_{3}g_{27} - f_{5}g_{16} = 0 \\
&& g_{7} + g_{9x} + f_{5}g_{20} - f_{3}g_{31} = 0 \\
\end{eqnarray}
\begin{eqnarray}
&& p_{1}\left(g_{25} + \frac{1}{3}f_{5}a_{3}\right) - p_{2}\left(g_{14} + \frac{1}{3}f_{3}a_{3}\right) = 0 \\
&& f_{1t} - g_{1x} + f_{3}g_{23} - f_{5}g_{12} = 0 \\
&& f_{7t} - g_{34x} + f_{5}g_{12} - f_{3}g_{23} = 0 \\
&& g_{5} + g_{7x} + f_{5}g_{18} - f_{3}g_{29} = 0 \\
&& g_{38} + g_{40x} + f_{3}g_{29} - f_{5}g_{18} = 0 \\
&& g_{40} + g_{42x} + f_{3}g_{31} - f_{5}g_{20} = 0 \\
&& (p_{2}a_{6})_{x} - g_{31} + p_{2}a_{6}(f_{1} - f_{7}) = 0 \\
&& g_{35x} + p_{1}g_{23} + f_{3}g_{24} - p_{2}g_{12} - f_{5}g_{13} = 0 \\
&& g_{36x} + p_{1}g_{24} + f_{3}g_{25} - p_{2}g_{13} - f_{5}g_{14} = 0 \\
&& g_{2x} + p_{2}g_{12} + f_{5}g_{13} - p_{1}g_{23} - f_{3}g_{24} = 0
\end{eqnarray}
\begin{eqnarray}
&& g_{3x} + p_{2}g_{13} + f_{5}g_{14} - p_{1}g_{24} - f_{3}g_{25} = 0 \\
&& g_{8x} + p_{2}g_{18} + f_{5}g_{19} - p_{1}g_{29} - f_{3}g_{30} = 0 \\
&& g_{41x} + p_{1}g_{29} + f_{3}g_{30} - p_{2}g_{18} - f_{5}g_{19} = 0 \\
&& g_{28x} + g_{28}(f_{1} - f_{7}) + f_{5}(g_{39} - g_{6}) = 0 \\
&& g_{17x} + g_{17}(f_{7} - f_{1}) + f_{3}(g_{6} - g_{39}) = 0 \\
&& g_{20} - (p_{1}a_{6})_{x} + p_{1}a_{6}(f_{1} - f_{7}) = 0 \\
&& \frac{1}{3}(p_{2}a_{3})_{x} + \frac{1}{3}p_{2}a_{3}(f_{1} - f_{7}) + p_{2}(g_{3} - g_{36}) = 0 \\
&& f_{5t} - g_{23x} + g_{23}(f_{7} - f_{1}) + f_{5}(g_{1} - g_{34}) = 0 \\
\end{eqnarray}
\begin{eqnarray}
&& g_{24} + g_{27x} + g_{27}(f_{1} - f_{7}) + f_{5}(g_{38} - g_{5}) = -p_{2}a_{8} \\
&& f_{3t} - g_{12x} + g_{12}(f_{1} - f_{7}) + f_{3}(g_{34} - g_{1}) = 0 \\
&& g_{29} + g_{31x} + g_{31}(f_{1} - f_{7}) + f_{5}(g_{42} - g_{9}) = -p_{2}a_{5} \\
&& \frac{1}{3}(p_{1}a_{3})_{x} + \frac{1}{3}p_{1}a_{3}(f_{7} - f_{1}) + p_{1}(g_{36} - g_{3}) = 0 \\
&& g_{18} + g_{20x} + g_{20}(f_{7} - f_{1}) + f_{3}(g_{9} - g_{42}) = -p_{1}a_{5} \\
&& g_{13} + g_{16x} + g_{16}(f_{7} - f_{1}) + f_{3}(g_{5} - g_{38}) = -p_{1}a_{8} \\
&& g_{16} + g_{18x} + g_{18}(f_{7} - f_{1}) + f_{3}(g_{7} - g_{40}) = 0 \\
&& g_{27} + g_{29x} + g_{29}(f_{1} - f_{7}) + f_{5}(g_{40} - g_{7}) = 0 \\
\end{eqnarray}
\begin{eqnarray}
&& g_{19x} + g_{19}(f_{7} - f_{1}) + p_{1}(g_{7} - g_{40}) + f_{3}(g_{8} - g_{41}) = 0 \\
&& g_{30x} + g_{30}(f_{1} - f_{7}) + p_{2}(g_{40} - g_{7}) + f_{5}(g_{41} - g_{8}) = 0 \\
&& g_{14x} + g_{14}(f_{7} - f_{1}) + p_{1}(g_{2} - g_{35}) + f_{3}(g_{3} - g_{36}) = 0 \\
&& g_{25x} + g_{25}(f_{1} - f_{7}) + p_{2}(g_{35} - g_{2}) + f_{5}(g_{36} - g_{3}) = 0 \\
&& p_{1t} - g_{13x} + g_{13}(f_{1} - f_{7}) + p_{1}(g_{34} - g_{1}) + f_{3}(g_{35} - g_{2}) = p_{1}a_{7} \\
&& p_{2t} - g_{24x} + g_{24}(f_{7} - f_{1}) + p_{2}(g_{1} - g_{34}) + f_{5}(g_{2} - g_{35}) = p_{2}a_{7}
\end{eqnarray}

\subsubsection{Deriving a relation between the $a_{i}$}
In this section we reduce the previous system down to equations which depend solely on the $a_{i}$'s. We find that
\newline
$g_{42} = g_{9} = g_{8} = g_{41} = g_{39} = g_{6} = g_{40} = g_{38} = g_{7} = g_{5} = g_{35} = g_{2} = g_{36} = g_{3} = 0,f_{7} = f_{1},f_{5} = f_{3},g_{23} = g_{12},g_{34} = g_{1},g_{30} = - g_{19} = p_{1}a_{1},g_{25} = g_{14} = - \frac{1}{2}p_{1}a_{4},g_{31} = - g_{20} =  - (p_{1}a_{6})_{x},g_{16} = - g_{27} = - g_{18x}, p_{2} = p_{1}$
\begin{eqnarray*}
&& g_{18} = - g_{29} = - p_{1}a_{5} - (p_{1}a_{6})_{xx} \\
&& g_{28} = - g_{17} = - \frac{1}{2}(H_{2}(t) - H_{1}(t)) \\
&& g_{24} = - g_{13} = -(p_{1}a_{5})_{xx} - (p_{1}a_{6})_{xxxx} - p_{1}a_{8} \\
&& p_{1} = \frac{H_{1}(t)}{a_{1}} \\
&& a_{2-4} = H_{2-4}(t)a_{1}
\end{eqnarray*}
which leads to the PDE

\begin{equation}
\left(\frac{H_{1}}{a_{1}}\right)_{t} + \left(\frac{H_{1}a_{5}}{a_{1}}\right)_{xxx} + \left(\frac{H_{1}a_{6}}{a_{1}}\right)_{xxxxx} + \left(\frac{H_{1}a_{8}}{a_{1}}\right)_{x} = \frac{H_{1}a_{7}}{a_{1}}
\end{equation}

for which one clear solution (for $H_{1} \neq 0$) is 

\begin{equation}
a_{7} = \frac{a_{1}}{H_{1}}\left(\left(\frac{H_{1}}{a_{1}}\right)_{t} + \left(\frac{H_{1}a_{5}}{a_{1}}\right)_{xxx} + \left(\frac{H_{1}a_{6}}{a_{1}}\right)_{xxxxx} + \left(\frac{H_{1}a_{8}}{a_{1}}\right)_{x}\right)
\end{equation}

where $a_{1},a_{5},a_{6},a_{8}$ and $H_{1-4}$ are arbitrary functions in their respective variables.

\newpage
\setcounter{equation}{0}
\renewcommand{\theequation}{C.\arabic{equation}}

\section{Appendix:  Intermediate Results for Fifth-Order Equation}

The intermediate results mentioned at the appropriate places in Section $3$ are given here, with the derivation
and the use of each detailed there. These intermediate results are:

\begin{eqnarray} 
&& \mathbb{X}_{1,t} + \mathbb{X}_{2,t}u - \mathbb{X}_{2}(a_{3}u^{2}u_{x} + a_{4}uu_{x} + a_{7}u + a_{8}u_{x}) - (a_{6}\mathbb{X}_{6})_{x}u_{x}^{2} + a_{1}\mathbb{X}_{4}u_{x}^{2} \nonumber \\
&& - (a_{6}\mathbb{X}_{2})_{xxxx}u_{x} - \frac{1}{2}(a_{1}\mathbb{X}_{2})_{x}u_{x}^{2} + \frac{1}{2}([\mathbb{X}_{2},(a_{6}\mathbb{X}_{2})_{x}])_{x}u_{x}^{2} + \frac{1}{2}(a_{2}\mathbb{X}_{2})_{x}u_{x}^{2} \nonumber \\
&& + (a_{6}\mathbb{X}_{4})_{xxx}u_{x} + ([\mathbb{X}_{1},(a_{6}\mathbb{X}_{2})_{x}])_{xx}u_{x} + ([\mathbb{X}_{2},(a_{6}\mathbb{X}_{2})_{x}])_{xx}uu_{x} - (a_{6}\mathbb{X}_{5})_{xx}u_{x} \nonumber \\
&& + (a_{1}\mathbb{X}_{4})_{x}uu_{x} - \frac{1}{2}(a_{6}\mathbb{X}_{6})_{x}u_{x}^{2} - ([\mathbb{X}_{2},(a_{6}\mathbb{X}_{4})_{x}])_{x}uu_{x} - (a_{5}\mathbb{X}_{2})_{xx}u_{x} - (a_{1}\mathbb{X}_{2})_{xx}uu_{x} \nonumber \\
&& + (a_{5}\mathbb{X}_{4})_{x}u_{x} + ([\mathbb{X}_{1},(a_{6}\mathbb{X}_{2})_{xx}])_{x}u_{x} + ([\mathbb{X}_{2},(a_{6}\mathbb{X}_{2})_{xx}])_{x}uu_{x} - ([\mathbb{X}_{1},(a_{6}\mathbb{X}_{4})_{x}])_{x}u_{x} \nonumber \\
&& - ([\mathbb{X}_{1},[\mathbb{X}_{1},(a_{6}\mathbb{X}_{2})_{x}]])_{x}u_{x} - ([\mathbb{X}_{1},[\mathbb{X}_{2},(a_{6}\mathbb{X}_{2})_{x}]])_{x}uu_{x} - ([\mathbb{X}_{2},[\mathbb{X}_{1},(a_{6}\mathbb{X}_{2})_{x}]])_{x}uu_{x} \nonumber \\
&& - ([\mathbb{X}_{2},[\mathbb{X}_{2},(a_{6}\mathbb{X}_{2})_{x}]])_{x}u^{2}u_{x} + (a_{6}\mathbb{X}_{10})_{x}u^{2}u_{x} - \mathbb{K}^{3}_{x} + ([\mathbb{X}_{2},(a_{6}\mathbb{X}_{2})_{x}])_{x}u_{x}^{2} + a_{6}\mathbb{X}_{9}u_{x}^{2} \nonumber \\
&& - [\mathbb{X}_{2},(a_{6}\mathbb{X}_{4})_{x}]u_{x}^{2} + [\mathbb{X}_{2},(a_{6}\mathbb{X}_{2})_{xx}]u_{x}^{2} + a_{6}\mathbb{X}_{8}u_{x}^{2} + (a_{6}\mathbb{X}_{9})_{x}uu_{x} + (a_{6}\mathbb{X}_{8})_{x}uu_{x} \nonumber \\
&& - [\mathbb{X}_{1},[\mathbb{X}_{2},(a_{6}\mathbb{X}_{2})_{x}]]u_{x}^{2} - [\mathbb{X}_{2},[\mathbb{X}_{1},(a_{6}\mathbb{X}_{2})_{x}]]u_{x}^{2} - (a_{1}\mathbb{X}_{2})_{x}u_{x}^{2} - (a_{6}\mathbb{X}_{6})_{xx}uu_{x} \nonumber \\
&& - 2[\mathbb{X}_{2},[\mathbb{X}_{2},(a_{6}\mathbb{X}_{2})_{x}]]uu_{x}^{2} + 2a_{6}\mathbb{X}_{10}uu_{x}^{2} - \mathbb{K}^{3}_{u}u_{x} + [\mathbb{X}_{1},(a_{6}\mathbb{X}_{5})_{x}]u_{x} - \frac{1}{2}a_{2}\mathbb{X}_{4}u_{x}^{2} \nonumber \\
&& + [\mathbb{X}_{1},(a_{6}\mathbb{X}_{2})_{xxx}]u_{x} + \frac{1}{2}a_{1}\mathbb{X}_{4}u_{x}^{2} - \frac{1}{2}[\mathbb{X}_{1},[\mathbb{X}_{2},(a_{6}\mathbb{X}_{2})_{x}]]u_{x}^{2} + [\mathbb{X}_{1},(a_{6}\mathbb{X}_{6})_{x}]uu_{x} \nonumber \\
&& - [\mathbb{X}_{1},(a_{6}\mathbb{X}_{4})_{xx}]u_{x} - [\mathbb{X}_{1},([\mathbb{X}_{1},(a_{6}\mathbb{X}_{2})_{x}])_{x}]u_{x} - [\mathbb{X}_{1},([\mathbb{X}_{2},(a_{6}\mathbb{X}_{2})_{x}])_{x}]uu_{x} \nonumber \\
&& - a_{1}\mathbb{X}_{5}uu_{x} + \frac{1}{2}a_{6}\mathbb{X}_{9}u_{x}^{2} + [\mathbb{X}_{1},[\mathbb{X}_{2},(a_{6}\mathbb{X}_{4})_{x}]]uu_{x} + [\mathbb{X}_{1},(a_{5}\mathbb{X}_{2})_{x}]u_{x} + (a_{6}\mathbb{X}_{7})_{x}u_{x} \nonumber \\
&& - a_{5}\mathbb{X}_{5}u_{x} - [\mathbb{X}_{1},[\mathbb{X}_{1},(a_{6}\mathbb{X}_{2})_{xx}]]u_{x} - [\mathbb{X}_{1},[\mathbb{X}_{2},(a_{6}\mathbb{X}_{2})_{xx}]]uu_{x} + [\mathbb{X}_{1},[\mathbb{X}_{1},(a_{6}\mathbb{X}_{4})_{x}]]u_{x} \nonumber \\
&& + [\mathbb{X}_{1},[\mathbb{X}_{1},[\mathbb{X}_{1},(a_{6}\mathbb{X}_{2})_{x}]]]u_{x} + [\mathbb{X}_{1},[\mathbb{X}_{1},[\mathbb{X}_{2},(a_{6}\mathbb{X}_{2})_{x}]]]uu_{x} + [\mathbb{X}_{1},[\mathbb{X}_{2},[\mathbb{X}_{1},(a_{6}\mathbb{X}_{2})_{x}]]]uu_{x} \nonumber \\
&& - a_{6}[\mathbb{X}_{1},\mathbb{X}_{7}]u_{x} - a_{6}[\mathbb{X}_{1},\mathbb{X}_{9}]uu_{x} - a_{6}[\mathbb{X}_{1},\mathbb{X}_{8}]uu_{x} + [\mathbb{X}_{2},[\mathbb{X}_{1},(a_{6}\mathbb{X}_{4})_{x}]]uu_{x} \nonumber \\
&& + [\mathbb{X}_{1},[\mathbb{X}_{2},[\mathbb{X}_{2},(a_{6}\mathbb{X}_{2})_{x}]]]u^{2}u_{x} - a_{6}[\mathbb{X}_{1},\mathbb{X}_{10}]u^{2}u_{x} + [\mathbb{X}_{1},\mathbb{K}^{3}] + [\mathbb{X}_{1},(a_{1}\mathbb{X}_{2})_{x}]uu_{x} \nonumber \\
&& + [\mathbb{X}_{2},(a_{6}\mathbb{X}_{2})_{xxx}]uu_{x} - \frac{1}{2}[\mathbb{X}_{2},[\mathbb{X}_{2},(a_{6}\mathbb{X}_{2})_{x}]]uu_{x}^{2} + [\mathbb{X}_{2},(a_{6}\mathbb{X}_{6})_{x}]u^{2}u_{x} + [\mathbb{X}_{2},(a_{6}\mathbb{X}_{5})_{x}]uu_{x} \nonumber \\
&& - [\mathbb{X}_{2},(a_{6}\mathbb{X}_{4})_{xx}]uu_{x} - [\mathbb{X}_{2},([\mathbb{X}_{1},(a_{6}\mathbb{X}_{2})_{x}])_{x}]uu_{x} - [\mathbb{X}_{2},([\mathbb{X}_{2},(a_{6}\mathbb{X}_{2})_{x}])_{x}]u^{2}u_{x} \nonumber \\
&& - a_{1}\mathbb{X}_{6}u^{2}u_{x} + \frac{1}{2}a_{6}\mathbb{X}_{10}uu_{x}^{2} + [\mathbb{X}_{2},[\mathbb{X}_{2},(a_{6}\mathbb{X}_{4})_{x}]]u^{2}u_{x} + [\mathbb{X}_{2},(a_{5}\mathbb{X}_{2})_{x}]uu_{x} \nonumber \\
&& - a_{5}\mathbb{X}_{6}uu_{x} - [\mathbb{X}_{2},[\mathbb{X}_{1},(a_{6}\mathbb{X}_{2})_{xx}]]uu_{x} - [\mathbb{X}_{2},[\mathbb{X}_{2},(a_{6}\mathbb{X}_{2})_{xx}]]u^{2}u_{x} \nonumber \\
&& + [\mathbb{X}_{2},[\mathbb{X}_{1},[\mathbb{X}_{1},(a_{6}\mathbb{X}_{2})_{x}]]]uu_{x} + [\mathbb{X}_{2},[\mathbb{X}_{1},[\mathbb{X}_{2},(a_{6}\mathbb{X}_{2})_{x}]]]u^{2}u_{x} + [\mathbb{X}_{2},(a_{1}\mathbb{X}_{2})_{x}]u^{2}u_{x} \nonumber \\
&& + [\mathbb{X}_{2},[\mathbb{X}_{2},[\mathbb{X}_{1},(a_{6}\mathbb{X}_{2})_{x}]]]u^{2}u_{x} - a_{6}[\mathbb{X}_{2},\mathbb{X}_{7}]uu_{x} - a_{6}[\mathbb{X}_{2},\mathbb{X}_{9}]u^{2}u_{x} - a_{6}[\mathbb{X}_{2},\mathbb{X}_{8}]u^{2}u_{x} \nonumber \\
&& + [\mathbb{X}_{2},[\mathbb{X}_{2},[\mathbb{X}_{2},(a_{6}\mathbb{X}_{2})_{x}]]]u^{3}u_{x} - a_{6}[\mathbb{X}_{2},\mathbb{X}_{10}]u^{3}u_{x} + [\mathbb{X}_{2},\mathbb{K}^{3}]u = 0, \label{KDV9}
\end{eqnarray}

\begin{eqnarray}
&& - \mathbb{X}_{2}(a_{3}u^{2} + a_{4}u + a_{8}) - (a_{6}\mathbb{X}_{2})_{xxxx} - \mathbb{K}^{3}_{u} + [\mathbb{X}_{1},(a_{6}\mathbb{X}_{5})_{x}] \nonumber \\
&& + (a_{6}\mathbb{X}_{4})_{xxx} + ([\mathbb{X}_{1},(a_{6}\mathbb{X}_{2})_{x}])_{xx} + ([\mathbb{X}_{2},(a_{6}\mathbb{X}_{2})_{x}])_{xx}u - (a_{6}\mathbb{X}_{5})_{xx} \nonumber \\
&& + (a_{1}\mathbb{X}_{4})_{x}u - ([\mathbb{X}_{2},(a_{6}\mathbb{X}_{4})_{x}])_{x}u - (a_{5}\mathbb{X}_{2})_{xx} - (a_{1}\mathbb{X}_{2})_{xx}u - a_{1}\mathbb{X}_{6}u^{2} \nonumber \\
&& + (a_{5}\mathbb{X}_{4})_{x} + ([\mathbb{X}_{1},(a_{6}\mathbb{X}_{2})_{xx}])_{x} + ([\mathbb{X}_{2},(a_{6}\mathbb{X}_{2})_{xx}])_{x}u - ([\mathbb{X}_{1},(a_{6}\mathbb{X}_{4})_{x}])_{x} \nonumber \\
&& - ([\mathbb{X}_{1},[\mathbb{X}_{1},(a_{6}\mathbb{X}_{2})_{x}]])_{x} - ([\mathbb{X}_{1},[\mathbb{X}_{2},(a_{6}\mathbb{X}_{2})_{x}]])_{x}u - ([\mathbb{X}_{2},[\mathbb{X}_{1},(a_{6}\mathbb{X}_{2})_{x}]])_{x}u \nonumber \\
&& + (a_{6}\mathbb{X}_{9})_{x}u + (a_{6}\mathbb{X}_{8})_{x}u - (a_{6}\mathbb{X}_{6})_{xx}u + [\mathbb{X}_{1},(a_{6}\mathbb{X}_{2})_{xxx}] + [\mathbb{X}_{1},(a_{6}\mathbb{X}_{6})_{x}]u \nonumber \\
&& - [\mathbb{X}_{1},(a_{6}\mathbb{X}_{4})_{xx}] - [\mathbb{X}_{1},([\mathbb{X}_{1},(a_{6}\mathbb{X}_{2})_{x}])_{x}] - [\mathbb{X}_{1},([\mathbb{X}_{2},(a_{6}\mathbb{X}_{2})_{x}])_{x}]u \nonumber \\
&& - a_{1}\mathbb{X}_{5}u + [\mathbb{X}_{1},[\mathbb{X}_{2},(a_{6}\mathbb{X}_{4})_{x}]]u + [\mathbb{X}_{1},(a_{5}\mathbb{X}_{2})_{x}] + (a_{6}\mathbb{X}_{7})_{x} + [\mathbb{X}_{2},[\mathbb{X}_{2},(a_{6}\mathbb{X}_{4})_{x}]]u^{2} \nonumber \\
&& - a_{5}\mathbb{X}_{5} - [\mathbb{X}_{1},[\mathbb{X}_{1},(a_{6}\mathbb{X}_{2})_{xx}]] - [\mathbb{X}_{1},[\mathbb{X}_{2},(a_{6}\mathbb{X}_{2})_{xx}]]u + [\mathbb{X}_{1},[\mathbb{X}_{1},(a_{6}\mathbb{X}_{4})_{x}]] \nonumber \\
&& + [\mathbb{X}_{1},[\mathbb{X}_{1},[\mathbb{X}_{1},(a_{6}\mathbb{X}_{2})_{x}]]] + [\mathbb{X}_{1},[\mathbb{X}_{1},[\mathbb{X}_{2},(a_{6}\mathbb{X}_{2})_{x}]]]u + [\mathbb{X}_{1},[\mathbb{X}_{2},[\mathbb{X}_{1},(a_{6}\mathbb{X}_{2})_{x}]]]u \nonumber \\
&& - a_{6}[\mathbb{X}_{1},\mathbb{X}_{7}] - a_{6}[\mathbb{X}_{1},\mathbb{X}_{9}]u - a_{6}[\mathbb{X}_{1},\mathbb{X}_{8}]u + [\mathbb{X}_{2},[\mathbb{X}_{1},(a_{6}\mathbb{X}_{4})_{x}]]u \nonumber \\
&& + [\mathbb{X}_{2},(a_{6}\mathbb{X}_{2})_{xxx}]u + [\mathbb{X}_{2},(a_{6}\mathbb{X}_{6})_{x}]u^{2} + [\mathbb{X}_{2},(a_{6}\mathbb{X}_{5})_{x}]u + [\mathbb{X}_{1},(a_{1}\mathbb{X}_{2})_{x}]u \nonumber \\
&& - [\mathbb{X}_{2},(a_{6}\mathbb{X}_{4})_{xx}]u - [\mathbb{X}_{2},([\mathbb{X}_{1},(a_{6}\mathbb{X}_{2})_{x}])_{x}]u - [\mathbb{X}_{2},([\mathbb{X}_{2},(a_{6}\mathbb{X}_{2})_{x}])_{x}]u^{2} \nonumber \\
&& - a_{5}\mathbb{X}_{6}u - [\mathbb{X}_{2},[\mathbb{X}_{1},(a_{6}\mathbb{X}_{2})_{xx}]]u - [\mathbb{X}_{2},[\mathbb{X}_{2},(a_{6}\mathbb{X}_{2})_{xx}]]u^{2} + [\mathbb{X}_{2},(a_{5}\mathbb{X}_{2})_{x}]u \nonumber \\
&& + [\mathbb{X}_{2},[\mathbb{X}_{1},[\mathbb{X}_{1},(a_{6}\mathbb{X}_{2})_{x}]]]u + [\mathbb{X}_{2},[\mathbb{X}_{1},[\mathbb{X}_{2},(a_{6}\mathbb{X}_{2})_{x}]]]u^{2} + [\mathbb{X}_{2},(a_{1}\mathbb{X}_{2})_{x}]u^{2} \nonumber \\
&& + [\mathbb{X}_{2},[\mathbb{X}_{2},[\mathbb{X}_{1},(a_{6}\mathbb{X}_{2})_{x}]]]u^{2} - a_{6}[\mathbb{X}_{2},\mathbb{X}_{7}]u - a_{6}[\mathbb{X}_{2},\mathbb{X}_{9}]u^{2} - a_{6}[\mathbb{X}_{2},\mathbb{X}_{8}]u^{2} = 0, \label{KDV13}
\end{eqnarray}

\noindent
and

\begin{eqnarray}
\mathbb{K}^{3} &=& - \frac{1}{3}a_{3}\mathbb{X}_{2}u^{3} - \frac{1}{2}a_{4}\mathbb{X}_{2}u^{2} - a_{8}\mathbb{X}_{2}u - (a_{6}\mathbb{X}_{2})_{xxxx}u + [\mathbb{X}_{1},(a_{6}\mathbb{X}_{5})_{x}]u - \frac{1}{2}a_{6}[\mathbb{X}_{2},\mathbb{X}_{7}]u^{2} \nonumber \\
&& + (a_{6}\mathbb{X}_{4})_{xxx}u + ([\mathbb{X}_{1},(a_{6}\mathbb{X}_{2})_{x}])_{xx}u + \frac{1}{2}([\mathbb{X}_{2},(a_{6}\mathbb{X}_{2})_{x}])_{xx}u^{2} - (a_{6}\mathbb{X}_{5})_{xx}u \nonumber \\
&& + \frac{1}{2}(a_{1}\mathbb{X}_{4})_{x}u^{2} - \frac{1}{2}([\mathbb{X}_{2},(a_{6}\mathbb{X}_{4})_{x}])_{x}u^{2} - (a_{5}\mathbb{X}_{2})_{xx}u - \frac{1}{2}(a_{1}\mathbb{X}_{2})_{xx}u^{2} - \frac{1}{3}a_{1}\mathbb{X}_{6}u^{3} \nonumber \\
&& + (a_{5}\mathbb{X}_{4})_{x}u + ([\mathbb{X}_{1},(a_{6}\mathbb{X}_{2})_{xx}])_{x}u + \frac{1}{2}([\mathbb{X}_{2},(a_{6}\mathbb{X}_{2})_{xx}])_{x}u^{2} - ([\mathbb{X}_{1},(a_{6}\mathbb{X}_{4})_{x}])_{x}u \nonumber \\
&& - ([\mathbb{X}_{1},[\mathbb{X}_{1},(a_{6}\mathbb{X}_{2})_{x}]])_{x}u - \frac{1}{2}([\mathbb{X}_{1},[\mathbb{X}_{2},(a_{6}\mathbb{X}_{2})_{x}]])_{x}u^{2} - \frac{1}{2}([\mathbb{X}_{2},[\mathbb{X}_{1},(a_{6}\mathbb{X}_{2})_{x}]])_{x}u^{2} \nonumber \\
&& + \frac{1}{2}(a_{6}\mathbb{X}_{9})_{x}u^{2} + \frac{1}{2}(a_{6}\mathbb{X}_{8})_{x}u^{2} - \frac{1}{2}(a_{6}\mathbb{X}_{6})_{xx}u^{2} + [\mathbb{X}_{1},(a_{6}\mathbb{X}_{2})_{xxx}]u + \frac{1}{2}[\mathbb{X}_{1},(a_{6}\mathbb{X}_{6})_{x}]u^{2} \nonumber \\
&& - [\mathbb{X}_{1},(a_{6}\mathbb{X}_{4})_{xx}]u - [\mathbb{X}_{1},([\mathbb{X}_{1},(a_{6}\mathbb{X}_{2})_{x}])_{x}]u - \frac{1}{2}[\mathbb{X}_{1},([\mathbb{X}_{2},(a_{6}\mathbb{X}_{2})_{x}])_{x}]u^{2} - \frac{1}{2}a_{1}\mathbb{X}_{5}u^{2} \nonumber \\
&& + \frac{1}{2}[\mathbb{X}_{1},[\mathbb{X}_{2},(a_{6}\mathbb{X}_{4})_{x}]]u^{2} + [\mathbb{X}_{1},(a_{5}\mathbb{X}_{2})_{x}]u + (a_{6}\mathbb{X}_{7})_{x}u + \frac{1}{3}[\mathbb{X}_{2},[\mathbb{X}_{2},(a_{6}\mathbb{X}_{4})_{x}]]u^{3} \nonumber \\
&& - a_{5}\mathbb{X}_{5}u - [\mathbb{X}_{1},[\mathbb{X}_{1},(a_{6}\mathbb{X}_{2})_{xx}]]u - \frac{1}{2}[\mathbb{X}_{1},[\mathbb{X}_{2},(a_{6}\mathbb{X}_{2})_{xx}]]u^{2} + [\mathbb{X}_{1},[\mathbb{X}_{1},(a_{6}\mathbb{X}_{4})_{x}]]u \nonumber \\
&& + [\mathbb{X}_{1},[\mathbb{X}_{1},[\mathbb{X}_{1},(a_{6}\mathbb{X}_{2})_{x}]]]u + \frac{1}{2}[\mathbb{X}_{1},[\mathbb{X}_{1},[\mathbb{X}_{2},(a_{6}\mathbb{X}_{2})_{x}]]]u^{2} + \frac{1}{2}[\mathbb{X}_{1},[\mathbb{X}_{2},[\mathbb{X}_{1},(a_{6}\mathbb{X}_{2})_{x}]]]u^{2} \nonumber \\
&& - a_{6}[\mathbb{X}_{1},\mathbb{X}_{7}]u - \frac{1}{2}a_{6}[\mathbb{X}_{1},\mathbb{X}_{9}]u^{2} - \frac{1}{2}a_{6}[\mathbb{X}_{1},\mathbb{X}_{8}]u^{2} + \frac{1}{2}[\mathbb{X}_{2},[\mathbb{X}_{1},(a_{6}\mathbb{X}_{4})_{x}]]u^{2} \nonumber \\
&& + \frac{1}{2}[\mathbb{X}_{2},(a_{6}\mathbb{X}_{2})_{xxx}]u^{2} + \frac{1}{3}[\mathbb{X}_{2},(a_{6}\mathbb{X}_{6})_{x}]u^{3} + \frac{1}{2}[\mathbb{X}_{2},(a_{6}\mathbb{X}_{5})_{x}]u^{2} + \frac{1}{2}[\mathbb{X}_{1},(a_{1}\mathbb{X}_{2})_{x}]u^{2} \nonumber \\
&& - \frac{1}{2}[\mathbb{X}_{2},(a_{6}\mathbb{X}_{4})_{xx}]u^{2} - \frac{1}{2}[\mathbb{X}_{2},([\mathbb{X}_{1},(a_{6}\mathbb{X}_{2})_{x}])_{x}]u^{2} - \frac{1}{3}[\mathbb{X}_{2},([\mathbb{X}_{2},(a_{6}\mathbb{X}_{2})_{x}])_{x}]u^{3} \nonumber \\
&& - \frac{1}{2}a_{5}\mathbb{X}_{6}u^{2} - \frac{1}{2}[\mathbb{X}_{2},[\mathbb{X}_{1},(a_{6}\mathbb{X}_{2})_{xx}]]u^{2} - \frac{1}{3}[\mathbb{X}_{2},[\mathbb{X}_{2},(a_{6}\mathbb{X}_{2})_{xx}]]u^{3} + \frac{1}{2}[\mathbb{X}_{2},(a_{5}\mathbb{X}_{2})_{x}]u^{2} \nonumber \\
&& + \frac{1}{2}[\mathbb{X}_{2},[\mathbb{X}_{1},[\mathbb{X}_{1},(a_{6}\mathbb{X}_{2})_{x}]]]u^{2} + \frac{1}{3}[\mathbb{X}_{2},[\mathbb{X}_{1},[\mathbb{X}_{2},(a_{6}\mathbb{X}_{2})_{x}]]]u^{3} + \frac{1}{3}[\mathbb{X}_{2},(a_{1}\mathbb{X}_{2})_{x}]u^{3} \nonumber \\
&& + \frac{1}{3}[\mathbb{X}_{2},[\mathbb{X}_{2},[\mathbb{X}_{1},(a_{6}\mathbb{X}_{2})_{x}]]]u^{3} - \frac{1}{3}a_{6}[\mathbb{X}_{2},\mathbb{X}_{9}]u^{3} - \frac{1}{3}a_{6}[\mathbb{X}_{2},\mathbb{X}_{8}]u^{3} + \mathbb{X}_{0}(x,t) \label{KDV14}
\end{eqnarray}

%\end{appendices}

\end{document}